\documentclass[aps,showpacs,amsmath,amssymb,twocolumn,pra]{revtex4-1}
\usepackage{bm}
\usepackage[dvips]{graphicx}
\usepackage{wrapfig,subfigure}
\usepackage{amssymb}
\usepackage{color}
\usepackage[normalem]{ulem}
\usepackage{amsmath}
\usepackage{epstopdf}
\def\rb{{\bf r}}
\def\pb{{\bf p}}
\def\ub{{\bf u}}

\def\eb{{\bf e}}
\def\vb{{\bf v}}
\def\rhob{{\bm \rho}}

\def\n{{\bm\nabla}}
\def\kb{{\bf k}}
\def\qb{{\bf q}}
\def\nb{{\bf n}}
\def\Rb{{\bf R}}
\def\Kb{{\bf K}}
\def\eh{{\hat\epsilon^{\rm (h)}}}
\def\es{{\hat\epsilon^{\rm (s)}}}
\def\eq{{\epsilon^{\rm (q)}}}
\def\ep{\varepsilon}

\begin{document}
\title{
Nonlinear damping and dephasing in nanomechanical systems}
\author{Juan Atalaya}
\altaffiliation{Current address: Department of Electrical Engineering, UC Riverside, Riverside, CA 92521}
\affiliation{Department of Physics and Astronomy, Michigan State University, East Lansing, Michigan 48824, USA }
\author{Thomas W. Kenny}
\affiliation{Department of Mechanical Engineering, Standford University, Stanford, California 94305, USA}
\author{M. L. Roukes}
\affiliation{Kavli Nanoscience Institute and Departments of Physics, Applied Physics, and Bioengineering, Caltech, Pasadena, California 91125, USA}
\author{M. I. Dykman}
\affiliation{Department of Physics and Astronomy, Michigan State University, East Lansing, Michigan 48824, USA}

\date{\today}

\begin{abstract}
We present a microscopic theory of nonlinear damping and dephasing of low-frequency eigenmodes in nano- and micro-mechanical systems. The mechanism of the both effects is scattering of thermally excited vibrational modes off the considered eigenmode. 
The scattering is accompanied by energy transfer of $2\hbar\omega_0$ for nonlinear damping and is quasieleastic for dephasing. We develop a formalism that allows studying both spatially uniform systems and systems with a strong nonuniformity, which is smooth on the typical wavelength of thermal modes but not their mean free path. The formalism accounts for the decay of thermal modes, which plays a major role in the nonlinear damping and dephasing. We identify the nonlinear analogs of the Landau-Rumer, thermoelastic, and Akhiezer mechanisms and find the dependence of the relaxation parameters on the temperature and the geometry of a system.

\end{abstract}


\maketitle

\section{Introduction}
\label{sec:Intro}

Nano- and micro-mechanical vibrational systems have been attracting much attention in recent years. Vibrational modes in such systems are not only  interesting on their own, but also provide a platform for studying  physics far from thermal equilibrium and quantum physics at the macroscale  \cite{Poot2012,Dykman2012b,Aspelmeyer2014a,Mahboob2013,Rimberg2014,Armour2015,Buters2015,Defoort2015,Cohen2015,Singh2016,Riedinger2016}. The typical spatial scale of the low-lying modes is the size of the system, and their frequencies are  $\sim 10^5 -10^8$~Hz. They have enabled record high force resolution \cite{Stowe1997,Rugar2004,Moser2013,Tao2014} and mass sensitivity \cite{Jensen2008,Chaste2012,Hanay2015} and have found applications in modern electronic devices such as accelerometers and gyroscopes. 

One of the most important characteristics of nano-scale vibrational modes is their decay rate, which can be much smaller than  the vibration frequency \cite{Poot2012,Dykman2012b,Aspelmeyer2014a,Stowe1997,Rugar2004,Moser2013,Tao2014,Jensen2008,Chaste2012,Hanay2015}. The relaxation mechanisms include radiation into the bulk modes of the medium surrounding the nanosystem \cite{Cross2001,Park2004,Judge2007,Wilson-Rae2008,Croy2012}, nonlinear coupling between the modes localized inside the system \cite{Lifshitz2000,DeMartino2009,Kunal2011,Ghaffari2015,Mahboob2015,Singh2016}, and coupling with electronic excitations\cite{Usmani2007,Steele2009,Lassagne2009,Lulla2013} and two-level defects \cite{Hoehne2010,Venkatesan2010,Faust2014}. 

Conventionally, decay of a vibrational mode is associated with an exponential fall-off of the vibration amplitude in time. This behavior is often observed in the experiment. In the phenomenological description of the mode dynamics, it comes from the friction force $-2\Gamma \dot q$, where $q$ is the vibration coordinate and $\Gamma$ is the friction coefficient. Such friction is called linear. Recent experiments have shown that, already for moderate vibration amplitudes, in various nano- and micromechanical systems the amplitude decay is nonexponential in time and/or the maximal amplitude of forced vibrations nonlinearly depends on the force amplitude \cite{Eichler2011a,Zaitsev2012,Imboden2013,Miao2014,Mahboob2015};  respectively, the friction force is a nonlinear function of the mode coordinate and velocity. Nonlinear friction (nonlinear damping) has recently attracted considerable attention also in the context of 
quantum information processing with microwave cavity modes, where it was engineered  to be strong \cite{Leghtas2015,A_Roy2015}.

\begin{figure}[t!]
\centering
\includegraphics[width=2.5in]{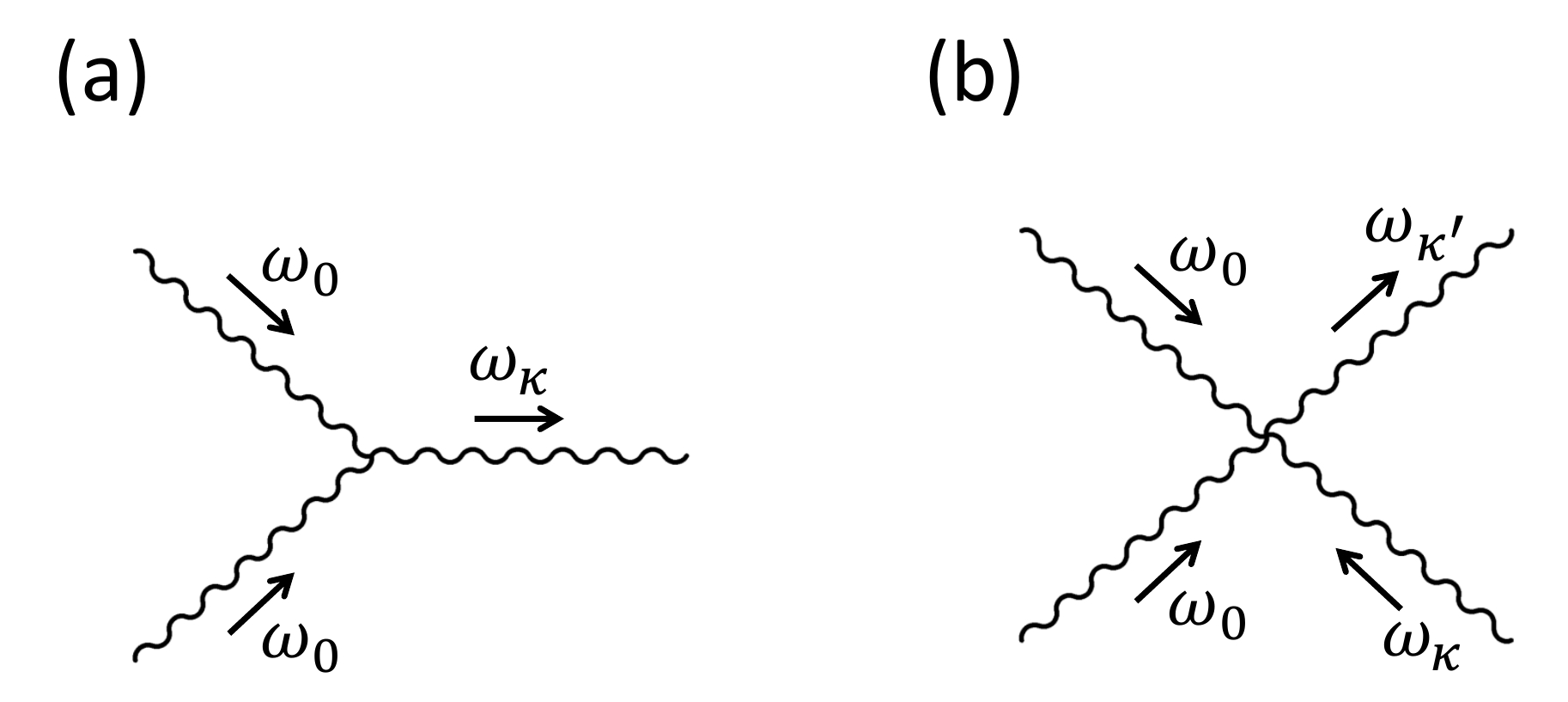}
\caption{Elementary phonon-phonon interaction processes leading to nonlinear damping. (a) Three phonon scattering: two quanta of the mode of interest, which has frequency $\omega_0$, scatter into a phonon of a continuous spectrum with frequency $\omega_{\kappa}\approx 2\omega_0$. (b) Four phonon scattering: a phonon of a continuous spectrum is scattered off the considered mode into another phonon, with the frequency change $\omega_{\kappa'}-\omega_{\kappa}\approx 2\omega_0$. The modes $\kappa,\kappa'$ are not assumed to be plane waves.}
 \label{fig:decay_processes}
\end{figure}

Nonlinear damping is well understood in quantum optics, where this is a major mechanism that limits the laser intensity \cite{Mandel1995}. Significantly less is known about mechanical systems. Here, a simple microscopic mechanism is a decay process in which two quanta of the considered mode scatter into a quantum of a mode of a continuous spectrum \cite{Dykman1975a}. The corresponding process is sketched in Fig.~\ref{fig:decay_processes}(a). It was implemented in \cite{Leghtas2015,A_Roy2015}. For graphene-based nanoresonators, it was discussed for nonlinear leakage of the resonator modes into bulk modes \cite{Croy2012}, and it should be relevant for the decay into bulk modes (clamping losses) in other systems. However, as we show, in many cases other mechanisms become more important.

In this paper we develop a theory of nonlinear damping due to the anharmonic coupling between vibrational modes of a nano/micro system, which serves as a resonator. We consider the damping of a low-lying eigenmode. The mode eigenfrequency $\omega_0$ is assumed to be small compared to  $k_BT/\hbar$ and the Debye frequency of a crystal. At the same time, the mode decay rate is much smaller than $\omega_0$, i.e., the mode has a high quality factor. 

An important feature of nano- and microsystems is nonuniformity. Familiar examples are ripples of the graphene sheets \cite{Meyer2007}, bending and twisting of nanotubes, or layer thickness variations in multi-layer nanobeams. The scale of the nonuniformity $l_{\rm sm}$ should be compared with the wavelength $\lambda_T$ and the mean free path $l_T$ of thermal phonons, i.e., of phonons with frequencies $\sim k_BT/\hbar$. If both $\lambda_T$ and $l_T$ in the absence of the nonuniformity are both small,
$\lambda_T \ll l_T\ll l_{\rm sm}$, phonon transport can be described in terms of coordinate-dependent transport coefficients, for example, a coordinate-dependent thermal conductivity \cite{Gurevich1988}. A qualitatively different situation occurs if $l_{\rm sm} < l_T$, as in rippled graphene membranes for low temperature \cite{Meyer2007,Nika2009}. One has to be careful in defining $l_T$ in this case, as the normal vibrational modes are no longer plain waves, and $l_T$ should be defined as the scattering rate of these modes due to their nonlinear coupling with each other. In this case one can no longer use locally defined transport coefficients even where $l_{\rm sm}\gg \lambda_T$.

The aim of this paper is twofold: first, to develop a theory of damping of low-frequency vibrational modes in nonuniform systems and second, to study the nonlinear damping and dephasing. 
The theory is based on the eikonal formulation of the kinetics of thermal phonons. We use the term ``phonon" loosely to describe normal modes of a nanosystem that belong to the quasi-continuous spectrum, even though they are not plane waves. The phonons are enumerated by a quantum number $\kappa$ which, for an ideal crystal, includes the wave vector and the branch number, and have frequencies $\omega_\kappa$.

The nonlinear damping with decay into a single phonon, see Fig.~\ref{fig:decay_processes}(a), comes primarily from the cubic anharmonicity of the vibrations \cite{Dykman1975a}. As mentioned earlier, this mechanism can be important for the nonlinear coupling to substrate phonons. However, it is less likely to work for decay into the modes localized within the system. This is because the localized low-frequency modes have a discrete spectrum, and the decay of the considered mode into one of these modes $\kappa$ requires resonance $\omega_\kappa \approx 2\omega_0$. 
It also requires that the decay rate of the mode $\kappa$ be
much larger than the coupling and the decay rate of the considered mode. Such fine tuning does not happen generically, but  if it happened, the resulting nonlinear damping would be strong.

A more important contribution to the nonlinear damping can come from the processes described by the quartic anharmonicity. In such processes, a  phonon with frequency $\omega_\kappa\gg 2\omega_0$ is scattered off the considered low-frequency mode into another high-frequency phonon, see Fig.~\ref{fig:decay_processes}(b). It is implied that the spectrum of the involved phonons $\kappa,\kappa'$ is quasi-continuous. This condition is met if the phonons are weakly confined to the resonator or $l_T$ is small compared to at least one of the resonator dimensions. The frequency change $2\omega_0$  can be smaller than the decay rate of the involved high-frequency modes, which complicates the analysis. 

The conventional mechanisms of linear damping for bulk modes \cite{LL_Elasticity,Lifshitz1981a} are  based on the processes similar to those in Fig.~\ref{fig:decay_processes}(b), except that the scattering of the high-frequency phonons is accompanied by absorption/emission of one quantum of the considered low-frequency mode. Linear damping comes from the cubic nonlinearity. The results of this paper describe such damping in nonuniform systems. 

The rates of linear and nonlinear damping can become comparable even for comparatively small vibration amplitudes as a consequence of symmetry restrictions or the restrictions imposed  by the energy and momentum conservation. The best known example where dissipation is due to quartic anharmonicity was proposed by Landau and Khalatnikov \cite{Landau1949,Landau1949a} to describe  phonon scattering in liquid helium. Another example is suppression of the cubic-anharmonicity induced linear damping for flexural modes in flat atomically-thin membranes. It is rooted in the reflection symmetry, which allows nonlinear coupling that involves only an even number of flexural modes. 
Ultimately, the linear damping is due to four-wave processes, with participation of four flexural modes or two flexural modes and two acoustic phonons \cite{DeMartino2009}. The nonlinear damping shown in Fig.~\ref{fig:decay_processes}(b) occurs in the same order in the anharmonicity and therefore has no additional smallness associated with the higher-order anharmonicity. This might be behind the numerically observed strong nonlinear damping in graphene nanomembranes \cite{Midtvedt2014}.

Along with damping, scattering of phonons off the considered mode leads to its dephasing, i.e., loss of coherence, as was first discussed for high-frequency vibrations localized near defects in solids \cite{Ivanov1965,Elliott1965}. In contrast to damping, the scattering that leads to dephasing is quasi-elastic, the energy of the mode is not changed. To the lowest order, the process comes from the quartic nonlinearity in the mode coupling, although the parameters are renormalized by the cubic nonlinearity, which is also the case for nonlinear friction. As we show, the formalism developed to describe nonlinear friction extends to the analysis of the dephasing rate.  The dephasing we consider is a $T_2$-process, in the language of nuclear magnetic resonance. We will not consider $1/f$-type frequency noise, which usually does not come from nonlinear mode coupling.

The problem of linear damping of low-frequency eigenmodes in nano- and micro-resonators overlaps with the problem of sound absorption in dielectrics, which has been intensely studied since 1930s \cite{Landau1937,Akhiezer1938}, see \cite{Maris1966,Gurevich1988,Garanin1992,Collins2013,Lindenfeld2013,Feng2015} and references therein for more recent work. However, to the best of our knowledge, the nonlinear damping has not been discussed nor has there been developed a theory for systems with the nonuniformity on the scale small compared to the mean free path of thermal phonons. 

\subsection{Outline}
\label{subsec:outline}

Our analysis of damping and dephasing of an underdamped low-frequency mode 
is based on the expressions that relate the relaxation rates to the correlators of the bath to which the mode is coupled, see Sec.~II. These correlators are the main object of interest. We consider the case where the bath is formed by phonons and the coupling to the bath is due to the vibration nonlinearity. Then the correlators of interest are 
those of the pairs of the phonon variables and are related to the phonon density matrix weighted with the interaction.
They are easy to calculate and to find the nonlinear damping where the frequency $\omega_0$ of the considered mode is large compared to the decay rates of thermal phonons, the Landau-Rumer limit, see Sec.~\ref{sec:LR}. The problem becomes more complicated where the phonon decay cannot be disregarded, and this is the central part of the paper.

The analysis of the phonon correlators is done for nonuniform systems with an arbitrarily strong nonuniformity, which is however smooth on the typical wavelength $\lambda_T$ of thermal modes. It is based on the eikonal approximation, as mentioned above. This approximation is developed in Sec.~\ref{sec:eikonal}. The thermal modes (phonons) are characterized by coordinate-dependent wave vectors, and there holds approximate local momentum conservation in phonon scattering.  The analysis of the dynamics can be conveniently done by writing the sought correlators in the Wigner representation. We formulate it in Sec.~\ref{sec:Green_defined} in terms of the eigenmodes of the nonuniform system (not plane waves). The central result of this section, Eq.~(\ref{eq:nonlin_friction_scalar}), relates the nonlinear decay rate of the considered system to the Wigner transform of the thermal phonon correlator $\Phi_\alpha(\Rb,\kb,t)$, which is a function of the position $\Rb$, the wave vector $\kb$, and the phonon mode index $\alpha$. 

In Sec.~\ref{sec:QKE} we discuss the quantum transport equation for function $\Phi_\alpha$. The derivation of this equation is somewhat cumbersome and is described in Appendices A to D. Our analysis suggests that, for smooth nonuniformity,  this equation is local in space even where the characteristic nonuniformity length $l_{\rm sm}$ is small compared to the phonon mean free path $l_T$, but still $l_{\rm sm}\gg (\l_T\lambda_T)^{1/2}$. The locality means that the scattering couples function  $\Phi_\alpha(\Rb,\kb,t)$ to functions $\Phi_{\alpha'}(\Rb,\kb',t)$ with different wave vectors and mode indices, but the same position. In Sec.~\ref{sec:disorder} and Appendix E we briefly consider the effect on the evolution of function $\Phi_\alpha$ of a weak short-range disorder.

The analysis of the quantum transport equation in uniform systems is done in Sec.~\ref{sec:uniform_systems}. Function $\Phi_\alpha$ is expanded in the eigenfunctions of the collision operator, which is advantageous, since this operator is independent of the coordinates in this case. The results are used in Sec.~\ref{sec:thermoelastic} to study nonlinear thermoelastic relaxation of a low-frequency mode. We show that this relaxation is of particular interest for Lam\'e modes in thin plates and also for flexural modes in thin nanobeams, where thermal diffusion transverse to the beam is fast compared to the mode vibrations; this limit is opposite to the one where the Zener linear thermoelastic damping is dominating. Along with the microscopic theory we develop a phenomenological theory of nonlinear thermoelastic damping and show that they lead to the same results. 

In Sec.~\ref{sec:Akhiezer} we briefly discuss the nonlinear Akhiezer relaxation. As in the case of linear Akhiezer relaxation, it is important where the mode frequencies are large compared to the reciprocal time of thermal diffusion. We show that the expressions for the Akhiezer decay rate are grossly simplified in the $\tau$-approximation for the dynamics of thermal phonons; however, the value of $\tau$ is generally different from that  used to describe thermal diffusion.

In Sec.~\ref{sec:dephasing}  we discuss the dephasing of low-frequency vibrational modes due to their coupling to thermal phonons. For high-frequency modes, where the decay of thermal phonons can be disregarded, the theory is similar to the theory of dephasing for localized vibrations in solids \cite{Ivanov1965,Elliott1965}. For smaller $\omega_0$  the decay of thermal phonons has to be taken into account. As we show, this can be done by simply extending the results on nonlinear friction. Finally, in Sec.~\ref{sec:conclusions}  we summarize the main conclusions of this work. 

{\it Notations:} Throughout the paper we use a hat to indicate tensors; for example, $\hat \epsilon$ is the strain tensor with components $\epsilon_{ij}$ and $\hat\theta$ is an auxiliary  tensor with components $\theta_{ij}$. We use a wide hat, like $\widehat\Lambda$, for operators  that describe the coupling between different phonon correlators .

\section{General expression for the nonlinear damping rate} 
\label{sec:nonlinear_general}

We consider a low-frequency mode of a mesoscopic resonator interacting with a thermal bath. The bath excitations are vibrational modes with a quasi-continuous frequency spectrum, which we call phonons. The total Hamiltonian $H = H_0  + H_{b} + H_{i}$ is the sum of the Hamiltonians of the mode $H_0$, the bath $H_b$, and their interaction $H_i$. We write $H$ in terms of the creation and annihilation operators of the mode, $a^\dagger$ and $a$, and the phonons, $b_{\kappa}^\dagger$ and $b_{\kappa}$, where $\kappa$ enumerates the phonons. 

The displacement field of  the mode of interest can be written as $q(t)\ub({}\rb)$. Function $\ub(\rb)$ describes the spatial structure of the mode, whereas $q(t)$ characterizes the displacement at the antinode.  We choose the normalization of $\ub(\rb)$ in such a way that the Lagrangian of the mode in the harmonic approximation is $(M_0 /2)(\dot q^2 -\omega_0^2 q^2)$, where $M_0$ and $\omega_0$ are the effective mass and the eigenfrequency of the mode. We then introduce the operator of the mode coordinate as
\begin{align}
\label{eq:zero_point}
q= q_0(a+a^\dagger), \qquad q_0=(\hbar/2M_0\omega_0)^{1/2}.
\end{align}
Length $q_0$ is the scaled amplitude of the zero-point vibrations at the mode antinode. The characteristic wavelength of the mode is of the order of the size of the system, and therefore $q_0$ usually (but not necessarily) decreases with the system size. The mode Hamiltonian $H_0$ is 
\begin{align}
\label{eq:H_0}
H_{0}= \hbar\omega_0 a^\dagger a + \frac{1}{2}\hbar Va^{\dagger 2}a^2.
\end{align}
The term $\propto V$  comes from the internal nonlinearity of the mode and from its nonlinear coupling to phonons, cf. \cite{Dykman1975a}. It plays an important role in the dynamics of mesoscopic  modes \cite{Dykman2012b}, even though typically $|V|\ll \omega_0$. The energy decay rate is not affected by this term, except where a mode is very strongly driven \cite{Antonio2012,Mahboob2015}; such driving is not considered here.

In the Hamiltonian of the bath $H_b$, along with the harmonic part it is necessary \cite{LL_Elasticity,Lifshitz1981a} to take into account the nonlinear mode coupling responsible for the phonon decay, in particular, the decay of the high-frequency phonons $\kappa,\kappa'$ involved in the processes shown in Fig.~\ref{fig:decay_processes}(b). Respectively, we write $H_b$ as
\begin{align}
\label{eq:H_ph}
H_{b} =&\hbar \sum_{\kappa} \omega_{\kappa} b^\dagger_{\kappa} b_{\kappa} + H_{\textrm{ph}-\textrm{ph}},\nonumber \\
H_{\textrm{ph}-\textrm{ph}} = &\frac{1}{2}\sum_{\kappa \kappa' \kappa''}v{}_{\kappa\kappa'\kappa''}b^{\dagger}_{\kappa}b^{\dagger}_{\kappa'}b_{\kappa''}\nonumber\\
&+ \frac{1}{3}\sum_{\kappa \kappa' \kappa''} v'_{\kappa\kappa'\kappa''}b_\kappa^\dagger b_{\kappa'}^\dagger b_{\kappa''}^\dagger +  \textrm{H.c}.
\end{align}
The phonons $\kappa$ are generally not plane waves. This is a consequence of the finite size of the resonator and also of the presence of built-in static deformation and/or stress and maybe defects. However, in a deformed finite-size system one can still find normal vibrational mode, and they are enumerated by $\kappa$. 

In the coupling of the considered mode to the bath we will take into account the terms linear and quadratic in the mode displacement $q\propto a+a^\dagger$. The expression for $H_i$ in terms of the operators $a,a^\dagger$ can be written as
\begin{align}
\label{eq:H_i}
H_{i} = ah_{1} + a^\dagger h_1^\dagger  + a^2h_2 + a^{\dagger 2}h_{2}^\dagger  +  a^\dagger ah_{3}.
\end{align}
Operators $h_{1,2,3}$ depend on the variables of the bath.

For weak coupling $H_i$, one can describe the dynamics of the considered mode by switching to the rotating frame with the canonical transformation $\exp(-i\omega_0a^\dagger a t)$.  In slow time compared to the vibration period $2\pi/\omega_0$ one then obtains a Markovian quantum kinetic equation \cite{Dykman1975a,A_Roy2015,Dykman2012}
\begin{equation}
\label{eq:QKE}
\dot{\rho} = -i\frac{1}{2}V[a^{\dagger2}a^2,\rho] + \Gamma_{}\mathcal{D}[a]\rho + \Gamma^{\textrm{(nl)}}\mathcal{D}[a^2]\rho + \Gamma^{(\varphi)}\mathcal{D}^{(\varphi)}\rho,
\end{equation}
where $\rho$ is the mode density matrix and the super-operators ${\cal D}[a]$ and  ${\cal D}[a^2]$ describe linear and nonlinear damping of the mode, 
\begin{align}
\label{eq:D_operators}
 \mathcal{D}[a^m]\rho& =- \left(\bar{n}^{(m)}+1\right)\left(a^{\dagger m}a^m\rho - 2a^m\rho a^{\dagger m} + \rho a^{\dagger m}a^m\right)  \nonumber \\ 
&- \bar{n}^{(m)} \left( a^ma^{\dagger m}\rho  -2a^{\dagger m} \rho a^m + \rho a^ma^{\dagger m}\right).
\end{align}
Here, $\bar{n}^{(m)}\equiv \bar n(m\omega_0)$ is the Planck number, $\bar n(\omega)=\left[\exp(\hbar\omega/k_BT)-1\right]^{-1}$. To the lowest order in $H_i$, the linear and nonlinear damping rates $\Gamma$ and $\Gamma^{(\rm nl)}$ are  
\begin{align}
\label{eq:Gnl}
&\Gamma_{} = \hbar^{-2}[\bar n(\omega_0)+1]^{-1}\textrm{Re} \int_0^{\infty} d{}t\, e^{i\omega_0 t} \langle h_1^{\dagger}(t) h_1(0)\rangle
, \nonumber \\
 &\Gamma^{\textrm{(nl)}} = \hbar^{-2}[\bar n(2\omega_0)+1]^{-1}\textrm{Re} \int_0^{\infty} d{}t\, e^{2i\omega_0 t} \langle h_2^{\dagger}(t) h_2(0)\rangle
. 
\end{align}
These rates are determined by the processes in which the considered mode makes a transition to the nearest energy level, in the case of linear damping, $|n\rangle \to |n-1\rangle$, or to the next nearest level, in the case of nonlinear damping, $|n\rangle \to |n-2\rangle$, with the energy going into the thermal bath ($|n\rangle$ is the mode Fock state).

The super-operator $\mathcal{D}^{(\varphi)}$  describes dephasing, 
\begin{align}
\label{eq:D_dephasing}
\mathcal{D}^{(\varphi)}\rho = 2a^{\dagger}a\rho a^{\dagger}a - (a^{\dagger}a)^2\rho - \rho (a^{\dagger}a)^2.
\end{align}
The dephasing rate $\Gamma^{(\varphi)}$ is
\begin{equation}
\label{eq:Gphi}
\Gamma^{(\varphi)} = \hbar^{-2}\,{\rm Re} \int_0^{\infty} d{}t\, \langle h_3(t) h_3(0)\rangle.
\end{equation}
As indicated above, the dephasing we consider is a $T_2$-type process; it is caused by quasi-elastic scattering of the bath excitations off the mode, with no transitions between the mode energy levels.

In Eqs.~\eqref{eq:Gnl} and \eqref{eq:Gphi}, the averaging $\langle \cdot \rangle$ is done over the thermal state of the bath. We assume that $\langle h_{1,2,3}\rangle =0$.  The quantum kinetic equation applies for the relaxation rates small compared to $\omega_0$ and to the reciprocal correlation time of the bath excitations, i.e., to the reciprocal time over which the correlators in Eqs.~(\ref{eq:Gnl}) and (\ref{eq:Gphi}) decay.  Also, it is required that the derivatives of $\Gamma, \Gamma^{(\rm nl)}$ over $\omega_0$ be small. These conditions are met if the density of states of the bath weighted with the appropriate interaction $h_{1,2,3}$ is smooth near $\omega_0$, $2\omega_0$, and $\omega=0$.

Along with  decay and dephasing, the coupling (\ref{eq:H_i})  leads to a renormalization of the mode eigenfrequency $\omega_0$ and the parameter $V$ that describes the nonequidistance of the mode energy levels. We assume this renormalization to have been done; we note that the renormalization of $\omega_0$ is temperature-dependent even to the second order in $H_i$. In addition, the linear in the oscillator coordinates coupling $h_1$ generally renormalizes the operator of the nonlinear coupling $h_2$. The corresponding contribution to nonlinear friction can be thought of as a result of a nonlinear response of the bath to the perturbation $ah_1 + a^\dagger h_1^\dagger$. It does not lead to new effects.

\subsection{Relation to the phenomenological nonlinear friction}
\label{subsec:phenomenology}

The nonlinear friction term of the form (\ref{eq:QKE}) corresponds to the nonlinear friction in the van der Pol equation, which is broadly used for describing various types of  self-excited modes, in particular, in laser physics \cite{Mandel1995}. This can be seen from the equation for $\langle a\rangle = {\rm Tr} (a\rho)$ that follows from Eq.~(\ref{eq:QKE}),
\begin{equation}
\label{eq:moment}
\dot{\langle a\rangle} = -\left(\Gamma_{} + \Gamma^{(\varphi)} - 4\bar n^{(2)} \Gamma^{\textrm{(nl)}}\right)\langle {a}\rangle -  (iV + 2 \Gamma^{\textrm{(nl)}}) \langle {{a}^\dagger a^2}\rangle,
\end{equation}
The standard van der Pol equation in the rotating wave approximation is obtained from this equation if one ignores fluctuations, i.e., formally replaces $\langle a^\dagger a^2\rangle \to \langle a^\dagger\rangle \langle a\rangle^2$ and disregards the thermal term $\propto \bar n^{(2)}$. The term $\langle a^\dagger\rangle \langle a\rangle^2$ describes losses which nonlinearly depend on the mode amplitude; because of this term, the decay of $\langle a(t)\rangle $ is nonexponential. We note that, in contrast to the nonlinear friction usually discussed in lasers, in our case the dissipation and the fluctuations in the quantum kinetic equation (\ref{eq:QKE}) come from the coupling to a bath in thermal equilibrium and are related by the fluctuation-dissipation relation.

Parameters $V$ and $\Gamma^{\rm (nl)}$ have dimension of frequency and describe, respectively, the change of the oscillator frequency and the decay rate, which occur for the oscillator displacement on the order of the zero-point vibration amplitude $q_0$, Eq.~(\ref{eq:zero_point}). In the lab frame, in the phenomenological classical description of the oscillator dynamics, with account taken of the fluctuations that come along with the linear and nonlinear damping, Eq.~(\ref{eq:moment}) corresponds to the equation of motion of the form 
\begin{align}
\label{eq:VdP}
\ddot{q} + \omega_0^2q + \gamma q^3= &-2\Gamma \dot q - 4\Gamma^{\rm (nl)}(q/q_0)^2\dot q\nonumber\\
& +\xi_{\rm lin}(t) + \xi(t)q/q_0
\end{align}
As seen from this equation, the friction force nonlinearly depends on the displacement $q(t)$. We note that, phenomenologically, the friction force could have been written as $-(4/3\omega_0^2q_0^2)\Gamma^{\rm (nl)}\dot q^3$. This would lead to the same equation of motion in the rotating frame. Parameter $\gamma=(2\omega_0/3q_0^2))V$ in Eq.~(\ref{eq:VdP}) is the coefficient of nonlinearity of the restoring force. Both $V$ and $\Gamma^{\rm (nl)}$ are proportional to the corresponding classical nonlinearity parameters in Eq.~(\ref{eq:VdP}) multiplied by $q_0^2\propto \hbar$.

The term $\xi_{\rm lin}(t)$ in Eq.~(\ref{eq:VdP}) represents the additive thermal Gaussian noise with the intensity $\propto \Gamma T$ related to the linear friction coefficient by the fluctuation-dissipation theorem. The term $\xi(t)$ is also noise, but it has two distinct parts. One is Gaussian noise at frequencies close to $2\omega_0$. It is related to the nonlinear friction, has intensity $\propto \Gamma^{\rm (nl)} T$, and the power spectrum, which is flat around $2\omega_0$ in the region much broader than $\Gamma, \Gamma^{\rm (nl)}$. The other part is noise with a bandwidth smaller than $\omega_0$ yet  white on the slow time scale $\sim 1/\Gamma, 1/\Gamma^{\rm (nl)}$; its intensity is given by $\Gamma^{(\varphi)}$, cf. Ref.~\onlinecite{Zhang2014} and papers cited therein. We note that, in contrast to the phenomenological model, Eq.~(\ref{eq:QKE}) does not assume that the dynamics of the oscillator is Markovian in the lab frame; it is Markovian only on the time scale that largely exceeds $\omega_0^{-1}$ and the correlation time of the thermal reservoir.

\section{Nonlinear damping in the Landau-Rumer limit}
\label{sec:LR}

The goal of this paper is to calculate the correlators (\ref{eq:Gnl}) and (\ref{eq:Gphi}) for the case where operators $h_{1,2,3}$ are determined by the coupling to phonons. In what follows we concentrate on the analysis of nonlinear damping. We then briefly mention the relation to linear damping, and also discuss dephasing. 

To calculate the rate of nonlinear damping, we keep in $h_2$, Eq.~(\ref{eq:H_i}),  the relevant terms of the first and second order in the phonon coordinates,
\begin{equation}
\label{eq:h_2}
h_2=\sum_{\kappa} v_{\kappa}b_{\kappa}^\dagger + \sum_{\kappa\kappa'}v_{\kappa' \kappa}b_{\kappa'}^\dagger b_{\kappa}.
\end{equation}
Here, the first term describes the decay of two quanta of the considered mode into a phonon of the continuous spectrum, Fig.~\ref{fig:decay_processes}(a). The second term describes scattering of phonon $\kappa$ and two quanta of the considered mode into a phonon $\kappa'$, Fig.~\ref{fig:decay_processes}(b). 

In  the single-phonon decay, the considered mode primarily  emits substrate phonons, as mentioned earlier. Parameters $v_{\kappa}$ in Eq.~(\ref{eq:h_2}) determine nonlinear coupling  to such phonons at the boundary of the resonator. In the case of the Fermi resonance between the considered mode and another intracavity mode, $2\omega_0\approx \omega_\kappa$ \cite{Eichler2012}, $v_\kappa$ determines the strength of the mode coupling. The analysis of nonlinear dynamics in this case is beyond the scope of this paper. 

The two-mode coupling parameters $v_{\kappa' \kappa}$ in Eq.~(\ref{eq:h_2}) are of particular interest for coupling to  high-frequency phonons, which usually have a high density of states. Since the considered mode is a standing wave and the coupling depends on the displacements of the modes, operator $\sum v{}_{\kappa\kappa'}b_\kappa^\dagger b_{\kappa'}$ can be made Hermitian,  $v{}_{\kappa\kappa'}= v{}^*_{\kappa'\kappa}$. 
In Eq.~(\ref{eq:h_2}) we disregarded processes where two quanta of the considered mode decay into two phonons, $\omega_{\kappa}+\omega_{\kappa'}\approx 2\omega_0$. For small $\omega_0$, such phonons have low density of states, and decay into them is generally less probable than the single-phonon decay or the decay via scattering of high-frequency phonons.

The expression for the nonlinear damping rate (\ref{eq:Gnl}) takes a simple form in the Landau-Rumer limit where one can disregard the decay of the phonons of the continuous spectrum, i.e., their decay rate is small compared to $2\omega_0$. The Landau-Rumer nonlinear damping rate is 
\begin{align}
\label{eq:Gamma_nl_LR}
 &\Gamma^{\textrm{(nl)}}_{\textrm{LR}} =\pi \hbar^{-2}\sum_{\kappa} |v_{\kappa}^2| \delta(\omega_{\kappa}-2\omega_0 )  \nonumber \\
&+ \pi \hbar^{-2}\sum_{\kappa\kappa'} |v_{\kappa' \kappa}^2|(\bar{n}_{\kappa}-\bar{n}_{\kappa'})\delta(\omega_{\kappa'}-\omega_{\kappa}-2\omega_0) ,  
\end{align}
where $\bar n_{\kappa}\equiv \bar n(\omega_{\kappa})$. The first term in Eq.~(\ref{eq:Gamma_nl_LR}) describes the single-phonon decay and is independent of temperature. The second term describes the decay due to scattering of high-frequency phonons; the scattering rate is determined by the thermal population of these phonons and thus depends on temperature. 

If the high-frequency phonons $\kappa,\kappa'$ that contribute to the second term in Eq.~(\ref{eq:Gamma_nl_LR}) are of acoustic type and $\omega_{\kappa},\omega_{\kappa'}\gg \omega_0$, the coupling parameters are $v_{\kappa' \kappa}\propto\hbar\omega_{\kappa}$. This relation applies also if one thinks of the coupling $\propto v_{\kappa\kappa'}$ as coming from a parametric modulation of the energy of high-frequency phonons by the strain from the considered mode \cite{Akhiezer1938}.  The squared displacement of the considered mode gives a factor $\propto \hbar/M_0\omega_0$ in $v_{\kappa\kappa'}$. For $\hbar\omega_0\ll k_BT$, in Eq.~(\ref{eq:Gamma_nl_LR}) $\bar{n}_{\kappa}-\bar{n}_{\kappa'}\approx (2\hbar\omega_0/k_BT)\bar{n}_{\kappa}(\bar{n}_{\kappa}+1)$. Therefore, for scattering by acoustic phonons, the second term in the nonlinear damping rate (\ref{eq:Gamma_nl_LR})  is  $\propto  C_vT/\rho_0c_s^2$, where $C_v$ is the specific heat of the nanomechanical resonator ,  $c_s$ is the speed of sound, and $\rho_0$ is the resonator mass density. Typically  this factor is small, $C_vT/\rho_0c_s^2\ll 1$.

In obtaining Eq.~(\ref{eq:Gamma_nl_LR}) we disregarded the rate of inelastic scattering of phonons $\kappa$ compared to $\omega_0$, but  the static disorder was not assumed weak. This disorder makes high-frequency vibrational modes different from plane waves. Equation~(\ref{eq:Gamma_nl_LR}) applies both for short- and long-wavelength disorder, one just has to use the normal modes $\kappa$ calculated with the disorder taken into account. However, the inelastic decay rate of the modes $\kappa$ should exceed the spacing of their frequencies. Equation~(\ref{eq:Gamma_nl_LR}) also applies if modes $\kappa,\kappa'$ are weakly decaying bulk modes coupled to the considered mode at the resonator boundary.

\section{The eikonal approximation}
\label{sec:eikonal}

We now consider how the nonlinear coupling between the modes of the quasi-continuous spectrum (phonons) affects the nonlinear damping rate  $\Gamma^{\rm (nl)}$. 
Of primary interest is the case of smooth spatial nonuniformity of the resonator, with the spatial scale $l_{\rm sm}$  large compared to the wavelength $\lambda_T$  of thermal modes, i.e., the modes with energy $ \sim k_BT$. We will concentrate on the nontrivial case where  the mean free path $l_T$ of the thermal modes due to their nonlinear coupling is large, so that for acoustic-type modes
\begin{equation}
\label{eq:smotthness}
l_T\gg l_{\rm sm} \gg \lambda_T, \qquad \lambda_T =2\pi \hbar c_s/k_BT
\end{equation}
($c_s$ is the speed of sound). We will further assume that the thermal phonons are not affected by a magnetic field and the crystal is nonpolar. To simplify notations we will assume that there is just one atom per unit cell. The analysis immediately extends to several atoms per cell. 

For smooth nonuniformity,  thermal phonons can be described in the eikonal approximation, see Appendix~\ref{sec:appendix_eikonal}.  The displacement operator of an $n$th atom (at equilibrium poistion $\rb_n$) can be written as
\begin{align}
\label{eq:atom_displacement}
\ub_n= \sum_\kappa [\hbar/2M_n\omega_\kappa]^{1/2}\ub_\kappa(\rb_n)b_\kappa + {\rm H.c. },
\end{align}
where $M_n$ is the atomic mass, which smoothly depends on $n$. Instead of being plane waves, the normal modes $\ub_\kappa(\rb)$ have the form 
\begin{align}
\label{eq:eikonal}
&\ub_\kappa(\rb) =  {\mathbb N}^{-1/2}\eb_\kappa(\rb)\exp[iS_\kappa(\rb)], \nonumber\\
&\n S_\kappa(\rb) = \kb_\kappa(\rb).
\end{align}
One can think of Eq.~(\ref{eq:eikonal}) as the analytic continuation of the displacement field of the mode from the discrete values $\ub_\kappa(\rb_n)$ defined on the lattice sites. 

Vector $\kb_\kappa$ plays the role of the coordinate-dependent wave vector of the mode $\kappa$, and $\eb_\kappa$ describes the polarization of the mode; $\mathbb N$ is the number of the unit cells. Generally, not only $|\eb_\kappa|^2$ depends on the coordinate $\rb$, but also the relation between different components of $\eb_\kappa$ varies with $\rb$. To the leading order, in the eikonal approximation the equation of motion for $\ub_\kappa(\rb)$ is an algebraic equation that gives $\kb_\kappa(\rb)$ for a given mode frequency and branch, cf. \cite{Born1999} (see Appendix~\ref{sec:appendix_eikonal}).

The eikonal approximation implies smoothness of the vectors $\kb_\kappa (\rb)$ and $\eb_\kappa(\rb)$. The spatial derivatives like $|\n \kb_\kappa|$ and $(\kb_\kappa\n)\eb_\kappa$ are small compared to $\kb_\kappa^2$ and $|\eb_\kappa|\,\kb_\kappa^2$, respectively. 

An important case is where one or two of the nanoresonator dimensions are comparable to or smaller than the thermal wavelength $\lambda_T$. This is of interest for  nanowires or thin membranes, including carbon nanotubes and graphene membranes. Here modes propagate within the resonator in one or two directions spanned by vector $\rb_\parallel$, whereas in the transverse direction the modes are ``quantized", forming different transverse branches.  In Eq.~(\ref{eq:eikonal}) the eikonal now is a function of the propagation direction, $S_\kappa\equiv S_\kappa(\rb_\parallel)$, with
\[\partial_{\rb_\parallel} S_\kappa(\rb_\parallel) = \kb_\kappa(\rb_\parallel).\]
The polarization vector $\eb_\kappa(\rb)$ smoothly depends on the coordinate $\rb_\parallel$. 

The mode number $\kappa$ has discrete and quasi-continuous components. For an ideal bulk system these are the phonon branch $\alpha_\kappa$ and the wave vector $\kb_\kappa$. For  thin nanoresonators, the discrete components of $\kappa$  also enumerate the transverse branches. It is important that $\kappa$  has a quasi-continuous component and, as mentioned above, the  frequency spectrum is quasi-continuous, at least in the range where $\hbar\omega_\kappa \sim k_BT$. 

One can find the quasi-continuous values of $\kappa$ by formally extending the system to make it possible to impose periodic boundary conditions. One then requires that the difference of $S_\kappa(\rb)$ [or $S_\kappa(\rb_\parallel)$] on the boundaries be a multiple of $2\pi$. The quasi-continuous part of $\kappa$ can be associated with the value of vector $\kb_\kappa$ at some arbitrary point $\rb_0$, 
\begin{equation}
\label{eq:kappa_meaning}
 [\kappa]_{\rm quasi-continuous} \Longleftrightarrow \kb_\kappa(\rb_0).
\end{equation}
Normal vibrational modes introduced in Eq.(\ref{eq:atom_displacement}) satisfy the standard orthonormality and completeness conditions; in particular, 
\begin{align}
\label{eq:orthonormality}
&\sum_n\ub_\kappa^* (\rb_n) \cdot\ub_{\kappa'}(\rb_n)=\delta_{\kappa\kappa'} .
\end{align}

The operator  of the nonlinear phonon coupling $H_{\rm ph-ph}$ (\ref{eq:H_ph})  is obtained from the cubic in atomic displacements term in the potential energy of the system. This term is a sum of local contributions that depend on the difference of displacements of a few neighboring atoms.
From the expansion (\ref{eq:atom_displacement}) we see that, if we disregard umklapp processes, the matrix elements of the nonlinear mode-mode coupling in Eq.~(\ref{eq:H_ph}) can be calculated in the stationary phase approximation. We change from summation over lattice sites to integration, 
\[\sum\nolimits_n\to \int v_{\rm c}^{-1}(\rb)d\rb,\] 
where $v_{\rm c}(\rb)$ is the volume of the unit cell that smoothly depends on $\rb$. Then 
\begin{align}
\label{eq:cubic_cplng}
&v{}_{\kappa_1\kappa_2\kappa_3} \propto \int d\rb 
\;e^{i[S_{\kappa_3}(\rb) - S_{\kappa_1}(\rb) - S_{\kappa_2}(\rb)]} \hat U_{\kappa_1\kappa_2\kappa_3}(\rb) \nonumber\\
&\propto e^{i[S_{\kappa_3}(\rb_*) - S_{\kappa_1}(\rb_*) - S_{\kappa_2}(\rb_*)]} \hat U_{\kappa_1\kappa_2\kappa_3}(\rb_*). 
\end{align}
Here $\rb_*\equiv \rb_*(\kappa_1,\kappa_2,\kappa_3)$ is given by equation 
\begin{equation}
\label{eq:momentum_conservation}
\kb_{\kappa_1}(\rb_*) + \kb_{\kappa_2}(\rb_*)-\kb_{\kappa_3}(\rb_*) =0
\end{equation}
and $\hat U_{\kappa_1\kappa_2\kappa_3}(\rb)$ is a tensor that describes the nonlinear phonon coupling.
In systems with translational symmetry, $\hat U$ is independent of $\rb$ as are also vectors $\kb_\kappa$ \cite{Lifshitz1981a}, and then Eq.~(\ref{eq:momentum_conservation}) becomes just the condition of the quasi-momentum conservation in phonon scattering. The matrix elements $v'_{\kappa_1\kappa_2\kappa_3}$ in Eq.~(\ref{eq:H_ph}) are given by an expression similar to (\ref{eq:cubic_cplng}).  In Eq.~(\ref{eq:cubic_cplng}) we skipped the smooth mode polarization factors and $v_{\rm c}(\rb)$; matrix elements $v_{\kappa_1\kappa_2\kappa_3}$ are determined by the values of these factors at $\rb_*$.

The stationary phase approximation applies provided the nonuniformity is sufficiently strong. The range $\delta r_{\rm q}$ of the values of $\rb$ that contribute to the integral over $\rb$ in the first line of Eq.~(\ref{eq:cubic_cplng}) can be estimated by expanding $S_{\kappa_n}(\rb)$ ($n=1,2,3$) in Eq.~(\ref{eq:cubic_cplng}) to second order in $\rb-\rb_*$. We assume that, for the values of $\kappa$ that correspond to thermal phonons, there holds the condition
\begin{equation}
\label{eq:coordinate_uncertainty}
\lambda_T\ll \delta r_{\rm q} \ll l_{\rm sm}, \quad \delta r_{\rm q} = |\partial^2_\rb S_\kappa|^{-1/2} \sim (\lambda_Tl_{\rm sm})^{1/2} .
\end{equation}
The above estimate of $\delta r_{\rm q}$ is obtained by noting that, for sufficiently strong nonuniformity, the position-dependent wave vector of thermal phonons $\kb_\kappa (\rb)$ changes by $\sim \lambda_T^{-1}$ on the length $l_{\rm sm}$. Parameter $\delta r_{\rm q}$ gives the momentum uncertainty in the momentum conservation condition (\ref{eq:momentum_conservation}), $\delta k \sim (\delta r_{\rm q})^{-1}$.

Similar arguments apply to the matrix elements $v_{\kappa\kappa'}$ of the coupling of the considered low-frequency mode to high-frequency modes $\kappa,\kappa'$, see Eq.~(\ref{eq:h_2}).  For the typical wavelength of the low-frequency mode larger than $l_{\rm sm}$,  from Eq.~(\ref{eq:coordinate_uncertainty}) we can estimate the range of the difference in the quasi-momenta of the high-frequency phonons  as
\[|\kb_\kappa(\rb)-\kb_{\kappa'}(\rb)| \lesssim (\delta r_{\rm q})^{-1}.\]

The transition $\sum_n\to \int v_{\rm c}^{-1}(\rb)d\rb$ and the stationary-phase approximation used in Eq.~(\ref{eq:cubic_cplng}) apply also if  
\[\kb_{\kappa_1}(\rb_*) + \kb_{\kappa_2}(\rb_*)-\kb_{\kappa_3}(\rb_*) =\Kb(\rb_*),\] 
where $\Kb(\rb)$ is a smoothly dependent on $\rb$ reciprocal lattice vector. This expression describes umklapp processes in a nonuniform medium.

\section{The two-phonon correlator}
\label{sec:Green_defined}

It is convenient to express the nonlinear damping rate $\Gamma^{\rm (nl)}$, Eqs.~(\ref{eq:Gnl}) and (\ref{eq:h_2}),  in terms of the two-phonon correlation function $\phi_{\kappa\kappa'}(t)$. For $\hbar\omega_0\ll k_BT$
\begin{align}
\label{eq:Gnl_v2}
\Gamma^{(\rm nl)} = & -\frac{2\omega_0}{k_BT}{\rm Im}\sum_{\kappa,\kappa'} v^*_{\kappa' \kappa}
\int_0^\infty dt e^{i(2\omega_0+i\varepsilon) t}\phi_{\kappa\kappa'}(t)
\end{align}
($\varepsilon \to +0$), where 
\begin{align}
\label{eq:GreenFunction}
\phi_{\kappa\kappa'}(t) =   \frac{-i}{\hbar}\sum_{\kappa_0,\kappa_0'}v_{\kappa_0'\kappa_0}
\langle b^{\dagger}_{\kappa}(t)b_{\kappa'}(t)b^{\dagger}_{\kappa_0'}(0)b_{\kappa_0}(0) \rangle. 
\end{align}
Functions $\phi_{\kappa\kappa'}$ that contribute to the nonlinear damping correspond to mode pairs $\kappa,\kappa'$ with close eigenfrequencies, $|\omega_{\kappa}-\omega_{\kappa'}|\sim 2\omega_0 \ll \omega_{\kappa}$. For the values of $\rb$ where $v_{\kappa'\kappa}$ are large for given $\kappa,\kappa'$, the wave vectors of the modes $\kappa,\kappa'$ are close, $|\kb_\kappa (\rb)- \kb_{\kappa'}(\rb)|\ll |\kb_\kappa(\rb)|, |\kb_{\kappa'}(\rb)|$. The major contribution to $\Gamma^{\rm (nl)}$ comes from the modes $\kappa,\kappa'$ that belong to the same vibrational branch, $\alpha_\kappa = \alpha_{\kappa'}$. Functions $\phi_{\kappa\kappa'}$ for different branches are fast oscillating on the scale $\sim\omega_0$, except for the region where the branches cross or touch, but this region is small, and therefore the overall contribution of terms with $\alpha_\kappa\neq \alpha_{\kappa'}$ is small.

The correlator $\phi_{\kappa\kappa'}$ is the weighted off-diagonal phonon density matrix $\langle b^{\dagger}_{\kappa}(t)b_{\kappa'}(t)\rangle$. The weighting factor is independent of time, as seen from Eq.~(\ref{eq:GreenFunction}). The reduction of the problem to a calculation of the weighted thermal-phonon density matrix significantly simplifies the analysis. Our formulation allows us to study spatially nonuniform systems, and, in the first place, the nonlinear friction and the dephasing. Our results for linear friction in the limit of a spatially uniform system and weak disorder coincide with the results based on the linearized Boltzmann equation. 

A natural approach to the analysis of the phonon dynamics for close $\kb_\kappa(\rb)$ and $\kb_{\kappa'}(\rb)$ is to use the Wigner transformation ${\mathbb W}$. This is a linear integral transformation. The Wigner transform of the correlator $\phi_{\kappa\kappa'}$ for a nonuniform system is tensor $\hat \Phi_\alpha$,  
\begin{align}
\label{eq:Wigner_defined}
\hat \Phi_\alpha(\Rb,\kb,t)& = {\mathbb W}_{\Rb,\kb,\alpha}[\phi_{\kappa\kappa'}]\nonumber\\
&\equiv \sum_{\kappa,\kappa'}\hat\theta(\Rb,\kb;\kappa,\kappa';\alpha) \phi_{\kappa\kappa'}(t).
\end{align}
The transformation should be constructed in such a way that correlator $[\Phi_\alpha(\Rb,\kb)]_{ij}$ describes the dynamics of a wave packet centered at $\Rb$ with the typical wave vector $\kb$ and that $\hat\Phi_\alpha$  smoothly depends on $\Rb$ on the scale $\lambda_T$. The kernel of such transformation for a nonuniform system is tensor $\hat\theta$ with components
\begin{align}
\label{eq:polarization_tensor}
\theta_{ij}(\Rb,\kb;\kappa,\kappa';\alpha)&=\int \frac{d\rhob}{v_{\rm c}(\Rb)}{}u_{\kappa' i}\left(\Rb+\frac{1}{2}\rhob\right) e^{-i\kb\rhob}\nonumber\\
&\times u_{\kappa}^*{}_j \left(\Rb-\frac{1}{2}\rhob\right)\delta_{\alpha_\kappa\alpha}\delta_{\alpha_{\kappa'}\alpha}.
\end{align}
Alternatively, in the eikonal approximation (\ref{eq:eikonal}) for the lattice displacements, tensor $\hat\theta$ could be defined as a sum over lattice sites $n,m$ of $u_{\kappa' i}(\rb_n) u^*_{\kappa j}(\rb_m)\exp[-i\kb(\rb_n-\rb_m)]$ calculated  for $\Rb=(\rb_n+\rb_m)/2$ and for $\kappa,\kappa'$ belonging to a branch $\alpha$. The typical value of $|\rhob|$ in Eq.~(\ref{eq:polarization_tensor}) is $\sim\lambda_T$; it is small compared to the nonuniformity length $l_{\rm sm}$. We have set $v_{\rm c}(\Rb+\frac{1}{2}\rhob)v_{\rm c}(\Rb-\frac{1}{2}\rhob)\approx v_{\rm c}^2(\Rb)$. 

In nanowires and thin membranes, vectors $\Rb,\kb,\rhob$ in Eq.~(\ref{eq:polarization_tensor}) are one- or two-dimensional, respectively. In this case Eq.~(\ref{eq:polarization_tensor}) implies summation over the lattice sites in the transverse direction for the corresponding $\alpha$; function $\theta_{ij}$ is independent of the transverse coordinates.
In a spatially uniform system where $\kb_\kappa$ is a good quantum number, from Eq.~(\ref{eq:polarization_tensor})  $\hat \theta\propto \delta [\frac{1}{2}(\kb_\kappa + \kb_{\kappa'})-\kb]\exp[i(\kb_{\kappa'}-\kb_{\kappa})\Rb]$; it has the form of the kernel of the standard Wigner transformation in the continuous limit. 

The orthonormality of functions $\ub_\kappa(\rb)$, Eq.~(\ref{eq:orthonormality}), leads to the relation
\begin{align}
\label{eq:psi_orthonormality}
\int\frac{d\Rb d\kb}{(2\pi)^d} &{\rm Tr}\left[\hat \theta^\dagger(\Rb,\kb;\kappa_1,\kappa'_1; \alpha_1)\hat \theta(\Rb,\kb;\kappa,\kappa'; \alpha)\right] \nonumber\\
&=\delta_{\kappa\kappa_1}\delta_{\kappa'\kappa'_1}
\delta_{\alpha_\kappa\alpha}\delta_{\alpha_{\kappa'}\alpha}
\delta_{\alpha \alpha_1},
\end{align}
where $d$ is the dimension of vectors $\Rb$ and $\kb$ and the trace is taken over the tensor indices.

\subsection{Scalar Wigner function in the eikonal approximation}
\label{subsec:Green_eikonal}

For smooth nonuniformity, one can simplify tensor $\hat\theta$ by integrating over $\rhob$ in Eq.~(\ref{eq:polarization_tensor})  with the account taken of the eikonal form of function $\ub_\kappa$, Eq.~(\ref{eq:eikonal}). Expanding the exponents $ S_\kappa(\Rb -\frac{1}{2}\rhob)$ and $S_{\kappa'}(\Rb + \frac{1}{2}\rhob)$ to first order in $\rhob$, we obtain
\begin{align}
\label{eq:epsilon_eikonal}
\hat\theta(\Rb,\kb;\kappa,\kappa';\alpha)\propto \tilde\delta\left\{\tfrac{1}{2}[\kb_\kappa(\Rb)+\kb_{\kappa'}(\Rb)]-\kb\right\},
\end{align}
where $\tilde\delta(\kb)$ is a sharp $\delta$-like peak that describes the approximate momentum conservation; its typical width is $\sim (\delta r_{\rm q})^{-1} \ll \lambda_T^{-1}$ and is determined by the second-order term in the expansion of $S_\kappa, S_{\kappa'}$ in $\rhob$. 

If $\hat\theta$ were defined as a lattice sum, which is of interest for high temperatures, one would have quasi-momentum conservation in the form $\kb_{\kappa}(\Rb)+ \kb_{\kappa'}(\Rb) \approx \kb +\Kb(\Rb)$. The analysis of this case is a straightforward extension of the present analysis. 

Since $|\kb_\kappa(\Rb) - \kb_{\kappa'}(\Rb)|\ll |\kb_\kappa(\Rb)|$ and $\alpha_\kappa=\alpha_{\kappa'}$, we introduce a  quasi-continuous component $\tilde\kappa$, which is the ``center" of $\kappa,\kappa'$ for given $\Rb,\kb,\alpha$ and is defined by equation
\begin{align}
\label{eq:bar_kappa}
\kb_{\tilde\kappa}(\Rb) = \kb, \qquad \tilde\kappa \equiv\tilde\kappa(\Rb,\kb,\alpha).
\end{align}
Keeping in mind that the polarization vectors $\eb_\kappa(\rb)$ in Eq.~(\ref{eq:eikonal}) are smooth functions of the quasi-continuous part of $\kappa$, we can change from  tensor $\hat \theta$ to  a scalar ${}\theta$,
\begin{align}
\label{eq:kernel_scalar}
&\hat\theta(\Rb,\kb,;\kappa,\kappa';\alpha) \approx \hat M(\Rb,\kb;\alpha){}\theta_\alpha(\Rb,\kb;\kappa,\kappa'),\nonumber\\
&{}\theta_\alpha(\Rb,\kb;\kappa,\kappa')=\frac{(2\pi)^d}{{\mathbb N}v_{\rm c}(\Rb)}
\tilde\delta\left[\frac{\kb_\kappa(\Rb)+\kb_{\kappa'}(\Rb)}{2}-\kb\right]\nonumber\\
&\times |\eb_{\tilde\kappa}|^2\exp[iS_{\kappa'}(\Rb)- iS_{\kappa}(\Rb)]\delta_{\alpha_{\kappa},\alpha} \delta_{\alpha_{\kappa'},\alpha}.
\end{align}
Here, $\eb_{\tilde\kappa} \equiv \eb_{\tilde\kappa}(\Rb)$, whereas  $\hat M(\Rb,\kb;\alpha)$ is a unit-trace tensor, 
\begin{align}
\label{eq:M_tensor}
M_{ij}(\Rb,\kb;\alpha)={}e_{\tilde\kappa i}e_{\tilde\kappa}{}^*_j/|\eb_{\tilde\kappa}|^2,
\end{align}

Using Eq.~(\ref{eq:Wigner_defined}), we can now introduce a scalar correlation function ${}\Phi_\alpha(\Rb,\kb,t)$,
\begin{align}
\label{eq:scalar_Green}
&\hat \Phi_\alpha (\Rb,\kb,t)]\approx \hat M(\Rb,\kb;\alpha){}\Phi_\alpha(\Rb,\kb,t),  \nonumber\\
&{}\Phi_\alpha(\Rb,\kb,t)= \sum_{\kappa\kappa'}{}\theta_\alpha(\Rb,\kb;\kappa,\kappa')\phi_{\kappa\kappa'}(t).
\end{align}
Similarly we introduce a scalar parameter of the coupling to the low-frequency mode ${}V_\alpha(\Rb,\kb)$ (the vertex in the Wigner representation),
\begin{align}
\label{eq:scalar_coupling}
&{}V_\alpha(\Rb,\kb)= \sum_{\kappa,\kappa'}{}\theta_\alpha(\Rb,\kb;\kappa,\kappa')v_{\kappa'\kappa}
\end{align}
Using the orthonormality of tensors $\hat\theta$, Eq.~(\ref{eq:psi_orthonormality}), after simple algebra one obtains the expression for the nonlinear damping rate that reads 
\begin{align}
\label{eq:nonlin_friction_scalar}
&\Gamma^{\rm (nl)}=-\frac{2\omega_0}{k_BT}{\rm Im}\,\int\frac{ d\Rb d\kb}{(2\pi)^d} \sum_\alpha {}V_{\alpha}^*(\Rb,\kb)
\Phi_{\alpha,2\omega_0} (\Rb,\kb),\nonumber\\ 
&\Phi_{\alpha,\omega}(\Rb,\kb) = \int\nolimits_0^\infty dt \Phi_\alpha(\Rb,\kb,t)\exp(i\omega t).
\end{align}

The correlator ${}\Phi_\alpha(\Rb,\kb,t)$ is the scalar Wigner transform of $\phi_{\kappa\kappa'}$ in a spatially nonuniform system, i.e., the Wigner transform of the off-diagonal phonon density matrix weighted with a time-independent operator. It is much more convenient for the analysis than the tensor. 
Equation (\ref{eq:nonlin_friction_scalar})  expresses the nonlinear damping parameter $\Gamma^{\rm (nl)}$ in terms of this scalar. We note that ${}V_\alpha$ in Eq.~(\ref{eq:nonlin_friction_scalar}) is real for a nonlinear coupling to a standing mode. Interestingly, as shown in Appendix \ref{sec:transport_eq_omega_full}, starting with the dynamical equations for the tensor $\hat \Phi_\alpha$,  one can obtain a closed-form equation for the scalar function ${}\Phi_\alpha(\Rb,\kb,t)$. It is discussed in the next section.

\section{Transport equation for the Wigner function}
\label{sec:QKE}

Function ${}\Phi_\alpha(\Rb,\kb,t)$ smoothly depends on the coordinate $\Rb$ and the wave vector $\kb$. 
As a function of $\Rb$, it varies on the distance  determined by the scale of the spatial nonuniformity and by the phonon scattering rate. This distance is much longer than $|\kb|^{-1}\sim \lambda_T$. The separation of the spatial scales makes it possible to derive a transport equation for function ${}\Phi_\alpha$. The derivation is somewhat involved, since the standard momentum conservation in phonon scattering does not apply. It is given in the Appendix for the Fourier transform of ${}\Phi_\alpha(\Rb,\kb,\omega)$. In the time domain the transport equation reads
\begin{align}
\label{eq:transport_Wigner}
&\partial_t {}\Phi_\alpha + 
\{{}\Omega_\alpha,{}\Phi_\alpha\}
 ={\rm St}[{}\Phi_\alpha],\nonumber\\
&{}\Phi_\alpha\equiv {}\Phi_\alpha(\Rb,\kb,t),\qquad {}\Omega_\alpha \equiv {}\Omega_\alpha(\Rb,\kb).
\end{align}
Here, 
\begin{align}
\label{eq:bar_omega}
{}\Omega_\alpha(\Rb,\kb)=\sum_\kappa\omega_\kappa{}\theta_\alpha(\Rb,\kb,;\kappa,\kappa)
\end{align}
is the effective position- and momentum-dependent frequency of the branch $\alpha$, and
\[ \{A,B\} = \partial_\kb A \partial_\Rb B  -\partial_\kb B\partial_\Rb A\]
is the Poisson bracket.

The effective collision term St~${}\Phi_\alpha$ is 
\begin{align}
\label{eq:stoss_scalar}
{\rm St}[{}\Phi_\alpha(\Rb,\kb,t)] &=\sum_{\alpha_0}\int d\Rb_0 d\kb_0  [{}\lambda^{\alpha_0}_\alpha (\Rb,\Rb_0,\kb,\kb_0)]_{\omega \to  +i0} \nonumber\\
&\times  {}\Phi_{\alpha_0}(\Rb_0,\kb_0,t).
\end{align}
The kernel ${}\lambda$ for phonon-phonon scattering is given by Eq.~(\ref{eq:scalar_kernel_eikonal}) with operator $\hat \Lambda$ calculated for $\omega \to  +i0$. The leading-order term in $[{}\lambda]_{\omega\to +i0}$ is real. It contains $\delta$-functions of energy conservation in the processes where modes $\kappa,\kappa'$ decay into other modes or are scattered by other modes. The imaginary terms in ${}\lambda$, which are quadratic in the small parameters of the phonon nonlinearity, lead to a small renormalization of ${}\Omega_\alpha$, which we assume to have been done. 

The initial condition for function ${}\Phi_\alpha$ follows from Eq.~(\ref{eq:scalar_Green}) in which one replaces $\phi_{\kappa\kappa'}(t)$ with $\phi_{\kappa\kappa'}(0)$,
\begin{align}
\label{eq:initial_condition}
&\phi_{\kappa\kappa'}(0)= -i \hbar^{-1}v_{\kappa'\kappa}\bar n_{\kappa} (\bar n_{\kappa'}+1),\nonumber\\
&\Phi_\alpha(\Rb,\kb,0) = -i\hbar^{-1}V_\alpha(\Rb,\kb)\bar n_{\tilde\kappa}(\bar n_{\tilde\kappa}+1),
\end{align}
where $\tilde\kappa \equiv \tilde\kappa(\Rb,\kb,\alpha)$ is given by Eq.~(\ref{eq:bar_kappa}).

A feature of central importance in Eq.~(\ref{eq:stoss_scalar}) is the {\em locality}. As shown in Appendix~\ref{subsec:collision_eikonal}, the kernel ${}\lambda_\alpha^{\alpha_0}(\Rb,\Rb_0,\kb,\kb_0)$ becomes small for $|\Rb-\Rb_0|$ exceeding the quantum scale $\delta r_{\rm q} = (\lambda_Tl_{\rm sm})^{1/2}$. On the other hand, for a given $\Rb-\Rb_0$ such that $|\Rb-\Rb_0|\lesssim r_{\rm q}$, function $\lambda_\alpha^{\alpha_0}$ varies with $\Rb$ on the nonuniformity scale $\sim l_{\rm sm}$. It is this variation that determines the spatial variation of $\Phi_\alpha(\Rb,\kb,t)$ (recall that $l_{\rm sm}$ is small compared to the phonon mean free path $l_T$). Therefore function ${}\Phi_\alpha$ remains nearly constant on the quantum spatial scale, $\delta r_{\rm q}\ll l_{\rm sm}$, and one can rewrite Eq.~(\ref{eq:stoss_scalar}) as 
\begin{align}
\label{eq:locality}
&{\rm St}[{}\Phi_\alpha(\Rb,\kb,t)] = \sum_{\alpha_0}\int d\kb_0 \tilde\lambda^{\alpha_0}_\alpha(\Rb,\kb,\kb_0){}\Phi_{\alpha_0}(\Rb,\kb_0,t),\nonumber\\
&\tilde\lambda^{\alpha_0}_\alpha(\Rb,\kb,\kb_0) = \int d\Rb_0 [{}\lambda^{\alpha_0}_\alpha(\Rb,\Rb_0,\kb,\kb_0)]_{\omega\to +i0},
\end{align}
which means that the evolution of $\Phi_\alpha(\Rb,\kb,t)$ described by Eq.~(\ref{eq:transport_Wigner})  depends on $\Phi_{\alpha_0}(\Rb,\kb_0,t)$, i.e., on the Wigner function calculated for the same $\Rb$ but different $\kb_0,\alpha_0$.

The physics behind the locality is as follows. 
In phonon scattering, the total phonon momentum is conserved to within $\sim \hbar/\delta r_{\rm q}$, cf. Eq.~(\ref{eq:coordinate_uncertainty}). Respectively, the region in which  the scattering occurs is $\delta r_{\rm q}$. On the other hand, the scattering rate changes on the nonuniformity scale $l_{\rm sm}$ that largely exceeds $\delta r_{\rm q}$. On the scale $l_{\rm sm}$, phonon scattering events look ``local".

In the Markovian approximation used to derive Eq.~(\ref{eq:transport_Wigner}) (the ladder diagram approximation),  this equation has the same form as the kinetic equation for the Wigner transform of the phonon density matrix. The difference is in the initial condition, which has the form (\ref{eq:initial_condition}) for function $\Phi_\alpha$. 
The decoupling used to derive Eq.~(\ref{eq:transport_Wigner}) disregards many-phonon correlations. This equation is an analog of the linearized Boltzmann equation for the phonon distribution function. However, here it is derived for a phonon correlation function starting from a microscopic model of a nonuniform system with the nonuniformity, which is smooth on phonon wavelength, but not on the phonon mean free path.

\section{Short-range disorder}
\label{sec:disorder}

The above analysis directly extends to the case where, along with smooth nonuniformity, the crystal has a short range disorder, examples being mass disorder or impurities, cf. Refs.~\cite{Gurevich1988,Garanin1992,Feng2015} and references therein. This disorder leads to scattering of thermal phonons, but plays practically no role in the dynamics of the considered long-wavelength mode itself, except that it may slightly change its waveform. We will describe the disorder by the extra term in the Hamiltonian of thermal phonons
\begin{align}
\label{eq:disorder_Hamiltonian}
H_d = \sum_{\kappa \kappa'}v^{(d)}_{\kappa\kappa'}b_\kappa^\dagger b_{\kappa'};
\end{align}
we disregard the terms with $b_{\kappa}b_{\kappa'}$ and $b_{\kappa}^\dagger b_{\kappa'}^\dagger$, which lead to frequency renormalization. 
We will further assume that the disorder is weak. The simplest model is a weak zero-mean Gaussian field, with correlations limited to a single lattice site (a $\delta$-correlated field, in the continuous limit). In this case parameters  $v^{(d)}_{\kappa\kappa'}$ for thermal phonons are sums over the lattice sites $n$ of the factors $\propto u_{\kappa'}^i(\rb_n) u^*_{\kappa j}(\rb_n) $ weighted with the perturbation on site $n$. Then the quadratic in the disorder terms, which determine phonon scattering in the second order of the perturbation theory, have the structure
\begin{align}
\label{eq:quadratic_short_range}
&\langle v_{\kappa\kappa'}^{(d)} v^{(d)}_{\kappa_1\kappa_1'}\rangle \propto \sum_n (\hat v^{(d)})_{\kappa'\kappa_1'}^{\kappa\kappa_1}(\rb_n) \nonumber\\
&\times \exp[iS_{\kappa'}(\rb_n)-iS_\kappa(\rb_n) +iS_{\kappa_1'}(\rb_n)-iS_{\kappa_1}(\rb_n)],
\end{align}
where tensors $\hat v^{(d)}$ smoothly depend on $n$ and on the mode indices. 

Similar to Eq.~(\ref{eq:cubic_cplng}), the sum (\ref{eq:quadratic_short_range}) can be written as an integral over $\rb_n$, because we are interested in the correlation functions $\phi_{\kappa\kappa'}$ with close $\kappa$ and $\kappa'$. This integral can be calculated in the stationary phase approximation. The major contribution to Eq.~(\ref{eq:quadratic_short_range}) comes from the vicinity of a point $\rb_{**}$ in which the total local quasi-momentum is conserved,
\begin{align}
\label{eq:short_range_conservation}
\kb_{\kappa}(\rb_{**}) -\kb_{\kappa'}(\rb_{**}) + \kb_{\kappa_1}(\rb_{**}) - \kb_{\kappa_1'}(\rb_{**}) = 0.
\end{align}

We note that, because of the smooth nonuniformity, the quasi-momentum of the system of phonons is not conserved even without short-range disorder and umklapp processes. Phonons do not propagate along straight lines. However, Eq.~(\ref{eq:momentum_conservation}) indicates that, locally, the quasi-momentum of the system of phonons is conserved. Short-range disorder breaks down this conservation; in each scattering event the local momentum of a phonon is transferred to a short-range scatterer. 

It is shown in Appendix~\ref{sec:appendix_disorder} that, in the presence of short-range disorder, one still obtains a transport equation for the Wigner function of the form of Eqs.~(\ref{eq:transport_Wigner}) and (\ref{eq:locality}), but now the kernel of the collision term ${}\lambda^{\alpha_0}_\alpha$ has an extra term. This term is quadratic in parameters $v^{(d)}_{\kappa\kappa'}$, see Eq.~(\ref{eq:disorder_scattering}). Effectively, the approximation corresponds to a generalized Matthiessen rule where the phonon-phonon and the short-range disorder scattering processes do not affect each other. However, both types of processes are affected by the smooth nonuniformity. Short-range scattering is particularly important for lower temperatures, where the rate of the umklapp processes becomes exponentially small.

\section{Nonlinear friction in spatially uniform systems}
\label{sec:uniform_systems}

Solutions of the transport equation can be obtained for specific types of spatial nonuniformity, and in most cases this requires numerical calculations. In what follows we illustrate the power of our approach by applying it to spatially uniform systems and calculating nonlinear damping in such systems. In uniform systems  $\kb_\kappa(\rb)$ and ${}\Omega_\alpha(\rb,\kb)$ are independent of the coordinate $\rb$. The spatial dependence remains only in the displacement of the considered mode and in its Wigner function $\Phi_\alpha$. We therefore switch to using $\rb$ instead of $\Rb$ for the coordinate of the Wigner function. This will make other notations more conventional; in particular, the displacement of the mode will be a function of $\rb$, not $\Rb$.

In a uniform system function ${}\theta_\alpha$, Eq.~(\ref{eq:kernel_scalar}), becomes
\begin{align*}
{}\theta_\alpha ({}\rb,\kb;\kappa,\kappa')=&\frac{(2\pi)^d}{\mathbb V}\delta \left[\frac{1}{2}(\kb_\kappa + 
\kb_{\kappa'})-\kb\right]\\
&\times \exp[i(\kb_{\kappa'}-\kb_\kappa){}\rb]\delta_{\alpha_{\kappa},\alpha}\delta_{\alpha_{\kappa'},\alpha},
\end{align*}
whereas in Eq.~(\ref{eq:stoss_scalar}) the collision operator takes the form
\begin{align}
\label{eq:bar_lambda_uniform}
{\rm St}[{}\Phi_\alpha({}\rb,\kb,t]=\frac{\mathbb V}{(2\pi)^d}\sum_{\alpha_0}\int d\kb_0 \tilde\Lambda_{\kb\alpha}^{\kb_0\alpha_0} {}\Phi_{\alpha_0}({}\rb,\kb_0,t).
\end{align}
Here, ${\mathbb V} = {\mathbb N}v_{\rm c}$ is the volume of the $d$-dimensional system; operator $\tilde\Lambda$ is given by Eq.~(\ref{eq:rates_introduced}) and is expressed in terms of the rates of phonon scattering (\ref{eq:scatter_rate}), (\ref{eq:scatter_rate1}), and (\ref{eq:short_range_trans_invariant}). In addition, in the transport equation (\ref{eq:transport_Wigner}) the Poisson bracket becomes $\{{}\Omega_\alpha,{}\Phi_\alpha\} = \vb_{\kb\alpha}\partial_\rb{}\Phi_\alpha$, where $\vb_{\kb\alpha} = \partial_{\kb}\omega_{\kb\alpha}$ is the group velocity of phonons of the branch $\alpha$ with the wave vector $\kb$. We note that the phonon-phonon collision term in Eq.~(\ref{eq:bar_lambda_uniform}) can be written in the same form as the collision term in the linearized phonon kinetic equation  for the correction to the phonon occupation number  \cite{Lifshitz1981a}.

We are looking for a nonstationary solution of Eq.~(\ref{eq:transport_Wigner}) with the initial condition (\ref{eq:initial_condition}) and for the Fourier transform of this solution convoluted with function $V^*_\alpha({}\rb,\kb)$. The expression for  $V^*_\alpha({}\rb,\kb)$ and the initial condition for $\Phi_\alpha$ are simplified in a spatially uniform system. Indeed, in the considered problem of decay of smoothly varying in space standing-wave type modes, of interest are the parameters of coupling to phonons $v_{\kappa'\kappa} = v_{\kb'\alpha',\kb\alpha}$  with $\alpha'=\alpha$ and $|\kb-\kb'|\ll |\kb|$. They are given by a Fourier transform $\zeta_\alpha({}\rb)$ of the squared strain tensor components of the considered mode convoluted with the polarization vectors of thermal phonons . Function $\zeta_\alpha ({}\rb)$ is calculated for the considered mode being in the ground vibrational state. It smoothly depends on ${}\rb$ on the scale $|\kb|^{-1}$. Thus $v_{\kb' \alpha,\kb \alpha}\approx {\mathbb V}^{-1}\tilde v_\alpha[(\kb +\kb')/2]\int d\rb \exp[i(\kb-\kb')\rb]\zeta_\alpha(\rb)$, and from Eq.~(\ref{eq:scalar_coupling}) the coupling vertex in the Wigner representation is
\begin{align}
\label{eq:V_uniform}
V_\alpha({}\rb,\kb)=\tilde v_\alpha(\kb)\zeta_\alpha({}\rb).
\end{align}
The coefficient $\tilde v_\alpha(\kb)$ is quadratic in $\kb$ for acoustic modes and is inversely proportional to $\omega_{\kb\alpha}$. One can think of $V_\alpha({}\rb,\kb)/\hbar$ as a scaled local (at a point $\rb$) change of the frequency of the mode $(\kb,\alpha)$ due to the squared strain from the considered low-frequency mode in its ground vibrational state.

In a uniform system, the density $\rho\equiv \rho_0$ is constant. Therefore, if the displacement field of the considered mode is  $q(t)\ub (\rb)$, we have the effective mass of the mode 
\[M_0 = \rho_0\int d\rb [\ub(\rb)]^2\]
[note that $\ub(\rb)$ is dimensionless].

\subsection{The eigenvalue problem for the collision term}
\label{subsec:eigenvalues}

The collision operator ${\rm St}[{}\Phi_\alpha({}\rb,\kb,t)$ is independent of ${}\rb$, it couples the values of ${}\Phi_\alpha({}\rb,\kb,t)$ with different $\kb$ and $\alpha$, but with the same ${}\rb$. One can therefore introduce right and left eigenvectors of operator St, which we denote by $\psi_\nu(\kb,\alpha)$ and $\Psi_\nu(\kb,\alpha)$, respectively, and which are independent of ${}\rb$,
\begin{align}
\label{eq:right_eigenvectors}
&{\rm St}[\psi_\nu(\kb,\alpha)]=-\ep_\nu \psi_\nu(\kb,\alpha),\nonumber\\
&\frac{\mathbb V}{(2\pi)^d}\sum_\alpha\int d\kb \Psi_{\nu'}(\kb,\alpha) \psi_\nu(\kb,\alpha) = \delta _{\nu,\nu'}.
\end{align}
Here, $\ep_\nu$ are eigenvalues of the collision operator ${\rm St}$. On physical grounds, since this operator describes phonon relaxation, Re~$\ep_\nu \geq 0$. The left eigenvectors $\Psi_\nu$ have the same eigenvalues as the right eigenvectors and are orthogonal to the right eigenvectors (or can be made orthogonal, for degenerate $\ep_\nu$), the condition indicated in Eq.~(\ref{eq:right_eigenvectors}).  The eigenvector with zero eigenvalue, $\ep_0= 0$,  is $\psi_0(\kb,\alpha)=c_0 \omega_{\kb\alpha}\bar n_{\kb \alpha}(\bar n_{\kb \alpha}+1)$ (for concreteness, in what follows  we set $c_0=\hbar$); the left eigenvector is $\Psi_0(\kb,\alpha)={\rm const}\times\omega_{\kb\alpha}$, where the constant is determined by the normalization condition (\ref{eq:right_eigenvectors}). 

Except for special cases, vectors $\psi_\nu(\kb,\alpha)$ provide a complete set, and therefore the solution of the transport equation can be sought in the form
\begin{align}
\label{eq:solution_eigenvectors}
{}\Phi_\alpha({}\rb,\kb,t) = \sum_\nu T_\nu({}\rb,t)\psi_\nu(\kb,\alpha).
\end{align} 
It follows from Eqs.~(\ref{eq:transport_Wigner}) and (\ref{eq:initial_condition}) that functions $T_\nu$ satisfy  equation
\begin{align}
\label{eq:R_dependent_functions}
&\partial_t T_\nu({}\rb,t) + \sum_{\nu'}\vb_{\nu\nu'}\partial_\rb T_{\nu'}({}\rb,t) = -\ep_\nu T_\nu ({}\rb,t),\nonumber\\
&\vb_{\nu\nu'}=\frac{\mathbb V}{(2\pi)^d}\sum_\alpha \int d\kb \Psi_{\nu}(\kb,\alpha)\vb_{\kb\alpha} \psi_{\nu'}(\kb,\alpha). 
\end{align}
with the initial condition
\begin{align}
\label{eq:initial_T_function}
T_\nu({}\rb,0)=&\frac{-i\mathbb V}{(2\pi)^d\hbar}\sum_\alpha \int d\kb \Psi_\nu(\kb,\alpha)
\nonumber\\
&
\times V_\alpha(\rb,\kb) \bar n_{\kb\alpha}(\bar n_{\kb\alpha}+1).
\end{align}

Both linear and nonlinear decay of low-frequency standing waves is determined by functions $T_\nu$, which correspond to even in $\kb$ eigenfunctions $\psi_\nu(\kb,\alpha)$, as $\Phi_\alpha({}\rb,\kb,0)$ in this case is even in $\kb$. The separation of eigenfunctions in even and odd in $\kb$ is a consequence of the symmetry of operator $\tilde\Lambda_{\kb\alpha}^{\kb_0\alpha_0}$ with respect to the simultaneous inversion $\kb\to -\kb, \kb_0 \to -\kb_0$.

\subsection{Thermal diffusion}
\label{subsec:diffusion}

We are interested in the evolution of ${}\Phi_\alpha({}\rb,\kb,t)$ on the time scale $\sim \omega_0^{-1}$, which largely exceeds the reciprocal frequencies of thermal phonon $\sim \hbar/k_BT$. Another time scale is the characteristic relaxation time of thermal phonons  $\tau_r= l_T/c_s$. 
 Yet another relevant time scale is the relaxation time of the spatial nonuniformity, which characterizes the decay of functions $T_\nu({}\rb,t)$. Of particular interest is the decay of function $T_0({}\rb,t)$, since this may be the slowest decay which determines the long-time behavior of ${}\Phi_\alpha({}\rb,\kb,t)$.

If the decay of $T_0({}\rb,t)$ is slow on the time scale $({\rm Re}~\ep_{\nu>0})^{-1}$, one can describe it in the adiabatic approximation familiar from the analysis of spatial diffusion in phonon or electron systems. Physically, it corresponds to the picture in which there is locally (for given ${}\rb$) formed a thermal distribution of phonons with a coordinate-dependent temperature. Formally, in Eq.~(\ref{eq:R_dependent_functions}) one assumes that functions $T_{\nu>0}$ adiabatically follow function $T_0$, so that $T_\nu({}\rb,t)\approx -\ep_\nu^{-1}\vb_{\nu 0}\partial_\rb T_0({}\rb,t)$ for $\nu > 0$. Then the equation for $T_0$ reads
\begin{align}
\label{eq:adiabatic_T_0}
&\partial_t T_0({}\rb,t) = \sum_{ij}\partial_{r_i} D_{ij} \partial_{r_j}T_0({}\rb,t), \nonumber\\
 &D_{ij}=\sum_{\nu>0}(\vb_{0\nu})_i(\vb_{\nu 0}))_j/\ep_\nu.
\end{align}

Function $T_0({}\rb,t)$ can be thought of as a scaled coordinate-dependent correction to the temperature of high-frequency phonons. This interpretation follows from $\Phi_\alpha({}\rb,\kb,t)$ being the weighted Wigner phonon density matrix, which comes to local thermal equilibrium for a given $T_0({}\rb,t)$. In our perturbation theory $T_0({}\rb,t)$  is generally complex and small in the absolute value; the occupation numbers of thermal phonons are close to their equilibrium values $\bar n_{\kb\alpha}$.  

Equation (\ref{eq:adiabatic_T_0}) has the form of the equation of thermal diffusion. It is easy to see that the diffusion coefficients $D_{ij}$ are real. Indeed, operator St in Eq.~(\ref{eq:bar_lambda_uniform}) is real, $\tilde \Lambda = \tilde \Lambda^*$. Therefore the eigenvalues $\ep_\nu$ are either real, and then $\psi_\nu$ and $\Psi_\nu$ can be chosen real (the extension to degenerate eigenvalues is trivial), or $\ep_\nu$ form pairs of complex conjugate values, and then $\psi_\nu$ form complex conjugate pairs as well, as do also $\Psi_\nu$. Moreover, using Eq.~(\ref{eq:right_eigenvectors}) $D_{ij}$ can be nicely expressed in terms of the propagator $\Pi_{\alpha\alpha'}(\kb,\kb';t)$ which satisfies equation $\partial_t \Pi_{\alpha\alpha'}(\kb,\kb';t) = {\rm St}[\Pi_{\alpha\alpha'}(\kb,\kb';t)]$ with initial condition $\Pi_{\alpha\alpha'}(\kb,\kb';0) = [(2\pi)^d/{\mathbb V}]\delta (\kb - \kb')\delta_{\alpha\alpha'}$,
\begin{align}
\label{eq:D_correlator}
&D_{ij}=\frac{\mathbb V^2}{(2\pi)^{2d}}\sum_{\alpha,\alpha'}\int d\kb d\kb'\int_0^\infty dt \Psi_0(\kb,\alpha)\Pi_{\alpha,\alpha'}(\kb,\kb';t)\nonumber\\
&\times \psi_0(\kb',\alpha')[(\vb_{\kb\alpha})_i(\vb_{\kb'\alpha'})_j +(\vb_{\kb\alpha})_j(\vb_{\kb'\alpha'})_i]/2.
\end{align}
Equation (\ref{eq:D_correlator}) relates the diffusion coefficients to an analog of the integral over time of the time correlation function of the group velocity. One can see that the coefficients $D_{ij}$ have the same form as the thermal diffusion coefficient calculated for thermal phonons \cite{Lifshitz1981a}. 

The expression for $D_{ij}$ simplifies for low temperatures and small densities of short-range scatterers. In this case  relaxation  of the total phonon momentum becomes slow. Formally, $D_{ij}$ diverges in an ideal crystal for $T\to 0$, the well-known divergence of the thermal diffusion coefficient in this limit \cite{Lifshitz1981a}. The slowness of the relaxation for nonzero $T$ and a nonideal crystal means that the eigenfunction of the St operator 
$\psi_1(\kb,\alpha)\propto \kb \bar n_{\kb\alpha}(\bar n_{\kb\alpha}+1)  $ has a small eigenvalue $\ep _1 \ll {\rm Re}~\ep_{\nu>1}$, which is given by the sum of the rates of umklapp processes and short-range scattering. 
The matrix element $\vb_{0\nu}$ with $\nu=1$ gives the major contribution to the matrix elements $D_{ij}$. The value $\ep_1^{-1}$ determines the phonon relaxation time $\tau_{\rm th}$ in the frequently used expression for the bulk thermal conductivity $\tfrac{1}{3}C_vc_s^2\tau_{\rm th}$ [the specific heat $C_v$ (per unit volume) is given in Eq.~(\ref{eq:Cv}) below].

\section{Nonlinear thermoelastic damping}
\label{sec:thermoelastic}

Conventionally, the damping mechanisms of low-frequency modes are separated into thermoelastic and Akhiezer relaxation.  In our formulation, the thermoelastic relaxation is described by the effective scaled temperature $T_0({}\rb,t)$. It evolves in time due to the spatial nonuniformity created by the considered slowly vibrating mode, with the decay rate determined by the thermal diffusion coefficients $D_{ij}$. The Akhiezer relaxation is described by functions $T_{\nu >0}$, which decay on times $\ep_{\nu>0}^{-1}$ determined by the scattering rates of thermal phonons.

In this section we study the thermoelastic mechanism of nonlinear damping. This mechanism is relevant if the frequency of the considered mode $\omega_0$ is close to the reciprocal characteristic time of thermal diffusion \cite{Zener1937} and $\omega_0\ll \ep_{\nu>0}$. Nonlinear thermoelastic damping is particularly interesting for systems where the conventional linear thermoelastic damping is small. To this end, we consider Lam\'e modes in micromechanical systems and flexural modes in nanotubes and thin membranes. To illustrate the approach, we limit the analysis to isotropic systems.

The thermoelastic contribution to the nonlinear decay rate is determined by function $\Phi_\alpha$ for times $t\gg \ep_{\nu>0}^{-1}$. In this time range $\Phi_\alpha({}\rb,\kb,t) \approx T_0({}\rb,t) \psi_0(\kb,\alpha)$. Function $T_0$ is described by the diffusion equation (\ref{eq:adiabatic_T_0}) with $D_{ij}=D\delta_{ij}$. The initial condition for this equation follows from Eqs.~\eqref{eq:initial_condition} and (\ref{eq:initial_T_function}). With the account taken of the explicit form of $\psi_0(\kb,\alpha)$ and $\Psi_0(\kb,\alpha)$  we have
\begin{align}
\label{eq:initial_T_0}
&\partial_tT_0({}\rb,t) = D\n^2 T_0({}\rb,t),\\
&T_0({}\rb,0)= -i \sum_\alpha\int \frac{\omega_{\kb\alpha}d\kb}{(2\pi)^d} \,\frac{V_\alpha({}\rb,\kb) \bar n_{\kb\alpha} (\bar n_{\kb\alpha} + 1)}{k_BC_vT^2}. \nonumber
\end{align}
Here, $\n \equiv \partial_{{}\rb}$. Parameter $C_v$ is the specific heat,
\begin{align}
\label{eq:Cv}
C_v = \sum_\alpha \int \frac{d\kb}{(2\pi)^d}\, \frac{\hbar^2\omega_{\kb\alpha}^2  \bar n_{\kb\alpha} (\bar n_{\kb\alpha} + 1)}{k_BT^2}. 
\end{align}

From Eq.~(\ref{eq:nonlin_friction_scalar}), the rate of nonlinear damping due to thermoelastic relaxation is
\begin{align}
\label{eq:therm_decay_general}
\Gamma^{(\rm nl)} _{\rm th} =& -\frac{2\omega_0}{k_BT} {\rm Im} \sum_\alpha\int \frac{d{}\rb d\kb}{(2\pi)^d}V^*_\alpha({}\rb,\kb)\hbar\omega_{\kb\alpha} \bar n_{\kb\alpha} (\bar n_{\kb\alpha} + 1)\nonumber\\
&\times\int_0^\infty dt\,T_0({}\rb,t)\exp(2i\omega_0t)
\end{align}

The boundary conditions for function $T_0({}\rb,t)$ have a simple form in the limiting cases of a free boundary and a boundary where the system is clamped and is strongly coupled to the environment. To establish them we will use that, in the kinetic-equation approximation, $\Phi_\alpha({}\rb,\kb,t)$ differs from the phonon density matrix in the Wigner representation $\rho_\alpha({}\rb,\kb,t)$ only by a time-independent factor. The energy flux at a given ${}\rb$ is determined by $\qb=(2\pi)^{-d}\sum_\alpha \int d\kb \,\hbar\omega_{\kb \alpha}(\partial \omega_{\kb \alpha}/\partial \kb)\rho_\alpha ({}\rb,\kb,t)$.  Function $\rho_\alpha$ can be expanded in the same way as $\Phi_\alpha$, Eq.~(\ref{eq:solution_eigenvectors}), $\rho_\alpha ({}\rb,\kb,t)=(k_B T^2)^{-1}\sum_\nu \tilde T_\nu({}\rb,t)\psi_\nu(\kb,\alpha)$. We then find that, in an isotropic system,  $\qb \approx - C_vD\n \tilde T_0({}\rb,t)$ for $t\gg \ep_{\nu>0}^{-1}$.  The scaled solution of the transport equation $\rho_\alpha(\rb,\kb,t) = \tilde T_0(\rb,t)\psi_0(\kb,\alpha)/k_BT^2$ has a form of the correction to the equilibrium phonon distribution $\bar n_{\kb\alpha}$; this correction is calculated for the temperature $T+\tilde T_0(\rb,t)$ to the first order in $\tilde T_0$, assuming that it smoothly depends on coordinate and time.  This provides a physical insight into the meaning of $\tilde T_0$.

At a free boundary,  there is no energy flux in the normal direction $\hat\nb$. This means $(\nb\cdot\n)\tilde T_0=0$ on the boundary. Functions $T_\nu$ differ from $\tilde T_\nu$ only by a time-independent coefficient. Then the boundary condition for $T_0$ is $(\nb\cdot\n) T_0=0$. A similar analysis can be done for the boundary where there is maintained thermal equilibrium with the environment. Here, the temperature of thermal phonons in the system is equal to the temperature of the environment, and then $T_0=0$.
 
Both for a thermally insulated and thermal equilibrium boundaries function $T_0({}\rb,t)$ can be expanded in the eigenfunctions $T_{0,\mu}({}\rb)$ of the diffusion equation complemented by the corresponding boundary conditions, $D\n^2 T_{0,\mu}= -\lambda_\mu T_{0,\mu}$. Functions $T_{0,\mu}(\rb)$ form a complete orthogonal set, and we can normalize them as $\int d{}\rb T_{0,\mu}^*({}\rb)T_{0,\mu'}({}\rb)=\delta _{\mu \mu'}$.  

From Eqs.~(\ref{eq:initial_T_0}) - \eqref{eq:therm_decay_general} 
\begin{align}
\label{eq:Gamma_nl_thermal}
&\Gamma^{(\rm nl)}_{\rm th} = 2\hbar \omega_0 C_vT\sum_\mu|A_\mu|^2 \frac{\lambda_\mu}{\lambda_\mu^2 + 4\omega_0^2},\nonumber\\
&A_\mu =  \int d{}\rb\, T_{0,\mu}^*({}\rb)T_0({}\rb,0).
\end{align} 
Equation~\eqref{eq:Gamma_nl_thermal} gives the thermoelastic nonlinear decay rate in a simple form. It relates  $\Gamma^{(\rm nl)}_{\rm th}$ to the eigenmodes of the thermal diffusion equation. An immediate consequence is the observation that the mechanism is relevant if the temperature relaxation rates $\lambda_\mu$ are of the same order as the mode frequency $\omega_0$, as is also the case for thermoelastic linear damping \cite{Zener1937,Lifshitz2000}. However, the structure of the coefficients $A_\mu$ is different for linear and nonlinear damping, and for some modes their rates are significantly different.

\subsection{Nonlinear Gr\"uneisen parameters}
\label{subsec:Gruneisen}
Nonlinear damping of low-frequency modes depends on their coupling to thermal phonons. A simple form of such coupling is described in terms of the nonlinear Gr\"uneisen parameters. It is an extension of the conventional deformation potential model of the linear coupling, in which the frequencies of thermal modes are assumed to linearly depend on the strain from the low-frequency modes, cf. \cite{Akhiezer1938,Gurevich1988}. For the nonlinear coupling we will assume that the frequency change is quadratic in the deformation. 

We will consider isotropic systems. Here, the quadratic invariants are the squares of the linear  hydrostatic strain tensor $\hat\epsilon^{\rm (h)}$ and the shear strain tensor $\hat\epsilon^{\rm (s)}$, and also the quadratic in the deformation term in the strain tensor; the latter is a scalar and has the form $\eq=\tfrac{1}{2}(\partial u_i/\partial r_j)^2$, where we assume summation over repeated indices \cite{LL_Elasticity}. Then using Eq.~(\ref{eq:V_uniform}) we can write
\begin{align}
\label{eq:Gruneisen_param}
V_\alpha({}\rb,\kb) =& -\hbar \omega_{\kb\alpha}\left\{\gamma_{\kb\alpha}^{(1)}\eq({}\rb) + 
\gamma_{\kb\alpha}^{(2)}[{\rm Tr}\,\eh({}\rb)]^2 \right.\nonumber\\
&\left. +\gamma_{\kb\alpha}^{(3)}{\rm Tr}\,[\es({}\rb)]^2\right\}.  
\end{align}
Here, $\gamma_{\kb\alpha}^{(1)}$ is the standard Gr\"uneisen parameter that relates the frequency shift of the mode to the volume change, except that here it multiplies the quadratic in the deformation term. The dimensionless parameters $\gamma_{\kb\alpha}^{\rm (2,3)}$ describe the relative change of the vibration frequency $\omega_{\kb\alpha}$ of a thermal phonon due to the hydrostatic and shear strain, respectively. 

The value of the coupling parameter $V_\alpha(\rb,\kb)$ is given by Eq.~(\ref{eq:Gruneisen_param}) in which $\eq, \eh, \es$ are calculated for the strain created by the considered low-frequency mode in the ground quantum state. The typical scale of the displacement field of the mode $\ub({}\rb)$ is the amplitude of zero-point vibrations $q_0$, Eq.~(\ref{eq:zero_point}). Equation (\ref{eq:Gruneisen_param}) allows one to express the coefficients $A_\mu$ in the expression for the nonlinear decay rate (\ref{eq:Gamma_nl_thermal}) in terms of the integral strain invariants,
\begin{align}
\label{eq:A_mu_Gruneisen}
A_\mu =& i\hbar^{-1}\int d\rb T_{0,\mu}^*({}\rb)[\left\{\tilde\gamma^{(1)}\eq({}\rb) + 
\tilde\gamma^{(2)}[{\rm Tr}\,\eh({}\rb)]^2 \right.\nonumber\\
&\left. +\tilde\gamma^{(3)}{\rm Tr}\,[\es({}\rb)]^2\right\}
\end{align}
The dimensionless constants $\tilde\gamma^{(j)}$ in this expression are determined by the nonlinear Gr\"uneisen parameters,
\begin{align}
\label{eq:Gruneisen_avg}
\tilde\gamma^{(j)} =
\sum_\alpha\int \frac{d\kb}{(2\pi)^d}\gamma_{\kb\alpha}^{(j)}\;\frac{\hbar^2\omega_{\kb\alpha}^2\bar{n}_{\kb\alpha}(\bar{n}_{\kb\alpha}+1)}{k_BC_vT^2}
\end{align}
where $j=1,2,3$.

\subsection{Nonlinear damping of Lam\'e modes}
\label{subsec:Lame}

We now consider the thermoelastic nonlinear damping of Lam\'e modes. Such modes emerge in thin rectangular plates with an integer ratio of the side lengths, and in particular, in thin square micro- or nanomechanical plates \cite{Graff1975}. An important feature of Lam\'e modes is that the displacement is in the plane and is not accompanied by volume dilatation, ${\rm Tr} \, \eh =0$. Therefore in isotropic systems there is no standard linear thermoelastic relaxation of these modes \cite{Chandorkar2009}.  This facilitates the use of Lam\'e modes in high-quality frequency generators. At the same time, this makes the nonlinear thermoelastic damping relatively strong even for small vibration amplitudes.

To analyze the nonlinear damping, we assume that the plate is a square of side $L$ in the $xy$-plane, $0\leq x,y \leq L$. The thickness of the plate is $H\ll L$. We consider the displacement with components $u_x$ and $u_y$ in the plane of the plate and seek a solution of the elasticity equations that is independent of the coordinate normal to the plane and has ${\rm Tr}\, \eh =0$. The lowest-frequency solution for the mode in the ground quantum state is
\begin{align}
\label{eq:u_Lame}
u_x({}\rb) &= \sqrt{2}q_0\cos(\pi x/L)\sin(\pi y/L), \\
u_y({}\rb) &= -\sqrt{2}q_0\sin(\pi x/L)\cos(\pi y/L),\nonumber 
\end{align}
where ${}\rb=(x,y)$ and we set the boundaries so that $0\leq x,y\leq L$. The mode eigenfrequency is $\omega_0=\sqrt2\pi c_t/L$, where $c_t$ is the velocity of transverse sound waves, $q_0$ is the amplitude of zero-point vibrations [cf. Eq.~(\ref{eq:zero_point})], and the effective mass of the vibrations is  $M=\rho_0 L^2H$, where $\rho_0$ is the density.

Since Lam\'e modes refer to systems with free boundaries, in solving the diffusion equation for the scaled temperature $T_0$ we will assume that $(\n T_0)\hat n =0$ at the boundaries. The diffusion equation for $T_0$ (\ref{eq:initial_T_0}) is two-dimensional, $T_0$ is independent of the coordinate normal to the plate. The normal modes $T_{0,\mu}$ that contribute to the thermoelastic damping have the form
\begin{align}
\label{eq:T_nm}
&T_{0,\mu}({}\rb ) = (2^\zeta/LH^{1/2}) \cos(2m\pi x/L)\cos(2n\pi y/L), \nonumber\\
&\lambda_{\mu}=\Lambda_D(m^2+n^2),\qquad \Lambda_D=4\pi^2DL^{-2}. 
\end{align}
The mode number $\mu\equiv (n,m)$ is a set of integers $n,m\geq 0, n+m>0$, and  $\zeta = 1-\tfrac{1}{2}(\delta_{n0} + \delta_{m0})$. The thermal diffusion coefficient $D$ takes on its bulk value if the thickness of the plate exceeds the phonon mean free path. Otherwise $D$ should be calculated with the account taken of phonon scattering at the surfaces.

Equations (\ref{eq:initial_T_0}), (\ref{eq:Gamma_nl_thermal}), and (\ref{eq:Gruneisen_param}) - (\ref{eq:T_nm}) give 
\begin{align}
\label{eq:damping_Lame}
\Gamma_{\rm th}^{(\rm nl)} =&\frac{q_0^2\omega_0^4}{16\rho_0c_t^4}C_vT\left[ (\tilde \gamma^{(1)} + \tilde\gamma^{(3)})^2\frac{2\Lambda_D}{4\Lambda_D^2 + 4\omega_0^2} 
\right. \nonumber \\
&\left. +4\tilde\gamma^{(3)\,2}\frac{\Lambda_D}{\Lambda_D^2 + 4\omega_0^2}\right].
\end{align}

Nonlinear thermoelastic damping of Lam\'e modes is determined by the diffusion of heat towards the sides of the plate. The characteristic diffusion rate $\Lambda_D$ depends on temperature and quickly falls off with the increasing size of the plate: $\Lambda_D\propto L^{-2}$, whereas the mode frequency $\omega_0\propto L^{-1}$. The condition that the mean free path of thermal phonons is small, $l_T\ll L$, which was used in obtaining Eq.~(\ref{eq:damping_Lame}), implies that $\omega_0\sim c_s/L \gg \Lambda_D \sim c_sl_T/L^2$.  For comparatively large systems and high temperatures the ratio $\omega_0/\Lambda_D$ is very large, making the damping rate (\ref{eq:damping_Lame}) small; for example, for Si plate of length $L=10\mu$m  and for room temperatures $\omega_0/\Lambda_D \approx 80$. Nonlinear friction of Lam\'e modes is easier to observe in smaller systems and for lower temperatures. In particular, the factor $\omega_0^4$ in $\Gamma_{\rm th}^{\rm(nl)}$ increases as $L^{-2}$ with decreasing $L$. 

The temperature dependence of the nonlinear damping is determined by  the factor $C_vT$ and the thermal conductivity that enters $\Lambda_D$. In 3D systems, for temperatures above the Debye temperature, $C_vT\propto T$, whereas for low temperatures $C_vT\propto T^4$ . For high temperatures, the phonon momentum relaxation rate is determined primarily by phonon-phonon collisions, and then $D\propto T^{-1}$, whereas for low temperatures the phonon momentum relaxation rate is determined primarily by scattering from defects, and then $D\propto T^{-4}$, cf.~\cite{Lifshitz1981a,Gurevich1988}.

\subsection{Nonlinear damping of flexural modes}
\label{subsec:flexural_mode_calculation}

Linear thermoelastic damping of flexural modes in thin clamped beams was the first example of thermoelastic relaxation \cite{Zener1937}. The relaxation comes from the  heat flux generated by the bending beam, which propagates transverse to the beam. If the beam is thin, so that the rate of thermal diffusion transverse to the beam exceeds the vibration frequency, this mechanism becomes inefficient, cf. Ref.~\onlinecite{Lifshitz2000}. However, as we show here, nonlinear thermoelastic damping can be relevant in this case.

The standard geometry for the analysis of damping of flexural modes \cite{Zener1937,Lifshitz2000} is a thin doubly-clamped beam with width $W$ and thickness $H$ small compared to the length $L$ and the curvature of the bending. We assume that all lengths exceed the phonon mean free path and choose the $x$-axis to be along the beam, $0\leq x\leq L$, and the $z$-axis to be in the bending direction; we set $z=0$ at the  center of the beam, $-H/2 \leq z \leq H/2$. 

The mode frequencies and the linear strain in thin beams are well known \cite{LL_Mechanics2004}. For the lowest-frequency mode the eigenfrequency is $\omega_0= a_{\rm fl}^2(H/L^2)(E/12\rho_0)^{1/2}$, where $E$ is the Young modulus and $a_{\rm fl}\approx 4.73$. The diagonal components of the linear strain in $\eh$ and the strain $\es$ are small. Respectively, in Eq.~(\ref{eq:Gruneisen_param}) for $V_\alpha$ the invariants $[{\rm Tr}\,\epsilon^{\rm (h)}]^2, \, [{\rm Tr}\,\epsilon^{\rm (s)}]^2$ can be disregarded. The major contribution to $V_\alpha$ comes from the invariant $\epsilon^{\rm (q)} \approx (\partial Z/\partial x)^2$, where $Z$ is the displacement in the bending direction.

The modes $T_{0,\mu}(\rb)$ that contribute to the nonlinear damping are even with respect to the middle plane of the beam and are also symmetric with respect to the reflection $x\to L-x$.   They solve the diffusion equation (\ref{eq:initial_T_0}) with the boundary conditions of the absence of heat flux on the sides of the beam, whereas $T_0=0$ at the clamping points $x=0$ and $x=L$,
\begin{align}
\label{eq:Tmn_flexural}
T_{0,\mu}({}\rb) = \frac{2^\zeta}{(LWH)^{1/2}}\sin\frac{(2n+1)\pi x}{L}\,\cos\frac{2m\pi z}{H},
\end{align}
where the mode number $\mu\equiv (n,m)$ is a set of integers, $n,m\geq 0$, and $\zeta = 1-\tfrac{1}{2}\delta_{m0}$. The corresponding eigenvalues of the thermal diffusion equation are 
\begin{align}
\label{eq:lambda_nm_flexural}
\lambda_{nm}=\Lambda_D\left[ \frac{1}{4}(2n+1)^2 +  m^2(L/H)^2\right]
\end{align}
[$\Lambda_D$ is given by Eq.~(\ref{eq:T_nm})].

For thin beams, the ratio $L/H$ is large and the mode eigenfrequency $\omega_0$ is small. Respectively, we consider slow  modes of the scaled temperature, for which $\lambda_{mn}$ is small. Such modes correspond to $m=0$ and refer to temperature propagation {\it along} the beam, rather than transverse to the beam, as considered by Zener \cite{Zener1937}. 

Equations (\ref{eq:initial_T_0}), (\ref{eq:Gamma_nl_thermal}), (\ref{eq:Gruneisen_param}), and (\ref{eq:Tmn_flexural}) describe the thermoelastic nonlinear damping of a flexural mode in the explicit form.  We will keep only the contribution to $\Gamma^{\rm (nl)}_{\rm th}$ from the thermal modes with $m=0$ and the term $\propto \epsilon^{\rm (q)}$ in the nonlinear Gr\"uneisen parameter.
The shape of the flexural mode $Z(x)$ for a beam is an elementary function of $a_{\rm fl}x/L$ \cite{LL_Mechanics2004}. We will normalize it so that $Z(x)=q_0\tilde Z(x/L)$, where $\int dx [\tilde Z(x/L)]^2 =1$; this gives the right description of the displacement in the ground vibrational state, with the mass of the beam $M_0=\rho_0 LHS$. Then
\begin{align}
\label{eq:Gamma_thermo_flex}
\Gamma_{\rm th}^{\rm (nl)} = &\frac{24q_0^2}{a_{\rm fl}^4H^2 E}\omega_0 C_vT \tilde\gamma^{(1)\,2}\nonumber\\
&\times \sum_n f_{\rm fl}(n)^2\omega_0\lambda_{n0}/(\lambda_{n0}^2 + 4\omega_0^2),
\end{align}
where $\lambda_{n0}$ is given by Eq.~(\ref{eq:lambda_nm_flexural}). Function $f_{\rm fl}(n)=\int_0^1 dy (d\tilde Z(y)/dy)^2\sin[(2n+1)\pi y]$ is equal to $\approx 7.9$ for $n=0$ and rapidly falls off for large $n$; $|f_{\rm fl}(n)| < 0.2$ for $n\geq 5$.

It is seen from Eq.~(\ref{eq:Gamma_thermo_flex}) that nonlinear damping is important if the rate of thermal diffusion along the beam $\Lambda_D$ is comparable to the frequency of the flexural mode $\omega_0$. The factor in front of the sum in  Eq.~(\ref{eq:Gamma_thermo_flex}) scales with the beam dimensions as $(LWH^{3})^{-1}$, whereas $\lambda_{n0}\propto \Lambda_D$ is independent of $H$. Therefore the role of nonlinear damping is increasing fast with the decreasing thickness of the beam. We note that the condition that $H$ largely exceeds the mean free path of thermal phonons $l_T$ is incompatible with $\omega_0 \sim c_s H/L^2$ being of the order of $\Lambda_D \sim c_sl_T/L^2$.  With decreasing $H$ the role of surface scattering increases, and ultimately $l_T$ may become $\sim H$, making $\omega_0$ and $\Lambda_D$ comparable. We expect the nonlinear friction to be comparatively strong also in very thin beams where phonon motion normal to the beam is quantized.

\begin{figure}[t!]
\centering
\includegraphics[width=\linewidth,trim=.cm 0.cm 0cm 0cm, clip=true]{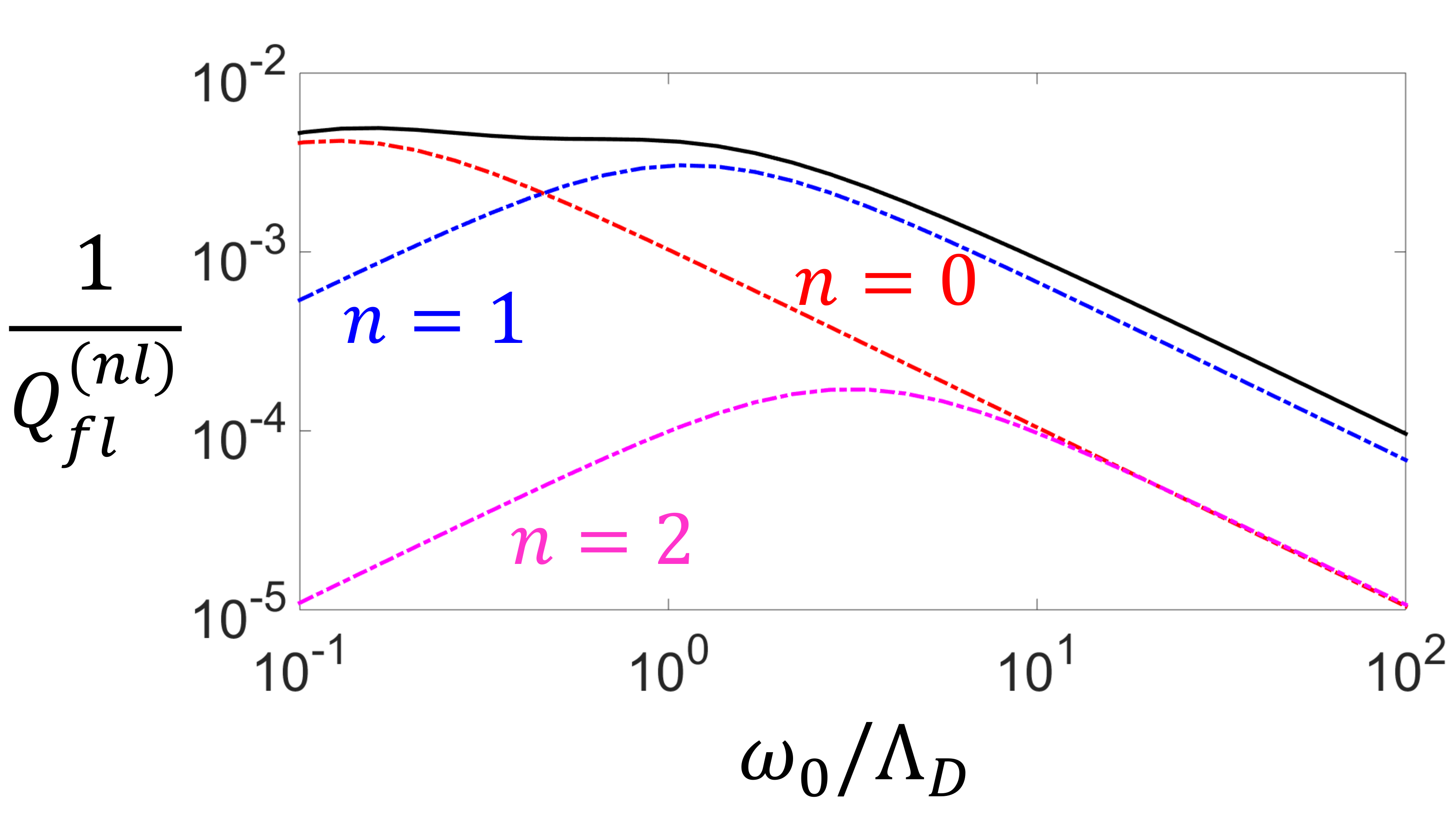}
\caption{The reciprocal effective quality factor for nonlinear decay $Q_{\rm th}^{(\rm nl)}$, Eq.~(\ref{eq:nonlinear_Q}), for the fundamental flexural mode of a doubly-clamped beam.  The material parameters refer to silicon at room temperature,  $D=1$~cm$^2$s$^{-1}$. The ratio $\omega_0/\Lambda_D$ can be controlled by varying the beam thickness $H$. The dashed lines show partial contributions from thermal diffusion modes, which are described by the individual terms in the sum in Eq.~\eqref{eq:Gamma_thermo_flex}. The solid line includes all terms in the sum.}
\label{fig:NL_flexural_thermoelastic_damping}
\end{figure}  

In Fig.~\ref{fig:NL_flexural_thermoelastic_damping} we show the evolution of the scaled reciprocal nonlinear damping rate
\begin{align}
\label{eq:nonlinear_Q}
Q^{\rm (nl)}_{\rm fl} = \frac{\omega_0}{2\Gamma^{\rm (nl)}}\,\frac{q_0^2}{H^2},
\end{align}
with the varying ratio $\omega_0/\Lambda_D$. The scaling in Eq.~(\ref{eq:nonlinear_Q}) is such as to make $Q^{\rm (nl)}_{\rm fl}$ depend on the geometry of the beam only in terms of $\omega_0/\Lambda_D$ which, in turn, depends only on the thickness $H$.  The scaled reciprocal damping rate $Q^{\rm (nl)}_{\rm fl}$ depends also on the material parameters, and in particular on the thermal expansion coefficient, which is determined by $\tilde\gamma^{(1)}$. 

From  Eq.~(\ref{eq:VdP}), $Q^{\rm (nl)}_{\rm fl}$ can be thought of as a nonlinear analog of the ``quality factor'', i.e., a measure of the nonlinear-damping induced energy dissipation per vibration period for the vibration amplitude $H$. The analysis of this section refers to  vibration amplitudes small compared to $H$, and therefore the nonlinear quality factor for them is higher than $Q^{\rm (nl)}_{\rm fl}$. Fig.~\ref{fig:NL_flexural_thermoelastic_damping} shows that $Q^{\rm (nl)}_{\rm fl}$ as a function of the mode frequency displays two well-resolved maxima in the region where $\Lambda_D = \pi^2D/L^2 $ is close to $\omega_0$. They correspond to the two lowest modes of thermal diffusion along the beam.

\subsection{Phenomenological theory of nonlinear thermoelastic decay}
\label{subsec:phenomen_thermoelastic}
The phenomenological theory of linear thermoelastic relaxation relates this relaxation to thermal expansion \cite{LL_Elasticity}.  In isotropic systems, the relaxation comes from the term $\propto - (T-T_0){\rm Tr}\,\eh$ in the free energy density, where $T_0$ is the ambient temperature and $T$ is the local temperature.  Because of this term, low-frequency vibrations produce heat, which is dissipated through thermal conductivity. The underlying assumption is that the vibration period is long compared to the time needed for high-frequency phonons to thermalize, locally. In other words, $\omega_0\ll \ep_{\nu}$ for $\nu>0$, where $\ep_\nu$ are the eigenvalues of the collision operator of high-frequency phonons, see Eq.~(\ref{eq:right_eigenvectors}). Respectively, the mean free path of high-frequency phonons is small compared to the characteristic wavelength of the low-frequency vibrations.

A phenomenological theory of nonlinear thermoelastic damping can be developed in a similar way. A correction to the free energy density $\delta F$ that describes such damping should be quadratic in strain. In an isotropic system it can be written in terms of the invariants introduced in Eq.~(\ref{eq:Gruneisen_param}). The model 
(\ref{eq:Gruneisen_param}) describes the shifts $\delta\omega_{\kb\alpha}= \hbar^{-1}V_\alpha(\rb,\kb)$ of the frequencies of high-frequencies phonons due to their coupling to the considered mode, which are quadratic in the displacement of this mode. 
The correction $\delta F$ due to such shifts 
can be obtained in the same way as for the shifts which are linear in the mode displacement \cite{LL_Elasticity,Gurevich1988}. For a small mode-induced temperature change the result reads
\begin{align}
\label{eq:delta_F}
\delta F=& -C_v\delta T f_F\bigl(\hat\epsilon(\rb)\bigr), \qquad f_F(\hat\epsilon) = \left\{\tilde\gamma^{(1)}\eq\right.\nonumber\\
&\left. 
 + 
\tilde\gamma^{(2)}[{\rm Tr}\,\eh]^2 +\tilde\gamma^{(3)}{\rm Tr}\,[\es]^2\right\},
\end{align}
where $\delta T=T-T_0$ and $\tilde\gamma^{(1,2,3)}$ are given by Eq.~(\ref{eq:Gruneisen_avg}). It should be noted that a quadratic in the strain term in the free energy density comes also from the linear in the strain coupling taken to the second order of the perturbation theory. For an isotropic system this term is propotional to the square of the linear hydrostatic strain $\eh$. For the modes that we consider it is small and will not be discussed.

The analysis of nonlinear damping described by Eq.~(\ref{eq:delta_F}) is similar to the corresponding analysis for linear damping \cite{LL_Elasticity,Lifshitz2000}. From Eq.~(\ref{eq:delta_F}), the extra term in the entropy density $\delta S=-\partial \delta F/\partial T$ is a function of time- and coordinate-dependent temperature increment $\delta T$ and the strain $\hat\epsilon$. Taking into account that $\partial S^{(0)}/\partial T=C_v/T$ ($S^{(0)}$ is the unperturbed entropy density; the difference between $C_p$ and $C_v$ can be disregarded within the perturbation theory), we write the heat balance equation as
\begin{align}
\label{eq:heat_balance}
\delta \dot T = D\n^2 \delta T - T\dot\epsilon_{ij}\,\partial f_F(\hat\epsilon)/\partial\epsilon_{ij}
\end{align}
(we have also disregarded the difference between $T$ and $T_0$ in the last term).

The equation of motion for the displacement $q(t)\ub(\rb)$ of the considered mode is
\begin{align}
\label{eq:eom_free_energy}
\rho_0\ddot q u_i = \frac{\partial \sigma^{(0)}_{ij}}{\partial r_j} -C_v \frac{\partial }{\partial r_j } \delta T\frac{\partial  f_F(\hat\epsilon)}{\partial \epsilon_{ij}}
\end{align}
where $\hat\sigma^{(0)}$ is the stress tensor calculated by disregarding the coupling to high-frequency phonons. To the leading order it is linear in $\hat\epsilon$; the mode displacement $q\ub$ is the eigenvector of $\partial \hat\sigma^{(0)}/\partial \rb$ with the eigenvalue $\propto \omega_0^2$. 

In Eqs.~(\ref{eq:heat_balance}) and (\ref{eq:eom_free_energy}) $\epsilon_{ij} = [q(t)/2](\partial u_i/\partial r_j + \partial u_j/\partial r_i)$. Given that $f_F(\hat\epsilon)$ is quadratic in $\epsilon$, we have the following useful relations: 
\begin{align*}
f_F(\hat\epsilon) = q^2(t) f_F(\hat\epsilon/q), \qquad \epsilon_{ij}\partial f_F/\partial \epsilon_{ij} = 2 f_F(\hat\epsilon).
\end{align*} 
Function $f_F(\hat\epsilon/q)$ is independent of $q$ and, thus, of time. Using these relations, we can expand the last term in Eq.~(\ref{eq:heat_balance}) in the eigenfunctions $T_{0,\mu}(\rb)$ of the diffusion operator, $D \n^2 T_{0,\mu}=-\lambda_\mu T_{0,\mu}$,  
\begin{align}
\label{eq:epsilon_to_A_mu}
\dot\epsilon_{ij}\partial f_F/\partial \epsilon_{ij} = -2i\hbar q\dot q q_0^{-2}\sum_\mu A_\mu T_{0,\mu}(\rb),
\end{align}
where we  used  Eq.~(\ref{eq:A_mu_Gruneisen}) for the coefficients $A_\mu$ and took into account that these coefficients were calculated in Eq.~(\ref{eq:A_mu_Gruneisen}) for $q=q_0$. 

Equation (\ref{eq:epsilon_to_A_mu})  can be also used to simplify Eq.~(\ref{eq:eom_free_energy}). If we multiply Eq.~(\ref{eq:eom_free_energy}) by $u_i(\rb)$, sum over $i$, and integrate over $\rb$ taking into account that $\int \rho_0[\ub(\rb)]^2 = M_0$, we can rewrite Eq.~(\ref{eq:eom_free_energy}) as
\begin{align}
\label{eq:eom_for_q}
M_0\ddot q =& -M_0\omega_0^2 q + 2i\hbar C_v q q_0^{-2}\nonumber\\
&\times\int d\rb \,\delta T(\rb,t)\sum_\mu A_\mu^* T_{0,\mu}^*(\rb)
\end{align}
(we have also used here that, since $q\ub(\rb)$ is an eigenmode with frequency $\omega_0$, we have $q^{-1}\int d\rb\, \epsilon_{ij}\sigma^{(0)}_{ij} = -M_0\omega_0^2q$).

 If we think of $q(t)$ as an almost periodic function with slowly varying amplitude and phase, we can write it as $q(t) = q_0[a(t)\exp(-i\omega_0 t) + {\rm c.c.}]$ with $a(t)$ slowly varying over the period $2\pi/\omega_0$. Function $a(t)$ has the meaning of the expectation value  $\langle a\rangle$ of the mode annihilation operator used in Sec.~\ref{subsec:phenomenology}, cf. Eq.~(\ref{eq:zero_point}); note that Eqs.~(\ref{eq:heat_balance}) and (\ref{eq:eom_free_energy}) are classical equations of motion that disregard fluctuations. 

From Eqs.~(\ref{eq:heat_balance}) and (\ref{eq:epsilon_to_A_mu}) one finds $\delta T$ as a series in $T_{0,\mu}(\rb)$ with coefficients $\propto (\lambda_\mu-2i\omega_0)^{-1} a^2\exp(-2i\omega_0t) + {\rm c.c.})$. Substituting this series into Eq.~(\ref{eq:eom_for_q}), writing $M_0(\ddot q + \omega_0^2 q)=-2iM_0q_0\omega_0 \dot a\exp(-i\omega_0t)$ and keeping in the last term in Eq.~(\ref{eq:eom_for_q}) only the term $\propto \exp(-i\omega_0t)$, we obtain
\[ \dot a = -(2\Gamma^{\rm (nl)}_{\rm th} + iV_{\rm th})a|a|^2,\]
where $\Gamma^{\rm (nl)}_{\rm th}$ is the rate of thermoelastic nonlinear damping given by Eq.~(\ref{eq:Gamma_nl_thermal}) and $V_{\rm th} = 8\hbar\omega_0^2C_vT\sum_\mu |A_\mu|^2(\lambda_\mu^2 + 4\omega_0^2)^{-1}$ (in $V_{\rm th}$ we have again disregarded the difference between $T$ and $T_0$). The above equation for $ a(t)$ coincides with Eq.~(\ref{eq:moment}) for $\langle a(t)\rangle$ if one drops in the latter equation the terms that describe linear decay and dephasing and replace $\langle a^\dagger a^2\rangle \to \langle a^\dagger\rangle  \langle a\rangle^2$. This approximation is in line with the analysis of this section where we did not consider linear friction and fluctuations. We note that, as indicated earlier, nonlinear coupling to high-frequency phonons leads not only to nonlinear decay, but also to a dependence of the vibration frequency of the mode on its amplitude, which is described by the term $V_{\rm th}$ in the phenomenological theory.

\section{The Akhiezer mechanism of viscous nonlinear damping}
\label{sec:Akhiezer}

When the frequency of the considered mode significantly exceeds the rate of thermal diffusion, $\omega_0\gg c_sl_T/L^2$, one should take into account the finite time it takes for the phonon gas to locally equilibrate (we remind that $l_T$ and $L$ are the phonon mean free path and the relevant size of the system, $l_T\ll L$). For linear damping, the corresponding mechanism was discussed by Akhiezer \cite{Akhiezer1938}. Here we consider the Akhiezer mechanism of nonlinear damping, which can be regarded as a nonlinear viscous friction experienced by a slow vibrational mode.

The Akhiezer damping is determined by the evolution of function $\Phi_\alpha(\rb,\kb,t)$ on times $\lesssim l_T/c_s$, i.e., on the phonon relaxation times $\ep_{\nu>0}^{-1}$.
On such times one can disregard the drift term in Eq.~(\ref{eq:R_dependent_functions}), which comes from the spatial nonuniformity of the phonon distribution. Indeed, the scale of the spatial nonuniformity is the characteristic wavelength of the mode $\sim L$, and therefore the drift term is $\propto c_s/L\ll c_s/l_T$. 

In the case of a spatially uniform system, functions $T_{\nu}$ in the expression  (\ref{eq:solution_eigenvectors}) for $\Phi_\alpha$ exponentially decay in time. For $\nu>0$ 
\begin{align}
\label{eq:T_nu_Akh}
T_\nu(\rb,t) \approx T_\nu(\rb,0)\exp(-\epsilon_\nu t), \qquad t\lesssim l_T/c_s.
\end{align}
Then, from Eqs.~(\ref{eq:nonlin_friction_scalar}) and (\ref{eq:initial_T_function}),  the viscous contribution to the nonlinear damping is 
\begin{align}
\label{eq:Gamma_Akh}
&\Gamma^{(\rm nl)}_{\rm Akh} = \frac{2\omega_0}{k_B T\hbar }\,\frac{\mathbb V}{(2\pi)^{2d}} {\rm Re}\sum_{\alpha,\alpha'}
\int d\rb d\kb d\kb'\; V^*_\alpha(\rb,\kb) \nonumber \\
& \times\sum_{\nu>0}\frac{\psi_\nu(\kb,\alpha)\Psi_{\nu}(\kb',\alpha')}{\epsilon_\nu-2i\omega_0} V_{\alpha'}(\rb,\kb')\bar n_{\kb'\alpha'}(\bar n_{\kb'\alpha'} + 1),
\end{align}
It is important that the mode $\nu=0$, that describes thermal diffusion, does not contribute to the damping $\Gamma^{\rm (nl)}_{\rm Akh}$.  We note that, at this point, no assumptions have been made about the type of the mode and the symmetry of the medium. 

Typically, $\omega_0\ll {\rm Re}\,\ep_{\nu>0}$, and then in Eq.~(\ref{eq:Gamma_Akh}) $2\omega_0$ can be disregarded compared to $\ep_\nu$.    Equation~(\ref{eq:Gamma_Akh}) describes also linear Akhiezer damping if one uses for $V_\alpha$ a vertex that corresponds to the coupling of the considered mode to thermal phonons, which is linear in the  displacement of the considered mode.

Equation (\ref{eq:Gamma_Akh}) simplifies if we use the approximation of a $\nu$-independent relaxation rate, $\ep_\nu = \tau_r^{-1}$. An approximation of this type is often used in the analysis of linear decay rate, cf. \cite{Iyer2016} and references therein. As mentioned above, it is seen from the general expression for the nonlinear decay rate in terms of $\Phi_\alpha$, Eq.~(\ref{eq:nonlin_friction_scalar}), that only even in $\kb$ components of $\Phi_\alpha(\rb,\kb,t)$ directly contribute to $\Gamma^{\rm (nl)}$. Respectively, Eq.~(\ref{eq:Gamma_Akh}) has a contribution of the eigenmodes $\nu$ related to these components only, and $\tau_r$ should characterize the decay of such components. The relaxation time defined this way may significantly differ from the relaxation time of the odd in $\kb$ components of the density matrix of thermal phonons, which determine, in particular, the thermal conductivity.  

With $\epsilon_\nu$ replaced by $\tau_r^{-1}$ in Eq.~(\ref{eq:Gamma_Akh}), using the completeness of the eigenfunctions of the collision operator one re-writes in this equation $\sum_{\nu>0} \psi_\nu(\kb,\alpha)\Psi_{\nu}(\kb',\alpha') = [(2\pi)^d/{\mathbb V}]\delta (\kb-\kb')\delta_{\alpha\alpha'} - \psi_0(\kb,\alpha)\Psi_0(\kb',\alpha')$. Then, if one denotes the averaging over the modes $\kb,\alpha$  by an overline,
\begin{align}
\label{eq:kb_averaging}
\overline{ B_\alpha(\rb,\kb)} = &\frac{\hbar^2}{(2\pi)^d C_vk_BT^2}\sum_\alpha\int d\kb B_\alpha(\rb,\kb)\nonumber\\
&\times\omega_{\kb\alpha}^2 \bar n_{\kb\alpha}(\bar n_{\kb\alpha} + 1)
\end{align}
[here $B_\alpha(\rb,\kb)$ is an arbitrary function of $\rb, \kb,\alpha$], one can rewrite Eq.~(\ref{eq:Gamma_Akh}) as
\begin{align}
\label{eq:Gamma_Akh_simple}
&\Gamma^{(\rm nl)}_{\rm Akh} = \frac{2\omega_0}{\hbar }C_vT   \frac{\tau_r}{1+4\omega_0^2\tau_r^2}
\nonumber\\
&\times\int d\rb  \left[\overline{|v_\alpha(\rb,\kb)|^2} - |\overline{v_\alpha(\rb,\kb)}|{}^{^2}\right]
\end{align}
where $v_\alpha(\rb,\kb) = V_\alpha(\rb,\kb)/\hbar\omega_{\kb\alpha}$.

In the deformation potential model of the coupling, which for an isotropic medium is described by Eq.~(\ref{eq:Gruneisen_param}), the integrand in Eq.~(\ref{eq:Gamma_Akh_simple}) would be zero if  the nonlinear Gr\"uneisen parameters $\gamma^{(1,2,3)}_{\kb\alpha}$ were independent of $\kb,\alpha$. The temperature dependence of the integrand is thus determined by the dependence  of  $\gamma^{(1,2,3)}_{\kb\alpha}$ on $\kb$ and the difference between the Gr\"uneisen parameters for different phonon branches $\alpha$. The prefactor $C_vT\tau_r$ weakly depends on temperature in 3D systems for  temperatures high compared to the Debye temperature, since in this case $\tau_r \propto T^{-1}$ for phonon-phonon scattering and $C_v$ is independent of $T$.  For low temperatures, if $\tau_r\propto T^{-5}$ \cite{Gurevich1988}, we have $\Gamma^{(\rm nl)}_{\rm Akh}$ approximately proportional to $T^{-1}$ for $\omega_0\tau_r \ll 1$. These results change for thin films or nanobeams, where phonon motion transverse to the film/nanobeam is quantized or the phonon mean free path exceeds the film/nanobeam thickness.


\section{Dephasing due to quasielastic phonon scattering}
\label{sec:dephasing}

An important effect of a nonlinear coupling of the considered low-frequency mode to thermal phonons is the mode dephasing.  To the lowest order of the perturbation theory, it comes from the quartic nonlinearity and is described by the last term in the coupling Hamiltonian $H_{i}$, Eq.~(\ref{eq:H_i}), with $h_3$ of the form
\begin{equation}
\label{eq:h_3}
h_3= \sum_{\kappa\kappa'}v'_{\kappa' \kappa}b_{\kappa'}^\dagger b_{\kappa}.
\end{equation}
One can think of the coupling given by Eqs.~(\ref{eq:H_i}) and (\ref{eq:h_3}) as a modulation of the frequency of the considered mode resulting from quasi-elastic scattering of thermal phonons off the mode. Parameters $v'_{\kappa'\kappa}$ are renormalized in the second order by the parameters of the cubic nonlinearity. Such renormalization can be particularly important for degenerate modes \cite{Dykman1990h}; however, here we consider dephasing of a  nondegenerate low-frequency mode, and the renormalization is small and does not change the temperature dependence of the dephasing rate $\Gamma^{(\varphi)}$.

For the coupling (\ref{eq:h_3}), the expression for the dephasing rate \eqref{eq:Gphi} is similar to the expression for the nonlinear damping parameter (\ref{eq:Gnl_v2}),
\begin{align}
\label{eq:Gphi2}
\Gamma^{(\varphi)} = -\hbar^{-1}{\rm Im}\sum_{\kappa,\kappa'} v'_{\kappa\kappa'}\int_0^{\infty}dt e^{-\ep t}\phi'_{\kappa\kappa'}(t)
\end{align}
($\ep \to +0$). Function $\phi'_{\kappa\kappa'}(t)$ has the same form as function $\phi_{\kappa\kappa'}(t)$, Eq.~(\ref{eq:GreenFunction}), the only difference being that the coupling parameters $v_{\kappa'_0\kappa_0}$ are replaced with $v'_{\kappa'_0\kappa_0}$. If the nonlinear coupling of the considered mode to phonons is due to the lattice nonlinearity, then
\begin{align}
\label{eq:v_vs_vprime}
v'_{\kappa'\kappa} = 2v_{\kappa'\kappa}.
\end{align} 
One can show that this relation is not changed if the renormalization of the parameters $v_{\kappa\kappa'}$ and $v'_{\kappa\kappa'}$ by the coupling linear in the displacement of the considered mode is taken into account.

It is important that, for the considered low-frequency mode, the factor $\hbar^{-1}$ in front of the sum in the expression (\ref{eq:Gphi2}) is much larger than the corresponding factor $2\omega_0/k_BT$ in Eq.~(\ref{eq:Gnl_v2}) for $\Gamma^{\rm (nl)}$. Therefore one may expect that the dephasing rate parameter $\Gamma^{(\varphi)}$ will be significantly larger than the nonlinear friction coefficient $\Gamma^{\rm (nl)}$.

\subsection{The Landau-Rumer dephasing rate}
\label{section:LR_Dephasing}

In the Landau-Rumer limit, where we can neglect the decay of the  vibrational modes belonging to the continuous spectrum and calculate function $\phi'_{\kappa\kappa'}(t)$ disregarding the interaction between these modes, from Eqs.~(\ref{eq:GreenFunction}) and (\ref{eq:Gphi2}) we obtain
\begin{align}
\label{eq:Dephasing_LR}
 \Gamma^{(\varphi)}_{\rm LR} =& \pi\hbar^{-2}\sum_{\kappa\kappa'} |v'_{\kappa \kappa'}|^2\bar{n}_{\kappa}(\bar{n}_{\kappa}+1)\delta(\omega_{\kappa'}-\omega_{\kappa}). 
\end{align}
Equation (\ref{eq:Dephasing_LR}) has the same form as the expression for the dephasing rate of the modes localized near defects in solids \cite{Ivanov1965,Elliott1965}. In Refs.~\cite{Ivanov1965,Elliott1965} there was studied the temperature dependence of the dephasing rate, which switches from $T^2$ for high temperatures to $T^7$ for low temperatures \cite{Ivanov1966}.

An important distinction of Eq.~(\ref{eq:Dephasing_LR}) and of the general expression (\ref{eq:Gphi2}) from the case of the modes localized near defects in solids is that the localization length of the considered low-frequency modes is the typical size of the system. Therefore the dephasing rate $\Gamma^{(\varphi)}$ generally depends on the system size. In large spatially uniform systems the condition of forward scattering of phonons, which is imposed by the smooth spatial dependence of the deformation field of the considered mode, sharply reduces the phase volume of the phonons that contribute to the rate (\ref{eq:Dephasing_LR}) and changes the temperature dependence of the dephasing rate. 

\subsection{Thermoelastic and Akhiezer dephasing rates}
\label{section:Akz_Th_dephasing}

Finite lifetime of thermal phonons strongly affects the dephasing rate of low-frequency modes. The effect can be taken into account in the same way as for nonlinear friction. For spatially uniform systems the rate $\Gamma^{(\varphi)}$ is given by the theory of nonlinear friction of Secs.~\ref{sec:uniform_systems} - \ref{sec:Akhiezer}, which has to be slightly modified to take into account that we are looking for the response of the phonon bath at zero frequency rather than at frequency $2\omega_0$. Respectively, to obtain $\Gamma^{(\varphi)}$ from the expressions for $\Gamma^{\rm(nl)}$ one has first to multiply these expressions by $k_BT/2\hbar\omega_0$. Then, with the account taken of  Eq.~(\ref{eq:v_vs_vprime}), one has to replace $V_\alpha(\rb,\kb)$ with $V'_\alpha(\rb,\kb)\approx 2V_\alpha(\rb,\kb)$.  Finally, in all denominators that contain the term $4\omega_0^2$, cf. Eqs.~(\ref{eq:Gamma_nl_thermal}), (\ref{eq:damping_Lame}),  (\ref{eq:Gamma_thermo_flex}), (\ref{eq:Gamma_Akh_simple}), this term should be dropped.

In spite of the similarity of the formal expressions for the rates of nonlinear damping and dephasing, the physics behind the corresponding processes is very different. This becomes apparent when one thinks of the phenomenological description of nonlinear damping in terms of the free energy, Sec.~\ref{subsec:phenomen_thermoelastic}. This description does not apply to dephasing. Dephasing is a fluctuation effect, and it is not described by the free energy in thermal equilibrium. However, the thermoelastic dephasing mechanism exists. It is described by the microscopic theory, as explained above. In terms of a phenomenological theory, it corresponds to fluctuations of the frequency of the considered mode due to fluctuations of the temperature. The temperature dependence of the frequency can be immediately seen from Eqs.~(\ref{eq:H_i}) and (\ref{eq:h_3}). It is important that, where the thermoelastic rate of nonlinear damping is suppressed by the large ratio of $\omega_0$ to the rate of thermal diffusion, such suppression does not occur for thermoelastic dephasing.

\section{Conclusions}
\label{sec:conclusions}

We have presented a microscopic theory of nonlinear decay and dephasing of low-frequency eigenmodes in nano- and micro-mechanical resonators. Both phenomena are well-known to play an important role in the dynamics of mesoscopic vibrational systems. The paper is focused on the dissipation and decoherence due to the nonlinear coupling of the considered eigenmodes to thermally excited vibrational modes of the resonators. The analysis is not limited to spatially uniform systems, it also includes strongly nonuniform systems, with the nonuniformity scale exceeding the characteristic thermal wavelength but small compared to the mean free path of the thermal modes.

The major steps of the analysis are as follows. First, we expressed the decay and dephasing rates in terms of the pair correlation functions of thermal vibrational modes weighted with the interaction. We then obtained a transport equation for these correlation functions. The decay and dephasing rates were expressed in terms of the eigenfunctions of this transport equation. Closed form expressions were obtained for spatially uniform systems assuming that the coupling of the low-frequency mode to thermal phonons can be described in terms of nonlinear Gr\"uneisen parameters, which is a natural extension to the nonlinear decay problem of the coupling used to describe  linear decay of low-frequency modes.

Both the nonlinear damping and the dephasing come from scattering of thermal vibrational modes off the considered low-frequency modes. The damping  is due to scattering with energy transfer $2\hbar\omega_0$, whereas the dephasing is due to quasielastic scattering. In the both cases the analysis could be done in terms of the Wigner representation of the correlation function. For spatially uniform systems, where thermal vibrational modes are plane waves,  the transport equation for the  Wigner function that we obtained can be shown to be equivalent to the linearized Boltzmann equation, except for the initial conditions. 

For systems with strong but smooth nonuniformity the modes can be described in the eikonal approximation. This makes it possible not only to introduce a scalar Wigner representation of the correlation function, but also to derive a Markovian kinetic equation for this function.  We found that, in the scattering of short-wavelength thermal modes off each other, the sum of the coordinate-dependent wave vectors (the gradients of the eikonal) is approximately locally conserved (except for the Umklapp processes). As a result, it turns out that the transport equation for the Wigner function as a function of a coordinate and a wave vector $\Phi_\alpha(\Rb,\kb,t)$ is local in space: this equation couples $\Phi_\alpha(\Rb,\kb,t)$ with  functions $\Phi_{\alpha'}(\Rb,\kb',t)$ with different wave numbers $\kb'$ and different mode branches $\alpha'$, but the same coordinate $\Rb$.

We have shown that, similar to the case of linear damping, the mechanisms of nonlinear damping and dephasing can be conditionally separated  into the Landau-Rumer, thermoelastic, and Akhiezer-type. In the Landau-Rumer limit one disregards decay of the thermal modes responsible for the relaxation of the considered low-frequency mode. However,  in the systems of interest the thermal mode decay plays a significant role, and we primarily concentrated on the effects of this decay. 

We found that nonlinear thermoelastic damping can be important even where linear thermoelastic damping is inefficient, for symmetry reasons or because the mode frequency is small compared to the rate of thermal diffusion transverse to the resonator. The rate of nonlinear damping is simply expressed in terms of the eigenmodes of the diffusion equation for temperature. We also developed a phenomenological formulation in terms of the free energy of the system, which takes into account the terms nonlinear in the strain produced by the considered low-frequency mode. The specific examples included nonlinear thermoelastic decay of Lam\'e and flexural modes, where the linear thermoelastic damping is suppressed. The nonlinear damping is due to thermal diffusion along rather than transverse to the system. We found the dependence of the damping on temperature and the geometric factors. 

The Akhiezer nonlinear damping is similar to the Akhiezer linear damping. This similarity is particularly clear in the developed formulation based on the eigenmodes of the transport equation. The Akhiezer damping is due to the decay of the eigenmodes that are even in the wave vector $\kb$. As a consequence, if one uses the $\tau$-approximation for the decay of thermal modes, one should keep in mind that the value of the corresponding decay rate is generally different from the one that defines the thermal conductivity. This strongly affects the temperature dependence of the rate of nonlinear damping. 

We found that the dephasing rate may be significantly larger than the nonlinear damping rate. It contains an extra factor $k_BT/\hbar\omega_0$, which is large if the considered low-frequency mode is thermally excited (the case of the primary interest in this work). In addition, the dephasing rate does not contain  terms $\propto \omega_0^2\tau_r^2$ in the denominator, where $\tau_r^{-1}$ is the decay rate of the appropriate mode of the transport equation for thermal phonons. This can significantly increase the thermoelastic dephasing rate, a mechanism that has not been previously discussed. The Akhiezer mechanism of dephasing has not been previously discussed either, to the best of our knowledge. 

An important contribution to nonlinear damping can come from the losses at the boundaries, which are associated with emission of phonons into the supporting structure. The corresponding mechanism leads to a temperature-independent damping rate \cite{Dykman1975a}, and thus it can be separated from the internal nonlinear damping we have considered. Based on our results we expect a comparatively strong nonlinear damping and dephasing in small and low-dimensional systems. The analysis of this effect as well as the detailed analysis of linear and nonlinear damping in systems with strong smooth nonuniformity will be done in the follow-up paper.

\acknowledgments
JA, TK, and MID gratefully acknowledge partial support from US Defense Advanced Research Projects Agency (FA8650-13-1-7301). JA was also supported in part by the Institute of Mathematical Physics at MSU. MR and MID were supported in part by the National Science Foundation ((DMR-1514591).

\appendix

\section{The eikonal approximation}
\label{sec:appendix_eikonal}

Here, for completeness, we provide a simple formulation of the eikonal approximation for smoothly varying lattice parameters. The lattice Hamiltonian in the harmonic approximation is
\begin{align}
\label{eq:harmonic_H}
H_{\rm harm} = \sum_n\frac{\pb_n^2}{2M_n} + \frac{1}{2}\sum_{mn}L^{mn}_{ij}u_{mi} u_{nj},
\end{align}
where $M_n$ is the mass of the atom at site $n$, $\pb_n$ is the momentum, and $\ub_n$ is the displacement from the equilibrium position $\rb_n$. Matrix $\hat L^{mn}$ quickly falls off with the increasing distance between the sites $m,n$. The key assumption is that $M_n$ and $\hat L^{mn}$ smoothly depend on the position of the lattice cell, i.e., on $\rb_n$ for a single-atomic lattice. For conciseness, we will consider such lattices, the extension to a more general case is straightforward. The smoothness of $M_n, \hat L^{mn}$ means that the spatial scale on which they vary $l_{\rm sm}$ is large compared to the wavelength of thermal vibrations. 

For the modes with wavelength small compared to $l_{\rm sm}$,  one can seek a solution of the equation of motion for the displacement $\ub_n$,
\begin{align}
\label{eq:eom_harmonic}
M_n\ddot u_{nj} + \sum_{mj'} L^{nm}_{jj'} u_{mj'} = 0,
\end{align}
in the eikonal approximation by setting $\ub_n(t) = M_n^{-1/2}\ub_\kappa(\rb_n)\exp(-i\omega_\kappa t)$ with 
\begin{align}
&\ub_\kappa(\rb_n)=  C_\kappa\eb_\kappa(\rb_n) \exp[ iS_\kappa(\rb_n) ] \nonumber
\end{align}
($C_\kappa$ is a constant that determines the normalization of $\eb_\kappa$). To the leading order in the nonuniformity Eq.~(\ref{eq:eom_harmonic}) gives
\begin{align}
\label{eq:eikonal_eigen}
\omega_\kappa^2e_{\kappa j}(\rb_n)\approx \sum_{mj'}\frac{e_{\kappa j'}(\rb_n)}{(M_nM_m)^{1/2}}L^{nm}_{jj'} e^{i\kb_\kappa(\rb_n)(\rb_m-\rb_n)}
\end{align}
where $\kb_\kappa(\rb) = \n S_\kappa(\rb)$ is the position-dependent wave vector. In the eikonal approximation it smoothly depends on $\rb_n$ as do also vectors $\eb(\rb_n)$. Respectively,  in Eq.~(\ref{eq:eikonal_eigen}) we replaced $\eb_\kappa(\rb_m)$ with $\eb_\kappa(\rb_n)$.

Equation (\ref{eq:eikonal_eigen}) is a simple algebraic equation that relates the polarization vectors $\eb_\kappa(\rb_n)$ and the wave vectors $\kb_\kappa(\rb_n)$ to the mode eigenfrequency $\omega_\kappa$. Index $\kappa$ has discrete components that enumerate the branch of the vibrational mode as well as the quasi-continuous components, which are determined by the boundary conditions discussed in the text. The matrix $\hat L^{mn}/(M_nM_m)^{1/2}$ is Hermitian, and its eigenvectors form a complete orthonormal set;  the eikonal form used above applies only to vectors $\ub_\kappa$ that are fast oscillating on the length $l_{\rm sm}$.

\section{Kinetic equation beyond the plane wave approximation} 
\label{sec:A}

In this section we derive, in the Born approximation, the equation for the two-phonon correlation function introduced in Sec.~\ref{sec:QKE}.  We consider the Fourier transform
%
%
%
\begin{equation}
\label{eq:A_FT}
\langle A|_{\omega} = -\frac{i}{\hbar}\sum_{\kappa_1,\kappa_1'} v_{\kappa_1'\kappa_1}\int_0^{\infty}d{}t\, e^{i\omega t} \langle A(t)b^{\dagger}_{\kappa_1'}(0)b_{\kappa_1}(0) \rangle
\end{equation}
with Im~$\omega \to +0$. We are interested in the operator $A$ of the form of $b^{\dagger}_{\kappa}b_{\kappa'}$ with $\kappa$ and $\kappa'$ such that $|\kb_\kappa (\rb) - \kb_{\kappa'}(\rb)|\ll |\kb_\kappa(\rb)|$ in a sufficiently large range $\delta\rb$ (in particular,  $|\delta\rb|\gg 1/|\kb_\kappa(\rb)|$) where the amplitude of the considered mode with frequency $\omega_0$ is large. The typical frequencies $\omega_\kappa,\omega_{\kappa'}$ are $\sim k_BT/\hbar$ and are close to each other, and in the problem of nonlinear friction $|\omega_\kappa-\omega_{\kappa'}|\sim 2\omega_0 \ll  \omega_\kappa$. Respectively, Re~$\omega \sim 2\omega_0$ is small compared to $k_BT/\hbar$. We will  disregard the terms in $H_{\rm ph-ph}$ that contain $v'_{\kappa\kappa'\kappa''}$; these terms lead to the (temperature-dependent) renormalization of the mode frequencies.

\begin{figure}
\centering
\includegraphics[width=\linewidth,trim=4cm 1.5cm 4cm 1cm, clip=true]{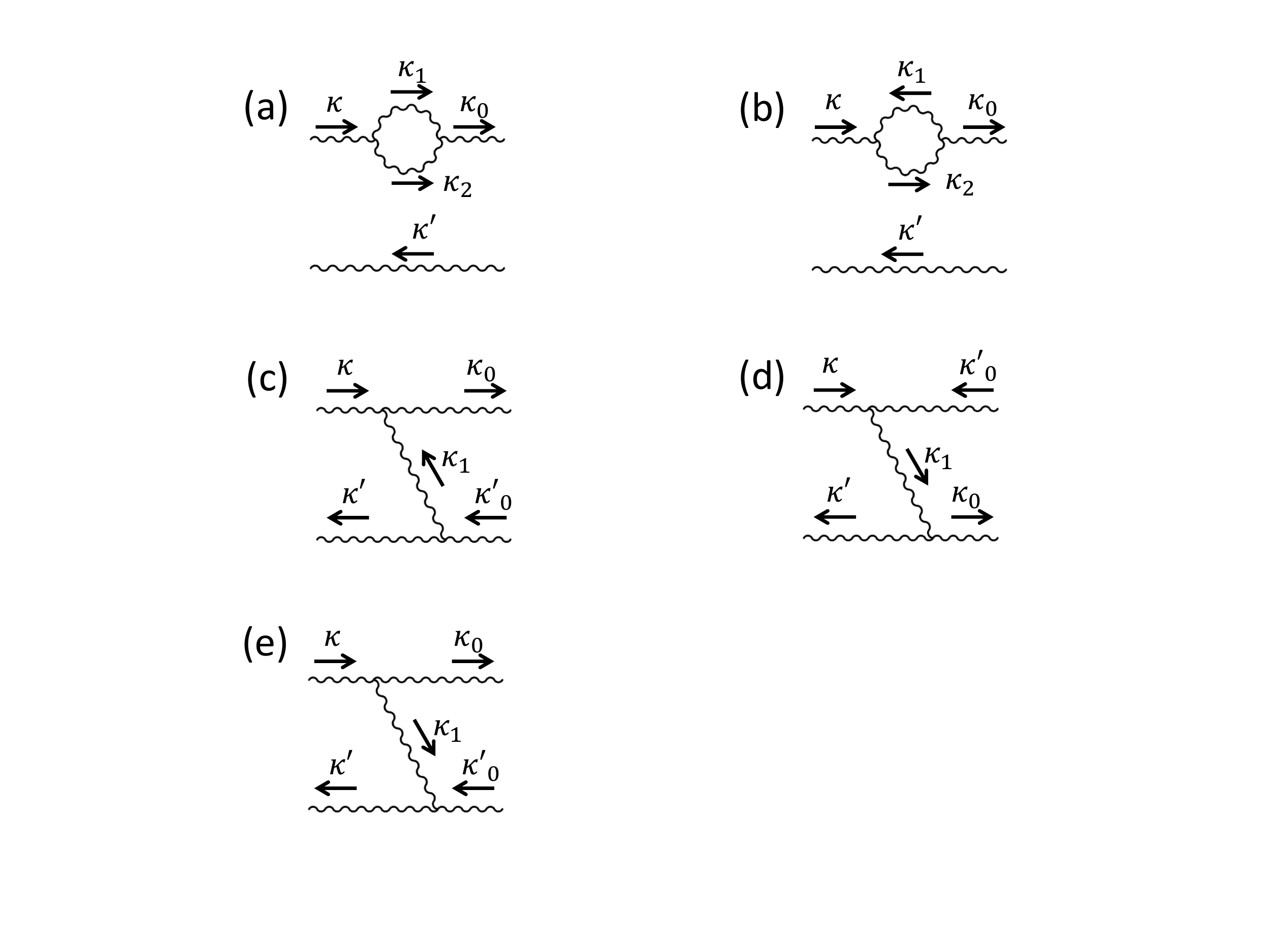}
\caption{A part of the diagrams that describe resonant mode scattering for the correlation function $\langle b_\kappa^\dagger b_{\kappa'}|_\omega$ for small frequency $\omega$. The upper and lower lines refer to the modes $\kappa$ and $\kappa'$, the arrow direction indicates the creation (right arrow) and annihilation (left arrow) operators in the correlation function. (a) and (b) show the self-energy contribution that comes from resonant scattering of the $\kappa$-mode only, whereas (c)-(e) show three types of the vertex terms that correspond to scattering processes in which both modes are involved. The diagrams for the mode $\kappa'$ are similar}
\label{fig:appendix_diagrams}
\end{figure}

To the lowest order in the mode interaction $H_{\rm ph-ph}$, one can obtain an equation for the correlation function by summing the simple sequence of diagrams shown in 
Fig.~\ref{fig:appendix_diagrams}. 
The result reads
\begin{align}
\label{eq:aa_v1}
& (\omega+{\omega}_{\kappa}-{\omega}_{\kappa'}) \langle b^{\dagger}_{\kappa}b_{\kappa'}|_{\omega}  - \hbar^{-1}\bar{n}_{\kappa}(\bar{n}_{\kappa'}+1) v_{\kappa' \kappa}  \nonumber \\
&=  {}\widehat{\Lambda}\langle b^{\dagger}_{\kappa}b_{\kappa'}|_{\omega}\equiv \sum_{\kappa_0,\kappa_0'}\Lambda_{\kappa\kappa'}^{\kappa_0\kappa_0'}\langle b^\dagger_{\kappa_0}b_{\kappa_0'}|_\omega \, . 
\end{align}
Here and below we use  wide hat, like ${}\widehat {\Lambda}$, to indicate that the symbol refers to an operator. We assume the nonlinear phonon-phonon coupling sufficiently weak, in particular compared to temperature, and in this approximation we decouple $\langle b^{\dagger}_{\kappa}b_{\kappa'}b^{\dagger}_{\kappa_1'}b_{\kappa_1}\rangle \approx \bar{n}_{\kappa}(\bar{n}_{\kappa'}+1)\delta _{\kappa \kappa_1}\delta_{\kappa'\kappa_1'}$.

Operator ${}\widehat{ \Lambda}\equiv {}\widehat{ \Lambda}(\omega)$ describes phonon scattering. It has two contributions, which come from the self-energy parts of the modes $\kappa$ and $\kappa'$ (${}\widehat{\Sigma}$)
and the vertex part $({}\widehat{\Gamma}$) in Fig.~\ref{fig:appendix_diagrams}, 
and can be written as 
\begin{align}
\label{eq:relaxation_symmetry}
&{}\widehat{ \Lambda} = {}\widehat{ \Sigma} + {}\widehat{\Gamma}, \qquad {}\widehat{\Gamma}(\omega) = {}\widehat{\gamma}(\omega) -{}\widehat{\gamma}^\dagger(-\omega^*),\nonumber\\
&{}\widehat{ \Sigma}(\omega) = {}\widehat{ \sigma}(\omega) - {}\widehat{ \sigma}^\dagger(-\omega^*), 
\end{align}
where $(\Lambda^\dagger)_{\kappa\kappa'}^{\kappa_0\kappa_0'} = (\Lambda_{\kappa'\kappa}^{\kappa_0'\kappa_0})^*$. The expression for  ${}\widehat{\sigma}$ reads
\begin{align}
\label{eq:sigma_operator}
&\sigma_{\kappa\kappa'}^{\kappa_0\kappa_0'} = \hbar^{-2}\sum_{\kappa_1,\kappa_2}\left(\frac{1}{2}v{}^*_{\kappa_1\kappa_2\kappa_0}v{}_{\kappa_1\kappa_2\kappa}
\frac{1+\bar n_{\kappa_1}+\bar n_{\kappa_2}}{\omega +\omega_{\kappa_1} + \omega_{\kappa_2}-\omega_{\kappa'}} \right. \nonumber\\ 
&\left.+ v{}^*_{\kappa\kappa_1\kappa_2}v{}_{\kappa_0\kappa_1\kappa_2} \frac{\bar n_{\kappa_1}-\bar n_{\kappa_2}}
{\omega +\omega_{\kappa_2} - \omega_{\kappa_1}-\omega_{\kappa'}}\right) \delta_{\kappa'\kappa_0'},
\end{align}
whereas ${}\widehat{\gamma}$ has the form
\begin{align}
\label{eq:Gamma_operator}
\gamma_{\kappa\kappa'}^{\kappa_0\kappa_0'}  &= \hbar^{-2}\sum_{\kappa_1}\left( - v{}^*_{\kappa\kappa_1\kappa_0}v{}_{\kappa'\kappa_1\kappa_0'}
\frac{1+\bar n_{\kappa_1}+\bar n_{\kappa'}}{\omega +\omega_{\kappa_0}-\omega_{\kappa_1}-\omega_{\kappa'}} \right.\nonumber\\
& + v{}^*_{\kappa\kappa_0'\kappa_1}v{}_{\kappa'\kappa_0\kappa_1} \frac{\bar n_{\kappa'}-\bar n_{\kappa_1}}
{\omega +\omega_{\kappa_1} - \omega_{\kappa_0'}-\omega_{\kappa'}} 
\nonumber\\
&\left. +  v{}^*_{\kappa_1\kappa_0'\kappa'}v{}_{\kappa_1\kappa_0\kappa}\frac{\bar n_{\kappa'}-\bar n_{\kappa_1}}
{\omega +\omega_{\kappa_0} +\omega_{\kappa_1}-\omega_{\kappa'}} 
\right).
\end{align}

In a system with smooth nonuniformity there is an approximate momentum conservation in phonon scattering, see Sec.~\ref{sec:eikonal}. If we do not consider umklapp processes, parameters $v{}_{\kappa'\kappa_0\kappa_1}$ are such that $|\kb_{\kappa'}(\rb) + \kb_{\kappa_0}(\rb) - \kb_{\kappa_1}(\rb)|\ll |\kb_{\kappa'}(\rb)|, |\kb_{\kappa_0}(\rb)|, |\kb_{\kappa_1}(\rb)|$  in a certain sufficiently broad range of $\rb$ with size $\delta r_{\rm q} \sim (\lambda_Tl_{\rm sm})^{1/2}$,  and similarly for other matrix elements $v, v{}^*$ in Eqs.~(\ref{eq:sigma_operator}) and (\ref{eq:Gamma_operator}). Of primary interest to us is the case where $\kb_\kappa(\rb)$ and $\kb_{\kappa'}(\rb)$ are  close in the same range of $\rb$, and thus the wave vectors of functions $\langle b_{\kappa_0}^\dagger b_{\kappa_0'}|_\omega$ in the right-hand side of Eq.~(\ref{eq:aa_v1}) are also close, 
$|\kb_{\kappa_0}(\rb) - \kb_{\kappa_0'}(\rb)|\ll |\kb_{\kappa_0}(\rb)|, |\kb_{\kappa_0'}(\rb)|$. In addition, the frequency difference $\omega_{\kappa_0} - \omega_{\kappa_0'}$ is small, of the order of $\omega_0$, otherwise functions $\langle b_{\kappa_0}^\dagger b_{\kappa_0'}|_\omega$ are nonresonant and small; nonresonant correlation functions have been disregarded in deriving Eq.~(\ref{eq:aa_v1}). The argument directly extends to the case where umklapp processes are present; vectors $\kb_{\kappa_0}(\rb)$ and  $\kb_{\kappa_0'}(\rb)$ are close to each other in this case, too.

In other terms, Eqs.~(\ref{eq:aa_v1}) describes scattering of a pair of phonons with close wave vector in a certain sufficiently large region of space into another pair of phonons with the wave vectors which are close in the same region. The frequency differences between the phonons in the pairs are also close.

For nanowires or membranes with transverse dimensions smaller or of the order of the thermal wavelength, $\kappa_0$ and $\kappa_0'$ should refer to the same subband of the quantized transverse motion, to meet the resonance condition. This means that the polarization and subband indices for the both phonons in a pair are the same, $\alpha_{\kappa_0} =\alpha_{\kappa_0'}$. An extension needed to include crossing or touching of the phonon branches is straightforward. We note that, conventionally, the kinetics of the phonon gas is described in terms of the mode occupation numbers \cite{Lifshitz1981a}. This corresponds to considering the evolution of the operator $b_\kappa^\dagger b_\kappa$, i.e., to setting $\kappa'=\kappa$.

\subsection{Zero-eigenvalue solution}
\label{subsec:zero_eigenvalue}

Function $\langle b_\kappa^\dagger b_{\kappa'}|_\omega$ is the Fourier transform of the correlation function $\phi_{\kappa\kappa'}(t)$ given by Eq.~(\ref{eq:GreenFunction}). From Eq.~(\ref{eq:aa_v1}), time evolution of $\phi_{\kappa\kappa'}\equiv \phi_{\kappa\kappa'}(t)$ is described by equation
\begin{align}
\label{eq:eom_for_g}
\dot \phi_{\kappa\kappa'} = i(\omega_\kappa-\omega_{\kappa'})\phi_{\kappa\kappa'} -i  \sum_{\kappa_0,\kappa_0'}\Lambda^{\kappa_0\kappa_0'}_{\kappa\kappa'}(+i0)\phi_{\kappa_0\kappa_0'},
\end{align}
where $\widehat\Lambda(+i0)$ indicates that the operator $\widehat\Lambda(\omega)$ is calculated for Re~$\omega = 0$ and Im~$\omega\to +0$. In the decoupling that underlies Eq.~(\ref{eq:aa_v1}) we have already used that operator $\widehat\Lambda(\omega)$ smoothly depends on $\omega$ in the range $|{\rm Re}\,\omega|\gtrsim \omega_0$.

One can show that matrix $\Lambda_{\kappa\kappa}^{\kappa_0\kappa_0'}(+i0)$ has a zero eigenvalue. The corresponding right eigenvector is 
\begin{align}
\label{eq:right_eigenvector}
X_{\kappa_0\kappa_0'} = \omega_{\kappa_0} \bar n_{\kappa_0}(\bar n_{\kappa_0} +1)\delta_{\kappa_0\kappa_0'}.
\end{align}
If one replaces the frequency denominators in Eqs.~(\ref{eq:sigma_operator}) and (\ref{eq:Gamma_operator}) for $\widehat\Lambda$ with $\delta$-functions, $X_{\kappa_0\kappa_0'}$ becomes an eigenvector of $\Lambda_{\kappa\kappa'}^{\kappa_0\kappa_0'}$.  This eigenvector has the same form as the well-known solution of the linearized kinetic equation for phonons in spatially uniform systems \cite{Lifshitz1981a}, which corresponds to the phonon gas remaining in thermal equilibrium if the temperature is changed.
The principal value of the denominators gives $\widehat\Lambda(+i0)X \propto |\omega_\kappa-\omega_{\kappa'}|$, which is small for the considered close $\kappa$ and $\kappa'$. Matrix $\Lambda_{\kappa\kappa'}^{\kappa_0\kappa_0'}(+i0)$  with the frequency denominators replaced by the $\delta$-functions also has left eigenvector $\omega_\kappa\delta_{\kappa\kappa'}$ with zero eigenvalue. The zero eigenvalue of the scattering operator plays an important role in the analysis of relaxation, as discussed in the main text. 

\section{Transport equation for the Wigner transform}
\label{sec:transport_eq_omega_full}

We now derive an equation of motion for the Wigner transform of  the correlation function, $[\Phi_{\alpha,\omega}(\Rb,\kb)]{}_{ij}\equiv \{ {\mathbb W}_{\Rb,\kb,\alpha}[\langle b_\kappa^\dagger b_{\kappa'}|_\omega]\}_{ij}$. Operator ${\mathbb W}$ is defined in Eq.~(\ref{eq:Wigner_defined}),
\begin{align}
\label{eq:Wigner_Fourier}
&\hat \Phi_{\alpha,\omega}(\Rb,\kb)= \sum_{\kappa,\kappa'}\hat\theta(\Rb,\kb; \kappa,\kappa';\alpha)\langle b_\kappa^\dagger b_{\kappa'} |_\omega,
\end{align}
where tensor $\hat \theta$ is given by Eq.~(\ref{eq:polarization_tensor}). 
Tensor $\hat \Phi_{\alpha,\omega}(\Rb,\kb)$ is the Fourier transform over time of the tensor $\hat \Phi_\alpha(\Rb,\kb,t)$ given by Eq.~(\ref{eq:Wigner_defined}).  As indicated in Sec.~\ref{sec:Green_defined}, $\hat \Phi_{\alpha,\omega}$ smoothly depends on $\Rb$ on the scale of the thermal wavelength $\lambda_T$. The typical values of $|\kb|$ are  $\sim \lambda_T^{-1}$.

To obtain an equation of motion for $\hat \Phi_{\alpha,\omega}$, we apply to Eq.~(\ref{eq:aa_v1}) the transformation ${\mathbb W}_{\Rb,\kb,\alpha}$, Eq.~(\ref{eq:Wigner_Fourier}). We start the analysis by looking at the transformation of the right-hand side of this equation, i.e., of the phonon-scattering induced term $\Lambda^{\kappa_0\kappa_0'}_{\kappa\kappa'}\langle b_{\kappa_0}^\dagger b_{\kappa_0'}|_\omega$. We note first that $\kappa_0$ and $\kappa_0'$ refer to the same branch $\alpha$. Next, we replace $\langle b_{\kappa_0}^\dagger b_{\kappa_0'}|_\omega$ with $\langle b_{\kappa_1}^\dagger b_{\kappa_1'}|_\omega$ and insert the $\delta$-symbol of the orthonormality of the $\hat\theta$-tensors (\ref{eq:psi_orthonormality}). Note that $\alpha_{\kappa_0}=\alpha_{\kappa_1}=\alpha_{\kappa_1'}$.

The outcome of the Wigner transformation of the phonon-scattering induced term is an integral transformation of $\hat \Phi_{\alpha,\omega}$, which gives the collision integral in the transport equation, as we show below,
\begin{equation}
\label{eq:stoss_defined}
{\rm St}[\hat \Phi_{\alpha,\omega}(\Rb,\kb)] = -i{\mathbb W}_{\Rb,\kb,\alpha}[{}\widehat{\Lambda}\langle b_\kappa^\dagger b_{\kappa'}|_\omega],
\end{equation}
where, in tensor notations,
\begin{widetext}
\begin{align}
\label{eq:lambda_transform}
&{\mathbb W}_{\Rb,\kb,\alpha}[{}\widehat{\Lambda}\langle b_\kappa^\dagger b_{\kappa'}|_\omega]= i \int d\kb_0 d\Rb_0 \sum_{\alpha{}_0}\hat \lambda_\alpha^{\alpha{}_0}(\Rb,\Rb_0,\kb,\kb_0) \hat \Phi_{\alpha{}_0,\omega}^{}(\Rb_0,\kb_0),
\nonumber\\
&\lambda_\alpha^{\alpha{}_0}{}^{i_1j_1}_{ij}(\Rb,\Rb_0,\kb,\kb_0)
= - i(2\pi)^{-d}\sum_{\kappa,\kappa',\kappa_0,\kappa_0'}\theta_{ij}(\Rb,\kb;\kappa,\kappa';\alpha)\Lambda^{\kappa_0\kappa_0'}_{\kappa\kappa'}\theta^*_{ i_ 1 j_1}(\Rb_0,\kb_0;\kappa_0,\kappa_0';\alpha{}_0).
\end{align}
Here, $d$ is the dimension of vectors $\Rb,\Rb_0,\kb,\kb_0$. In the first line, we imply a convolution of the fourth rank tensor $\hat\lambda^{\alpha_0}_\alpha$ with the second-rank tensor $\hat \Phi_{\alpha_0,\omega}$ [summation over the indices $i_1$ and $j_1$ in Eq.~(\ref{eq:lambda_transform})].

\end{widetext}

\subsection{The collision integral in the eikonal approximation}
\label{subsec:collision_eikonal}

The analysis of this paper relies on the smoothness of the nonuniformity. For a smooth nonuniformity,  tensor $\hat\theta$  can be simplified using the eikonal approximation, see Eq.~(\ref{eq:epsilon_eikonal}). Then the tensor correlation function $\hat \Phi_{\alpha,\omega}$ can be written in terms of a scalar correlation function ${}\Phi_{\alpha,\omega}$, similar to  Eq.~(\ref{eq:scalar_Green}) for the time-dependent correlation function,
\begin{align}
\label{eq:scalar_Green_omega}
&\hat \Phi_{\alpha,\omega}(\Rb,\kb)\approx \hat M(\Rb,\kb;\alpha){}\Phi_{\alpha,\omega}(\Rb,\kb), \quad \nonumber\\
&{}\Phi_{\alpha,\omega}(\Rb,\kb)= \sum_{\kappa,\kappa'}{}\theta_\alpha(\Rb,\kb;\kappa,\kappa')
\langle b^\dagger_\kappa b_{\kappa'}|_\omega,
\end{align}
where $\tilde\kappa\equiv \tilde\kappa(\Rb,\kb,\alpha)$, ${}\theta_\alpha(\Rb,\kb;\kappa,\kappa')$, and $\hat M(\Rb,\kb;\alpha)$ are given by Eqs.~(\ref{eq:bar_kappa}), (\ref{eq:kernel_scalar}), and (\ref{eq:M_tensor}).

Using the condition Tr~$\hat M^\dagger \hat M =1$, we can re-write the collision integral (\ref{eq:lambda_transform}) as
\begin{align}
\label{eq:collision_eikonal}
&{\mathbb W}_{\Rb,\kb,\alpha}[{}\widehat{\Lambda}\langle b_\kappa^\dagger b_{\kappa'}|_\omega]= i\hat M(\Rb,\kb;\alpha)\int d\kb_0 d\Rb_0 \nonumber\\
&\times \sum_{\alpha{}_0}{}\lambda_\alpha^{\alpha{}_0}(\Rb,\Rb_0,\kb,\kb_0) {}\Phi_{\alpha{}_0,\omega}(\Rb_0,\kb_0),
\end{align}
where the scalar kernel ${}\lambda$ has the form
\begin{align}
\label{eq:scalar_kernel_eikonal}
&{}\lambda_\alpha^{\alpha_0}(\Rb,\Rb_0,\kb,\kb_0)
= \frac{-i}{(2\pi)^d}
\sum_{\kappa,\kappa',\kappa_0,\kappa_0'}{}\theta_\alpha(\Rb,\kb;\kappa,\kappa')\nonumber\\
&\times \Lambda^{\kappa_0\kappa_0'}_{\kappa\kappa'}{}\theta_{\alpha_0}^*(\Rb_0,\kb_0;\kappa_0,\kappa_0').
\end{align}

The expression for ${}\lambda_\alpha^{\alpha_0}$ can be significantly simplified, because the major contributions to ${}\Phi_{\alpha,\omega}(\Rb,\kb)$ and  ${}\Phi_{\alpha_0,\omega}(\Rb_0,\kb_0)$ comes from $\kappa$ close to $\kappa'$ for given $\Rb$  and from $\kappa_0$ close to $\kappa_0'$ for given $\Rb_0$. 
In evaluating the matrix elements of $\widehat \Lambda$ in Eq.~(\ref{eq:scalar_kernel_eikonal}) one can set $\omega_\kappa=\omega_{\kappa'}$, $\omega_{\kappa_0} =\omega_{\kappa_0'}$, $\bar n_{\kappa}= \bar n_{\kappa'}$, $\bar n_{\kappa_0}= \bar n_{\kappa_0'}$ in Eqs.~(\ref{eq:sigma_operator}) and (\ref{eq:Gamma_operator}). One can also disregard Re~$\omega \sim 2\omega_0$ compared to $\omega_{\kappa}$ and to the frequencies of other modes involved in the scattering shown in Fig.~\ref{fig:appendix_diagrams}. Then the frequency denominators in the appropriate terms in ${}\widehat{ \gamma}$ and $\widehat\gamma^\dagger$ and in $\widehat\sigma$ and $\widehat\sigma^\dagger$ become complex conjugate and give $\delta$-functions of energy conservation in scattering. For example, $(\omega + \omega_{\kappa_0} - \omega_{\kappa_1}-\omega_{\kappa'})^{-1} \to -i\pi\delta(\omega_{\kappa_0} - \omega_{\kappa_1}-\omega_{\kappa})$. The disregarded terms are proportional to $\omega_\kappa-\omega_{\kappa'}$ and lead to a small renormalization of phonon frequencies.

We now show that function ${}\lambda_\alpha^{\alpha_0}(\Rb,\Rb_0,\kb,\kb_0)$ is large only for small $|\Rb-\Rb_0|$, i.e.,  that phonon scattering is local. We consider 
the first term in Eq.~(\ref{eq:Gamma_operator}) for the vertex $\widehat\Gamma$, but the result is the same for all other terms in the scattering operator $\widehat\Lambda$. The contribution of the considered term to ${}\lambda_\alpha^{\alpha_0}$ is proportional to
\begin{align}
\label{eq:product}
v^*_{\kappa\kappa_1\kappa_0} v_{\kappa'\kappa_1\kappa_0'} {}\theta_\alpha(\Rb,\kb;\kappa,\kappa'){}\theta_{\alpha_0}^*(\Rb_0,\kb_0;\kappa_0,\kappa_0').
\end{align}

The matrix elements $v$ are given by Eq.~(\ref{eq:cubic_cplng}). We
denote  $\rb_{*1}$ and $\rb_{*2}$ the extreme spatial points  in Eq.~(\ref{eq:momentum_conservation}) that determine  $v^*_{\kappa\kappa_1\kappa_0}$ and $ v_{\kappa'\kappa_1\kappa_0'}$, respectively. These points are close to each other. We assume, and will later confirm, that they are also close to $\Rb, \Rb_0$. 
We denote $\delta k= \max[|\kb_\kappa(\rb_{*1})-\kb_{\kappa'}(\rb_{*1})|, |\kb_{\kappa_0}(\rb_{*2})-\kb_{\kappa_0'}(\rb_{*2})|] $. Noting that mode $\kappa_1$ is involved in the both matrix elements $v$ in (\ref{eq:product}), we obtain from Eq.~(\ref{eq:momentum_conservation}) $|\rb_{*1}-\rb_{*2}|\lesssim \delta k(\delta r_{\rm q})^2= \lambda_Tl_{\rm sm}\,\delta k$. Using the estimate of Sec.~\ref{sec:eikonal} $\delta k\sim (\delta r_{\rm q})^{-1}$, we see that $|\rb_{*1}-\rb_{*2}|\sim  \delta r_{\rm q} \ll l_{\rm sm}$. This estimate coincides with the uncertainty in the values of $\rb_{*1,*2}$, Eq.~(\ref{eq:coordinate_uncertainty}). 

From the definition (\ref{eq:kernel_scalar}), ${}\theta_\alpha(\Rb,\kb;\kappa,\kappa') \propto \exp[iS_{\kappa'}(\Rb)-iS_\kappa(\Rb)]$, and similarly for ${}\theta_{\alpha_0}^*$ in Eq.~(\ref{eq:scalar_kernel_eikonal}). The exponents of the matrix elements $v$ in Eq.~(\ref{eq:product}) are also sums of actions $S$ calculated at $\rb_{*1}$ and $\rb_{*2}$, see Eq.~(\ref{eq:cubic_cplng}). The overall exponential factor in Eq.~(\ref{eq:product}) is fast oscillating with varying $\kappa,\kappa',\kappa_0$ and $\kappa_0'$ unless the points $\rb_{*1}, \rb_{*2}, \Rb$, and $\Rb_0$ are close to each other.  If these points are close, as we assume, we can expand
$S_\kappa(\rb_{*1}) - S_\kappa(\Rb) \approx \kb_\kappa(\rb_{*1})(\rb_{*1} - \Rb) $ and similarly for all other terms. 

Using the ``local quasi-momentum conservation" condition (\ref{eq:momentum_conservation}), we find that, as a result of the expansion, the exponential factor in Eq.~(\ref{eq:product}) becomes $ \exp[-i(\kb_{\kappa} -\kb_{\kappa'})(\Rb -\Rb_0)]$, where $\kb_\kappa,\kb_{\kappa'}$ are calculated at  $\rb_{*1}$ or, to the same accuracy, at $\rb_{*2}$ (we have also used that $\kb_{\kappa} -\kb_{\kappa'} \approx \kb_{\kappa_0} -\kb_{\kappa_0'}$). This imposes a limitation $|\Rb-\Rb_0|\lesssim 1/\delta k \sim \delta r_{\rm q}$. 
The analysis of the quadratic in $\rb_{*1,*2}-\Rb, \rb_{*1,*2}-\Rb_0$ terms in the exponent in Eq.~(\ref{eq:product}) shows that the distance between all points $\rb_{*1},\rb_{*2},\Rb,\Rb_0$ should be $\lesssim \delta r_{\rm q}$ for the exponential factor in Eq.~(\ref{eq:product}) to be smooth and thus to contribute to the collision term (\ref{eq:collision_eikonal}).

The same approach can be applied to other terms in ${}\lambda_\alpha^{\alpha_0}$ in Eq.~(\ref{eq:scalar_kernel_eikonal}). It also applies if we take into account umklapp processes. As explained in the text, this agrees with the physical picture in which phonons are scattered off each other with approximate ``local" momentum conservation. The scattering that determines the evolution of function ${}\Phi_{\alpha\omega}(\Rb,\kb)$ occurs within a distance $\delta r_q$ from point $\Rb$. This distance is much smaller than the nonuniformity length $l_{\rm sm}$ on which  ${}\Phi_{\alpha\omega}(\Rb,\kb)$ manifestly depends on $\Rb$. Therefore in the integral over $\Rb_0$ in Eq.~(\ref{eq:collision_eikonal}) one can replace ${}\Phi_{\alpha\omega}(\Rb_0,\kb_0)$ with ${}\Phi_{\alpha_0\omega}(\Rb,\kb_0)$, the locality condition, see Eq.~(\ref{eq:locality}).

\subsection{Spatially uniform systems}
\label{subsec:trans_symmetry}

The approximation (\ref{eq:lambda_transform}) applies also to the case of spatially uniform systems. We note that this case is not a special case of smooth nonuniformity discussed earlier: there, the nonuniformity was assumed sufficiently strong, so that the distance on which $\kb_\kappa(\rb)$ varies is small compared to the mean-free path. In spatially uniform systems the latter condition is obviously inapplicable. 

In a spatially uniform system the mode index $\kappa$ can be written as $(\kb_\kappa,\alpha_\kappa)$, i.e., as the wave vector and the polarization/subband index. The displacement in a normal mode is $\ub_\kappa(\rb) = {\mathbb N}^{-1/2}\eb_\kappa\exp(i\kb_\kappa \rb)$ with unit polarization vector, $|\eb_\kappa|=1$. Therefore $\hat\theta (\Rb,\kb;\kappa,\kappa';\alpha) \propto \delta [\kb - (\kb_\kappa+\kb_{\kappa'})/2]\exp[i(\kb_\kappa-\kb_{\kappa'})\Rb]$. The momentum conservation in a three-phonon process imposes the condition $\kb_{\kappa_0}-\kb_\kappa = \kb_{\kappa_0'}-\kb_{\kappa'}$;
three-phonon scattering can be also accompanied by umklapp processes.

The sum over $\kappa,\kappa'$ in Eq.~(\ref{eq:lambda_transform}) can be replaced by the integral over vectors $(\kb_\kappa+\kb_{\kappa'})/2,\kb_\kappa - \kb_{\kappa'}$, and similarly for the sum over $\kappa_0,\kappa_0'$. 
Using the smallness of  $|\kb_\kappa -\kb_{\kappa'}|,  |\kb_{\kappa_0} - \kb_{\kappa_0'}|$, in calculating $\Lambda_{\kappa\kappa'}^{\kappa_0\kappa_0'}$ one can set $\kappa=\kappa'$ and $\kappa_0=\kappa_0'$. Then integration over $\kb_\kappa -\kb_{\kappa'}$ in Eq.~(\ref{eq:lambda_transform}) gives $\delta(\Rb-\Rb_0)$, and this equation takes form
\begin{align}
\label{eq:trans_invariant}
{\mathbb W}_{\Rb,\kb,\alpha}&[{}\widehat{ \Lambda}\langle b^{\dagger}_{\kappa}b_{\kappa'}|_{\omega}] = i\int d\kb_0 
\sum_{\alpha{}_0} \hat\lambda_\alpha^{\alpha{}_0}(\kb,\kb_0) \hat \Phi_{\alpha{}_0,\omega}(\Rb,\kb_0). 
\end{align}
The scattering operator (\ref{eq:trans_invariant}) is local, it depends on the value of $\hat \Phi_{\alpha,\omega}$ at the point $\Rb$ where the collision integral is evaluated. 

The components of the tensor $\hat\lambda$ in Eq.~(\ref{eq:trans_invariant}) are
\begin{align}
\label{eq:lambda_invariant}
\lambda_\alpha^{\alpha{}_0}{}^{i_1j_1}_{ij}(\kb,\kb_0)=& \frac{\mathbb V}{(2\pi)^{d}}
(\eb_{\kb\alpha})_i (\eb^*_{\kb\alpha})_j \tilde\Lambda_{\kb\alpha}^{\kb_0\alpha{}_0}\nonumber\\
& \times(\eb^*_{\kb_0\alpha_0})_{i_1} (\eb_{\kb_0\alpha{}_0})_{j_1}.
\end{align}
Here ${\mathbb V}= v_{\rm c}{\mathbb N}$ is the volume of the system and 

\begin{widetext}
\noindent
\begin{align}
\label{eq:rates_introduced}
&\tilde\Lambda_{\kb\alpha}^{\kb_0\alpha{}_0} = w_{\kb\alpha}^{\kb_0\alpha{}_0} + {\tilde w}_{\kb\alpha}^{\kb_0\alpha{}_0}
-{\tilde{\tilde w}}_{\kb\alpha}^{\kb_0\alpha{}_0}-\delta (\kb_0-\kb)\delta_{\alpha\alpha{}_0}\sum_{\alpha_2} \int d \kb_2 \left( \frac{1}{2} w_{\kb_2\alpha_2}^{\kb\alpha} +{\tilde w}_{\kb_2\alpha_2}^{\kb\alpha}\right)
\end{align}
The expressions for $w, {\tilde w}, {\tilde{\tilde w}}$ are the  phonon scattering rates,
\begin{align}
\label{eq:scatter_rate}
&w_{\kb\alpha}^{\kb_0\alpha{}_0}=\frac{2\pi}{\hbar^2}\sum_{\beta}
|v{}_{\kb\alpha,\kb_0-\kb\,\beta,\kb_0\alpha{}_0}|^2
\delta (\omega_{\kb_0\alpha{}_0}-\omega_{\kb\alpha}-\omega_{\kb_0-\kb\,\beta})(1+\bar n_{\kb\alpha}+\bar n_{\kb_0-\kb\,\beta})  \nonumber\\
& 
{\tilde w}_{\kb\alpha}^{\kb_0\alpha{}_0}=\frac{2\pi}{\hbar^2}\sum_{\beta}
|v{}_{\kb -\kb_0\,\beta, \kb_0\alpha{}_0,\kb\alpha}|^2 \delta (\omega_{\kb_0\alpha{}_0} - \omega_{\kb\alpha} + \omega_{\kb-\kb_0\,\beta}) 
(\bar n_{\kb-\kb_0\,\beta}-\bar n_{\kb\alpha} ),
\end{align}
whereas
\begin{align}
\label{eq:scatter_rate1}
&{\tilde{\tilde w}}_{\kb\alpha}^{\kb_0\alpha{}_0}=\frac{2\pi}{\hbar^2}\sum_{\beta}
|v{}_{\kb\alpha, \kb_0\alpha{}_0, \kb_0+\kb\,\beta}|^2 \delta (\omega_{\kb_0\alpha{}_0}+\omega_{\kb\alpha}-\omega_{\kb_0+\kb\,\beta})
(\bar n_{\kb\alpha} -\bar n_{\kb_0+\kb\,\beta}).
\end{align}
\end{widetext}
We note that factors $v{}_{\kb\alpha,\kb_1\alpha_1,\kb_2\alpha_2}$ are $\propto {\mathbb V}^{-1/2}$, and therefore the volume of the system drops out of Eq.~(\ref{eq:trans_invariant}). The coefficients in Eqs.~(\ref{eq:scatter_rate}) and (\ref{eq:scatter_rate1}) coincide with the coefficients in the quantum kinetic equation for the phonon gas in a spatially uniform system \cite{Lifshitz1981a}, which is linearized in the deviations of the phonon occupation numbers from their values in thermal equilibrium. To allow for umklapp processes, one should replace $\kb - \kb_0$ in the matrix elements $v{}_{\kb\alpha,\kb_0 -\kb \beta,\kb_0\alpha_0}$ with $\kb -\kb_0 + \Kb$, where $\Kb$ is the reciprocal lattice vector and add a sum over $\Kb$; similarly, $\kb+\kb_0$ should be replaced with $\kb+\kb_0 + \Kb$.

\section{The generalized drift term}
\label{sec:drift}

We now consider the Wigner transformation of the left-hand side of equation (\ref{eq:aa_v1}) for the correlation function $\langle b_\kappa^\dagger b_{\kappa'}|_\omega$, and specifically of the term $(\omega_\kappa-\omega_{\kappa'})\langle b_\kappa^\dagger b_{\kappa'}|_\omega$. It gives the drift term in the kinetic equation for the correlation function in the Wigner representation
\begin{align}
\label{eq:drift}
\hat{\cal T}[\hat \Phi_{\alpha,\omega}(\Rb,\kb)] = -i{\mathbb W}_{\Rb,\kb,\alpha}[(\omega_\kappa-\omega_{\kappa'})\langle b_\kappa^\dagger b_{\kappa'}|_\omega]
\end{align}

 Following the procedure used to obtain Eq.~(\ref{eq:lambda_transform}), we can write 
\begin{align}
\label{eq:one_frequency_term}
&{\mathbb W}_{\Rb,\kb,\alpha}[\omega_\kappa\langle b_\kappa^\dagger b_{\kappa'}|_\omega]= 2^d\int \frac{d\kb_0 d\Rb_0}{(2\pi)^d} \int \frac{d\kb_2d\rhob}{(2\pi)^d}e^{i(\kb_2 -\kb)\rhob} \nonumber\\
&\times\hat \Phi_{\alpha,\omega}\left(\Rb_0+\frac{\rhob}{2},\kb_0\right)\hat\Omega_\alpha(\Rb_0,\kb_2)
e^{2i(\kb_0 -\kb_2)(\Rb -\Rb_0)} 
\end{align}
where tensor $\hat \Omega_\alpha$ is the Wigner transform of the mode eigenfrequencies,
\begin{align}
\label{eq:frequency_transformed}
[\Omega_{\alpha}(\Rb,\kb)]_{ij}&= \sum_\kappa \theta_{ij}(\Rb,\kb;\kappa,\kappa;\alpha)\omega_\kappa
\end{align}
Clearly, matrix $\hat\Omega_\alpha$ is Hermitian. 

The expression for the Wigner transform ${\mathbb W}_{\Rb,\kb,\alpha}[\omega_{\kappa'}\langle b_\kappa^\dagger b_{\kappa'}|_\omega]$ is similar to Eq.~(\ref{eq:one_frequency_term}), except that $\hat\Omega_\alpha$ and $\hat \Phi_{\alpha,\omega}$ are interchanged and the coefficients are complex-conjugate.
Expanding $\hat \Phi_{\alpha,\omega}(\Rb_0 +\rhob/2,\kb_0)$ in Eq.~(\ref{eq:one_frequency_term}) to the first order in $\rhob,\kb_0-\kb$, and performing a similar expansion in the conjugate term, we obtain the drift term (\ref{eq:drift}) in the form
\begin{align}
\label{eq:full_frequency_term}
\hat{\cal T}&[\hat \Phi_{\alpha,\omega}(\Rb,\kb)] =
i[\hat\Omega_\alpha(\Rb,\kb),\hat \Phi_{\alpha,\omega}(\Rb,\kb)] \nonumber\\
&+\frac{1}{2}\left\{\partial_\kb \hat\Omega_\alpha(\Rb,\kb), \partial_\Rb\hat \Phi_{\alpha,\omega}(\Rb,\kb)\right\}_+ \nonumber\\
& - \frac{1}{2} \left\{  \partial_\Rb\hat\Omega_\alpha(\Rb,\kb) , \partial_\kb \hat \Phi_{\alpha,\omega}(\Rb,\kb)\right\}_+,
\end{align}
where $\{\hat A,\hat B\}_+ = \hat A\hat B+\hat B\hat A$. Equation (\ref{eq:full_frequency_term}) describes the evolution of the Wigner transform $\hat \Phi_{\alpha,\omega}$ in the absence of phonon-phonon coupling. As seen from this equation, such evolution is local, it is determined by the values of $\hat \Phi_{\alpha,\omega}$ and the Wigner transform of the mode frequencies evaluated for the same $\Rb,\kb$. 

The commutator in Eq.~(\ref{eq:full_frequency_term}) (the first term in the right-hand side) is a consequence of the difference of the interrelation between the components of tensors $[\Omega_\alpha]{}_{ij}$ and $[\Phi_{\alpha,\omega}]{}_{ij}$. For translationally invariant systems, the interrelation between the components of the displacement $\ub_{\kb\alpha}(\rb) = {\mathbb N}^{-1/2}\eb_{\kb\alpha}\exp(i\kb\rb)$ is independent of $\rb$, but generally depends on $\kb$ (we remind that in such systems it is convenient to specify explicitly the wave vector and the branch of a mode, $\kb$ and $\alpha$, rather than to use $\kappa$). Therefore $\hat \Omega_\alpha (\Rb,\kb)$ is independent of $\Rb$. Using the condition $\eb\cdot \eb^* = 1$ one can show from Eq.~(\ref{eq:full_frequency_term}) that $\hat{\cal T}[\hat \Phi_{\alpha,\omega}(\Rb,\kb)]= \partial_\kb\omega_{\kb\alpha}\partial_\Rb \hat \Phi_{\alpha,\omega}(\Rb,\kb)$, as one would expect from the expansion $\omega_\kappa-\omega_{\kappa'}\approx (\kb-\kb')\partial_\kb\omega_{\kb\alpha}$ in Eq.~(\ref{eq:drift}) written for a translationally invariant system. An additional step involved in the derivation of this relation is the expansion of the product ${}(\eb_{\kb'\alpha})_i (\eb_{\kb\alpha})^*_j$ in the expression for $[\Phi_{\alpha,\omega}]_{ij}$ in $\kb-\kb'$.
\subsection{The drift term in the eikonal approximation}
\label{subsec:drift_eikonal}

The expression for the drift term is simplified in the eikonal approximation. In this approximation we write the frequency tensor (\ref{eq:frequency_transformed}) as 
\begin{align}
\label{eq:scalar_frequency}
\hat\Omega_\alpha (\Rb,\kb) =\hat M(\Rb,\kb;\alpha){}\Omega_\alpha(\Rb,\kb), 
\end{align}
where tensor $\hat M$ is given by Eq.~(\ref{eq:scalar_Green}) and  $
{}\Omega_\alpha(\Rb,\kb)= [(2\pi)^d/{\mathbb N}v_{\rm c}(\Rb)]\sum_\kappa   |\eb_{\kappa}|^2 \omega_\kappa\delta [\kb_\kappa(\Rb) -\kb] \delta_{\alpha_\kappa,\alpha}$ is  a scalar phonon frequency as a function of coordinate and wave vector, for a given phonon branch $\alpha$, cf. Eq.~(\ref{eq:bar_omega}).

To evaluate the drift term (\ref{eq:full_frequency_term}) in the eikonal approximation, one should go beyond the approximate expression for the tensor correlation function in terms of the scalar correlation function, Eq.~(\ref{eq:scalar_Green_omega}). Vectors $\eb_\kappa$ and $\eb_{\kappa'}$ that enter the expression for the tensor correlation function $\hat \Phi_{\alpha,\omega}$,  Eqs.~(\ref{eq:polarization_tensor}) and (\ref{eq:Wigner_Fourier}), are different. A somewhat cumbersome and lengthy calculation allows one to show  that
\begin{align}
{\rm Tr}~\hat{\cal T}[\hat \Phi_{\alpha,\omega}(\Rb,\kb)]& =
\partial_\kb {}\Omega_\alpha(\Rb,\kb) \partial_\Rb{}\Phi_{\alpha,\omega}(\Rb,\kb)\nonumber\\
 &- \partial_\Rb{}\Omega_\alpha(\Rb,\kb)  \partial_\kb {}\Phi_{\alpha,\omega}(\Rb,\kb)
\end{align}  
This equation reduces the trace of the drift term to a simple expression in terms of the scalar functions ${}\Omega_\alpha$ and ${}\Phi_{\alpha,\omega}$. This form is used in the transport equation for the scalar Wigner transform in Sec.~\ref{sec:QKE}.

\section{Phonon scattering by short-range disorder}
\label{sec:appendix_disorder}

Here we discuss phonon scattering by weak short-range disorder in the presence of smooth nonuniformity. The disorder Hamiltonian is given by Eq.~(\ref{eq:disorder_Hamiltonian}). To the second order in the phonon-phonon coupling and the disorder, their contributions to the collisions operator are additive, as in the case of systems  with no smooth nonuniformity (the extension of the Matthiessen rule). A straightforward calculation shows that the self-energy operators $\widehat \sigma$ and the vertex part $\widehat\gamma$ in Eqs.~(\ref{eq:relaxation_symmetry}) - (\ref{eq:Gamma_operator}) acquire extra terms $\widehat\sigma^{(d)}$ and $\widehat\gamma^{(d)}$, respectively,
\begin{align}
\label{eq:disorder_scattering}
&\widehat\sigma \to \widehat\sigma + \widehat\sigma^{(d)}, \qquad \widehat\gamma\to \widehat\gamma+ \widehat\gamma^{(d)},
\nonumber\\
&(\sigma^{(d)})_{\kappa\kappa'}^{\kappa_0\kappa_0'}(\omega)= \hbar^{-2}\sum_{\kappa_1}\frac {\langle v^{(d)}_{\kappa_0\kappa_1}v^{(d)}_{\kappa_1\kappa}\rangle}{\omega+\omega_{\kappa_1}-\omega_{\kappa'}}\delta_{\kappa'\kappa_0'},\nonumber\\
&(\gamma^{(d)})_{\kappa\kappa'}^{\kappa_0\kappa_0'}(\omega)= -\hbar^{-2}\frac {\langle v^{(d)}_{\kappa_0\kappa}v^{(d)}_{\kappa'\kappa_0'}\rangle}{\omega+\omega_{\kappa}-\omega_{\kappa_0'}}
\end{align}

In the eikonal approximation, the scattering described by Eq.~(\ref{eq:disorder_scattering}) can be analyzed in the same way as the phonon-phonon scattering in Appendix~\ref{subsec:collision_eikonal}. In the both cases the scattering is ``local''. It couples function ${}\Phi_{\alpha\omega}(\Rb,\kb)$ to  function  ${}\Phi_{\alpha_0\omega}(\Rb_0,\kb)$ with $|\Rb_0 - \Rb|\ll l_{\rm sm}$ for modes with the wavelength small compared to $l_{\rm sm}$. Moreover, also as in the case of phonon-phonon scattering, the points $\Rb,\Rb_0$ are close to the point $\rb_{**}$ given by Eq.~(\ref{eq:short_range_conservation}) for the relevant phonons. 

In the absence of smooth spatial nonuniformity, the collision integral is described by Eqs.~(\ref{eq:stoss_defined}) and (\ref{eq:trans_invariant}) - (\ref{eq:rates_introduced}) in which the collision rate ${\tilde w}_{\kb\alpha}^{\kb_0\alpha_0}$ is replaced with
\begin{align}
\label{eq:short_range_trans_invariant}
&{\tilde w}_{\kb\alpha}^{\kb_0\alpha_0} \to {\tilde w}_{\kb\alpha}^{\kb_0\alpha_0} + (w^{(d)})_{\kb\alpha}^{\kb_0\alpha_0}, \nonumber\\
& (w^{(d)})_{\kb\alpha}^{\kb_0\alpha_0} = \frac{2\pi}{\hbar^2} \langle |v^{(d)}_{\kb\alpha,\kb_0\alpha_0}|^2\rangle \delta(\omega_{\kb\alpha} - \omega_{\kb_0\alpha_0}).
\end{align}
The $\delta$-function reflects the fact that the scattering is elastic.


\begin{thebibliography}{72}%
\makeatletter
\providecommand \@ifxundefined [1]{%
 \@ifx{#1\undefined}
}%
\providecommand \@ifnum [1]{%
 \ifnum #1\expandafter \@firstoftwo
 \else \expandafter \@secondoftwo
 \fi
}%
\providecommand \@ifx [1]{%
 \ifx #1\expandafter \@firstoftwo
 \else \expandafter \@secondoftwo
 \fi
}%
\providecommand \natexlab [1]{#1}%
\providecommand \enquote  [1]{``#1''}%
\providecommand \bibnamefont  [1]{#1}%
\providecommand \bibfnamefont [1]{#1}%
\providecommand \citenamefont [1]{#1}%
\providecommand \href@noop [0]{\@secondoftwo}%
\providecommand \href [0]{\begingroup \@sanitize@url \@href}%
\providecommand \@href[1]{\@@startlink{#1}\@@href}%
\providecommand \@@href[1]{\endgroup#1\@@endlink}%
\providecommand \@sanitize@url [0]{\catcode `\\12\catcode `\$12\catcode
  `\&12\catcode `\#12\catcode `\^12\catcode `\_12\catcode `\%12\relax}%
\providecommand \@@startlink[1]{}%
\providecommand \@@endlink[0]{}%
\providecommand \url  [0]{\begingroup\@sanitize@url \@url }%
\providecommand \@url [1]{\endgroup\@href {#1}{\urlprefix }}%
\providecommand \urlprefix  [0]{URL }%
\providecommand \Eprint [0]{\href }%
\providecommand \doibase [0]{http://dx.doi.org/}%
\providecommand \selectlanguage [0]{\@gobble}%
\providecommand \bibinfo  [0]{\@secondoftwo}%
\providecommand \bibfield  [0]{\@secondoftwo}%
\providecommand \translation [1]{[#1]}%
\providecommand \BibitemOpen [0]{}%
\providecommand \bibitemStop [0]{}%
\providecommand \bibitemNoStop [0]{.\EOS\space}%
\providecommand \EOS [0]{\spacefactor3000\relax}%
\providecommand \BibitemShut  [1]{\csname bibitem#1\endcsname}%
\let\auto@bib@innerbib\@empty
\bibitem [{\citenamefont {Poot}\ and\ \citenamefont {van~der
  Zant}(2012)}]{Poot2012}%
  \BibitemOpen
  \bibfield  {author} {\bibinfo {author} {\bibfnamefont {M.}~\bibnamefont
  {Poot}}\ and\ \bibinfo {author} {\bibfnamefont {H.~S.}\ \bibnamefont {van~der
  Zant}},\ }\href@noop {} {\bibfield  {journal} {\bibinfo  {journal} {Physics
  Reports}\ }\textbf {\bibinfo {volume} {511}},\ \bibinfo {pages} {273}
  (\bibinfo {year} {2012})}\BibitemShut {NoStop}%
\bibitem [{\citenamefont {Dykman}(2012{\natexlab{a}})}]{Dykman2012b}%
  \BibitemOpen
  \bibinfo {editor} {\bibfnamefont {M.~I.}\ \bibnamefont {Dykman}},\ ed.,\
  \href@noop {} {\emph {\bibinfo {title} {Fluctuating Nonlinear Oscillators:
  from Nanomechanics to Quantum Superconducting Circuits}}}\ (\bibinfo
  {publisher} {OUP, Oxford},\ \bibinfo {year} {2012})\BibitemShut {NoStop}%
\bibitem [{\citenamefont {Aspelmeyer}\ \emph {et~al.}(2014)\citenamefont
  {Aspelmeyer}, \citenamefont {Kippenberg},\ and\ \citenamefont
  {Marquardt}}]{Aspelmeyer2014a}%
  \BibitemOpen
  \bibfield  {author} {\bibinfo {author} {\bibfnamefont {M.}~\bibnamefont
  {Aspelmeyer}}, \bibinfo {author} {\bibfnamefont {T.~J.}\ \bibnamefont
  {Kippenberg}}, \ and\ \bibinfo {author} {\bibfnamefont {F.}~\bibnamefont
  {Marquardt}},\ }\href@noop {} {\bibfield  {journal} {\bibinfo  {journal}
  {Rev. Mod. Phys.}\ }\textbf {\bibinfo {volume} {86}},\ \bibinfo {pages}
  {1391} (\bibinfo {year} {2014})}\BibitemShut {NoStop}%
\bibitem [{\citenamefont {Mahboob}\ \emph {et~al.}(2013)\citenamefont
  {Mahboob}, \citenamefont {Nishiguchi}, \citenamefont {Fujiwara},\ and\
  \citenamefont {Yamaguchi}}]{Mahboob2013}%
  \BibitemOpen
  \bibfield  {author} {\bibinfo {author} {\bibfnamefont {I.}~\bibnamefont
  {Mahboob}}, \bibinfo {author} {\bibfnamefont {K.}~\bibnamefont {Nishiguchi}},
  \bibinfo {author} {\bibfnamefont {A.}~\bibnamefont {Fujiwara}}, \ and\
  \bibinfo {author} {\bibfnamefont {H.}~\bibnamefont {Yamaguchi}},\ }\href
  {\doibase 10.1103/PhysRevLett.110.127202} {\bibfield  {journal} {\bibinfo
  {journal} {Phys. Rev. Lett.}\ }\textbf {\bibinfo {volume} {110}},\ \bibinfo
  {pages} {127202} (\bibinfo {year} {2013})}\BibitemShut {NoStop}%
\bibitem [{\citenamefont {Rimberg}\ \emph {et~al.}(2014)\citenamefont
  {Rimberg}, \citenamefont {Blencowe}, \citenamefont {Armour},\ and\
  \citenamefont {Nation}}]{Rimberg2014}%
  \BibitemOpen
  \bibfield  {author} {\bibinfo {author} {\bibfnamefont {A.~J.}\ \bibnamefont
  {Rimberg}}, \bibinfo {author} {\bibfnamefont {M.~P.}\ \bibnamefont
  {Blencowe}}, \bibinfo {author} {\bibfnamefont {A.~D.}\ \bibnamefont
  {Armour}}, \ and\ \bibinfo {author} {\bibfnamefont {P.~D.}\ \bibnamefont
  {Nation}},\ }\href {\doibase 10.1088/1367-2630/16/5/055008} {\bibfield
  {journal} {\bibinfo  {journal} {NJP}\ }\textbf {\bibinfo {volume} {16}},\
  \bibinfo {pages} {055008} (\bibinfo {year} {2014})}\BibitemShut {NoStop}%
\bibitem [{\citenamefont {Armour}\ \emph {et~al.}(2015)\citenamefont {Armour},
  \citenamefont {Kubala},\ and\ \citenamefont {Ankerhold}}]{Armour2015}%
  \BibitemOpen
  \bibfield  {author} {\bibinfo {author} {\bibfnamefont {A.~D.}\ \bibnamefont
  {Armour}}, \bibinfo {author} {\bibfnamefont {B.}~\bibnamefont {Kubala}}, \
  and\ \bibinfo {author} {\bibfnamefont {J.}~\bibnamefont {Ankerhold}},\ }\href
  {\doibase 10.1103/PhysRevB.91.184508} {\bibfield  {journal} {\bibinfo
  {journal} {Phys. Rev. B}\ }\textbf {\bibinfo {volume} {91}},\ \bibinfo
  {pages} {184508} (\bibinfo {year} {2015})}\BibitemShut {NoStop}%
\bibitem [{\citenamefont {Buters}\ \emph {et~al.}(2015)\citenamefont {Buters},
  \citenamefont {Eerkens}, \citenamefont {Heeck}, \citenamefont {Weaver},
  \citenamefont {Pepper}, \citenamefont {de~Man},\ and\ \citenamefont
  {Bouwmeester}}]{Buters2015}%
  \BibitemOpen
  \bibfield  {author} {\bibinfo {author} {\bibfnamefont {F.~M.}\ \bibnamefont
  {Buters}}, \bibinfo {author} {\bibfnamefont {H.~J.}\ \bibnamefont {Eerkens}},
  \bibinfo {author} {\bibfnamefont {K.}~\bibnamefont {Heeck}}, \bibinfo
  {author} {\bibfnamefont {M.~J.}\ \bibnamefont {Weaver}}, \bibinfo {author}
  {\bibfnamefont {B.}~\bibnamefont {Pepper}}, \bibinfo {author} {\bibfnamefont
  {S.}~\bibnamefont {de~Man}}, \ and\ \bibinfo {author} {\bibfnamefont
  {D.}~\bibnamefont {Bouwmeester}},\ }\href {\doibase
  10.1103/PhysRevA.92.013811} {\bibfield  {journal} {\bibinfo  {journal} {Phys.
  Rev. A}\ }\textbf {\bibinfo {volume} {92}},\ \bibinfo {pages} {013811}
  (\bibinfo {year} {2015})}\BibitemShut {NoStop}%
\bibitem [{\citenamefont {Defoort}\ \emph {et~al.}(2015)\citenamefont
  {Defoort}, \citenamefont {Puller}, \citenamefont {Bourgeois}, \citenamefont
  {Pistolesi},\ and\ \citenamefont {Collin}}]{Defoort2015}%
  \BibitemOpen
  \bibfield  {author} {\bibinfo {author} {\bibfnamefont {M.}~\bibnamefont
  {Defoort}}, \bibinfo {author} {\bibfnamefont {V.}~\bibnamefont {Puller}},
  \bibinfo {author} {\bibfnamefont {O.}~\bibnamefont {Bourgeois}}, \bibinfo
  {author} {\bibfnamefont {F.}~\bibnamefont {Pistolesi}}, \ and\ \bibinfo
  {author} {\bibfnamefont {E.}~\bibnamefont {Collin}},\ }\href {\doibase
  10.1103/PhysRevE.92.050903} {\bibfield  {journal} {\bibinfo  {journal} {Phys.
  Rev. E}\ }\textbf {\bibinfo {volume} {92}},\ \bibinfo {pages} {050903}
  (\bibinfo {year} {2015})}\BibitemShut {NoStop}%
\bibitem [{\citenamefont {Cohen}\ \emph {et~al.}(2015)\citenamefont {Cohen},
  \citenamefont {Meenehan}, \citenamefont {MacCabe}, \citenamefont
  {Groblacher}, \citenamefont {Safavi-Naeini}, \citenamefont {Marsili},
  \citenamefont {Shaw},\ and\ \citenamefont {Painter}}]{Cohen2015}%
  \BibitemOpen
  \bibfield  {author} {\bibinfo {author} {\bibfnamefont {J.~D.}\ \bibnamefont
  {Cohen}}, \bibinfo {author} {\bibfnamefont {S.~M.}\ \bibnamefont {Meenehan}},
  \bibinfo {author} {\bibfnamefont {G.~S.}\ \bibnamefont {MacCabe}}, \bibinfo
  {author} {\bibfnamefont {S.}~\bibnamefont {Groblacher}}, \bibinfo {author}
  {\bibfnamefont {A.~H.}\ \bibnamefont {Safavi-Naeini}}, \bibinfo {author}
  {\bibfnamefont {F.}~\bibnamefont {Marsili}}, \bibinfo {author} {\bibfnamefont
  {M.~D.}\ \bibnamefont {Shaw}}, \ and\ \bibinfo {author} {\bibfnamefont
  {O.}~\bibnamefont {Painter}},\ }\href@noop {} {\bibfield  {journal} {\bibinfo
   {journal} {Nature}\ }\textbf {\bibinfo {volume} {520}},\ \bibinfo {pages}
  {522} (\bibinfo {year} {2015})}\BibitemShut {NoStop}%
\bibitem [{\citenamefont {Singh}\ \emph {et~al.}(2016)\citenamefont {Singh},
  \citenamefont {Shevchuk}, \citenamefont {Blanter},\ and\ \citenamefont
  {Steele}}]{Singh2016}%
  \BibitemOpen
  \bibfield  {author} {\bibinfo {author} {\bibfnamefont {V.}~\bibnamefont
  {Singh}}, \bibinfo {author} {\bibfnamefont {O.}~\bibnamefont {Shevchuk}},
  \bibinfo {author} {\bibfnamefont {Y.~M.}\ \bibnamefont {Blanter}}, \ and\
  \bibinfo {author} {\bibfnamefont {G.~A.}\ \bibnamefont {Steele}},\ }\href
  {\doibase 10.1103/PhysRevB.93.245407} {\bibfield  {journal} {\bibinfo
  {journal} {Phys. Rev. B}\ }\textbf {\bibinfo {volume} {93}},\ \bibinfo
  {pages} {245407} (\bibinfo {year} {2016})}\BibitemShut {NoStop}%
\bibitem [{\citenamefont {Riedinger}\ \emph {et~al.}(2016)\citenamefont
  {Riedinger}, \citenamefont {Hong}, \citenamefont {Norte}, \citenamefont
  {Slater}, \citenamefont {Shang}, \citenamefont {Krause}, \citenamefont
  {Anant}, \citenamefont {Aspelmeyer},\ and\ \citenamefont
  {Gr\"oblacher}}]{Riedinger2016}%
  \BibitemOpen
  \bibfield  {author} {\bibinfo {author} {\bibfnamefont {R.}~\bibnamefont
  {Riedinger}}, \bibinfo {author} {\bibfnamefont {S.}~\bibnamefont {Hong}},
  \bibinfo {author} {\bibfnamefont {R.~A.}\ \bibnamefont {Norte}}, \bibinfo
  {author} {\bibfnamefont {J.~A.}\ \bibnamefont {Slater}}, \bibinfo {author}
  {\bibfnamefont {J.}~\bibnamefont {Shang}}, \bibinfo {author} {\bibfnamefont
  {A.~G.}\ \bibnamefont {Krause}}, \bibinfo {author} {\bibfnamefont
  {V.}~\bibnamefont {Anant}}, \bibinfo {author} {\bibfnamefont
  {M.}~\bibnamefont {Aspelmeyer}}, \ and\ \bibinfo {author} {\bibfnamefont
  {S.}~\bibnamefont {Gr\"oblacher}},\ }\href@noop {} {\bibfield  {journal}
  {\bibinfo  {journal} {Nature}\ }\textbf {\bibinfo {volume} {530}},\ \bibinfo
  {pages} {313} (\bibinfo {year} {2016})}\BibitemShut {NoStop}%
\bibitem [{\citenamefont {Stowe}\ \emph {et~al.}(1997)\citenamefont {Stowe},
  \citenamefont {Yasumura}, \citenamefont {Kenny}, \citenamefont {Botkin},
  \citenamefont {Wago},\ and\ \citenamefont {Rugar}}]{Stowe1997}%
  \BibitemOpen
  \bibfield  {author} {\bibinfo {author} {\bibfnamefont {T.~D.}\ \bibnamefont
  {Stowe}}, \bibinfo {author} {\bibfnamefont {K.}~\bibnamefont {Yasumura}},
  \bibinfo {author} {\bibfnamefont {T.~W.}\ \bibnamefont {Kenny}}, \bibinfo
  {author} {\bibfnamefont {D.}~\bibnamefont {Botkin}}, \bibinfo {author}
  {\bibfnamefont {K.~.}\ \bibnamefont {Wago}}, \ and\ \bibinfo {author}
  {\bibfnamefont {D.}~\bibnamefont {Rugar}},\ }\href@noop {} {\bibfield
  {journal} {\bibinfo  {journal} {Applied Physics Letters}\ }\textbf {\bibinfo
  {volume} {71}},\ \bibinfo {pages} {288} (\bibinfo {year} {1997})}\BibitemShut
  {NoStop}%
\bibitem [{\citenamefont {Rugar}\ \emph {et~al.}(2004)\citenamefont {Rugar},
  \citenamefont {Budakian}, \citenamefont {Mamin},\ and\ \citenamefont
  {Chui}}]{Rugar2004}%
  \BibitemOpen
  \bibfield  {author} {\bibinfo {author} {\bibfnamefont {D.}~\bibnamefont
  {Rugar}}, \bibinfo {author} {\bibfnamefont {R.}~\bibnamefont {Budakian}},
  \bibinfo {author} {\bibfnamefont {H.~J.}\ \bibnamefont {Mamin}}, \ and\
  \bibinfo {author} {\bibfnamefont {B.~W.}\ \bibnamefont {Chui}},\ }\href@noop
  {} {\bibfield  {journal} {\bibinfo  {journal} {Nature}\ }\textbf {\bibinfo
  {volume} {430}},\ \bibinfo {pages} {329} (\bibinfo {year}
  {2004})}\BibitemShut {NoStop}%
\bibitem [{\citenamefont {Moser}\ \emph {et~al.}(2013)\citenamefont {Moser},
  \citenamefont {G\"uttinger}, \citenamefont {Eichler}, \citenamefont
  {Esplandiu}, \citenamefont {Liu}, \citenamefont {Dykman},\ and\ \citenamefont
  {Bachtold}}]{Moser2013}%
  \BibitemOpen
  \bibfield  {author} {\bibinfo {author} {\bibfnamefont {J.}~\bibnamefont
  {Moser}}, \bibinfo {author} {\bibfnamefont {J.}~\bibnamefont {G\"uttinger}},
  \bibinfo {author} {\bibfnamefont {A.}~\bibnamefont {Eichler}}, \bibinfo
  {author} {\bibfnamefont {M.~J.}\ \bibnamefont {Esplandiu}}, \bibinfo {author}
  {\bibfnamefont {D.~E.}\ \bibnamefont {Liu}}, \bibinfo {author} {\bibfnamefont
  {M.~I.}\ \bibnamefont {Dykman}}, \ and\ \bibinfo {author} {\bibfnamefont
  {A.}~\bibnamefont {Bachtold}},\ }\href@noop {} {\bibfield  {journal}
  {\bibinfo  {journal} {Nat. Nanotech.}\ }\textbf {\bibinfo {volume} {8}},\
  \bibinfo {pages} {493} (\bibinfo {year} {2013})}\BibitemShut {NoStop}%
\bibitem [{\citenamefont {Tao}\ \emph {et~al.}(2014)\citenamefont {Tao},
  \citenamefont {Boss}, \citenamefont {Moores},\ and\ \citenamefont
  {Degen}}]{Tao2014}%
  \BibitemOpen
  \bibfield  {author} {\bibinfo {author} {\bibfnamefont {Y.}~\bibnamefont
  {Tao}}, \bibinfo {author} {\bibfnamefont {J.~M.}\ \bibnamefont {Boss}},
  \bibinfo {author} {\bibfnamefont {B.~A.}\ \bibnamefont {Moores}}, \ and\
  \bibinfo {author} {\bibfnamefont {C.~L.}\ \bibnamefont {Degen}},\ }\href
  {http://dx.doi.org/10.1038/ncomms4638} {\bibfield  {journal} {\bibinfo
  {journal} {Nat Commun}\ }\textbf {\bibinfo {volume} {5}},\ \bibinfo {pages}
  {3638} (\bibinfo {year} {2014})}\BibitemShut {NoStop}%
\bibitem [{\citenamefont {Jensen}\ \emph {et~al.}(2008)\citenamefont {Jensen},
  \citenamefont {Kim},\ and\ \citenamefont {Zettl}}]{Jensen2008}%
  \BibitemOpen
  \bibfield  {author} {\bibinfo {author} {\bibfnamefont {K.}~\bibnamefont
  {Jensen}}, \bibinfo {author} {\bibfnamefont {K.}~\bibnamefont {Kim}}, \ and\
  \bibinfo {author} {\bibfnamefont {A.}~\bibnamefont {Zettl}},\ }\href@noop {}
  {\bibfield  {journal} {\bibinfo  {journal} {Nature Nanotech.}\ }\textbf
  {\bibinfo {volume} {3}},\ \bibinfo {pages} {533} (\bibinfo {year}
  {2008})}\BibitemShut {NoStop}%
\bibitem [{\citenamefont {Chaste}\ \emph {et~al.}(2012)\citenamefont {Chaste},
  \citenamefont {Eichler}, \citenamefont {Moser}, \citenamefont {Ceballos},
  \citenamefont {Rurali},\ and\ \citenamefont {Bachtold}}]{Chaste2012}%
  \BibitemOpen
  \bibfield  {author} {\bibinfo {author} {\bibfnamefont {J.}~\bibnamefont
  {Chaste}}, \bibinfo {author} {\bibfnamefont {A.}~\bibnamefont {Eichler}},
  \bibinfo {author} {\bibfnamefont {J.}~\bibnamefont {Moser}}, \bibinfo
  {author} {\bibfnamefont {G.}~\bibnamefont {Ceballos}}, \bibinfo {author}
  {\bibfnamefont {R.}~\bibnamefont {Rurali}}, \ and\ \bibinfo {author}
  {\bibfnamefont {A.}~\bibnamefont {Bachtold}},\ }\href
  {http://dx.doi.org/10.1038/nnano.2012.42} {\bibfield  {journal} {\bibinfo
  {journal} {Nat. Nano}\ }\textbf {\bibinfo {volume} {7}},\ \bibinfo {pages}
  {301} (\bibinfo {year} {2012})}\BibitemShut {NoStop}%
\bibitem [{\citenamefont {Hanay}\ \emph {et~al.}(2015)\citenamefont {Hanay},
  \citenamefont {Kelber}, \citenamefont {O'Connell}, \citenamefont {Mulvaney},
  \citenamefont {Sader},\ and\ \citenamefont {Roukes}}]{Hanay2015}%
  \BibitemOpen
  \bibfield  {author} {\bibinfo {author} {\bibfnamefont {M.~S.}\ \bibnamefont
  {Hanay}}, \bibinfo {author} {\bibfnamefont {S.~I.}\ \bibnamefont {Kelber}},
  \bibinfo {author} {\bibfnamefont {C.~D.}\ \bibnamefont {O'Connell}}, \bibinfo
  {author} {\bibfnamefont {P.}~\bibnamefont {Mulvaney}}, \bibinfo {author}
  {\bibfnamefont {J.~E.}\ \bibnamefont {Sader}}, \ and\ \bibinfo {author}
  {\bibfnamefont {M.~L.}\ \bibnamefont {Roukes}},\ }\href
  {http://dx.doi.org/10.1038/nnano.2015.32} {\bibfield  {journal} {\bibinfo
  {journal} {Nat Nano}\ }\textbf {\bibinfo {volume} {10}},\ \bibinfo {pages}
  {339} (\bibinfo {year} {2015})}\BibitemShut {NoStop}%
\bibitem [{\citenamefont {Cross}\ and\ \citenamefont
  {Lifshitz}(2001)}]{Cross2001}%
  \BibitemOpen
  \bibfield  {author} {\bibinfo {author} {\bibfnamefont {M.~C.}\ \bibnamefont
  {Cross}}\ and\ \bibinfo {author} {\bibfnamefont {R.}~\bibnamefont
  {Lifshitz}},\ }\href@noop {} {\bibfield  {journal} {\bibinfo  {journal}
  {Phys. Rev. B}\ }\textbf {\bibinfo {volume} {64}},\ \bibinfo {pages} {085324}
  (\bibinfo {year} {2001})}\BibitemShut {NoStop}%
\bibitem [{\citenamefont {Park}\ and\ \citenamefont {Park}(2004)}]{Park2004}%
  \BibitemOpen
  \bibfield  {author} {\bibinfo {author} {\bibfnamefont {Y.-H.}\ \bibnamefont
  {Park}}\ and\ \bibinfo {author} {\bibfnamefont {K.~C.}\ \bibnamefont
  {Park}},\ }\href@noop {} {\bibfield  {journal} {\bibinfo  {journal} {JMEMS}\
  }\textbf {\bibinfo {volume} {13}},\ \bibinfo {pages} {238} (\bibinfo {year}
  {2004})}\BibitemShut {NoStop}%
\bibitem [{\citenamefont {Judge}\ \emph {et~al.}(2007)\citenamefont {Judge},
  \citenamefont {Photiadis}, \citenamefont {Vignola}, \citenamefont {Houston},\
  and\ \citenamefont {Jarzynski}}]{Judge2007}%
  \BibitemOpen
  \bibfield  {author} {\bibinfo {author} {\bibfnamefont {J.~A.}\ \bibnamefont
  {Judge}}, \bibinfo {author} {\bibfnamefont {D.~M.}\ \bibnamefont
  {Photiadis}}, \bibinfo {author} {\bibfnamefont {J.~F.}\ \bibnamefont
  {Vignola}}, \bibinfo {author} {\bibfnamefont {B.~H.}\ \bibnamefont
  {Houston}}, \ and\ \bibinfo {author} {\bibfnamefont {J.}~\bibnamefont
  {Jarzynski}},\ }\href {\doibase http://dx.doi.org/10.1063/1.2401271}
  {\bibfield  {journal} {\bibinfo  {journal} {J. Appl. Phys.}\ }\textbf
  {\bibinfo {volume} {101}} (\bibinfo {year} {2007}),\
  http://dx.doi.org/10.1063/1.2401271}\BibitemShut {NoStop}%
\bibitem [{\citenamefont {Wilson-Rae}(2008)}]{Wilson-Rae2008}%
  \BibitemOpen
  \bibfield  {author} {\bibinfo {author} {\bibfnamefont {I.}~\bibnamefont
  {Wilson-Rae}},\ }\href@noop {} {\bibfield  {journal} {\bibinfo  {journal}
  {Phys. Rev. B}\ }\textbf {\bibinfo {volume} {77}},\ \bibinfo {pages} {245418}
  (\bibinfo {year} {2008})}\BibitemShut {NoStop}%
\bibitem [{\citenamefont {Croy}\ \emph {et~al.}(2012)\citenamefont {Croy},
  \citenamefont {Midtvedt}, \citenamefont {Isacsson},\ and\ \citenamefont
  {Kinaret}}]{Croy2012}%
  \BibitemOpen
  \bibfield  {author} {\bibinfo {author} {\bibfnamefont {A.}~\bibnamefont
  {Croy}}, \bibinfo {author} {\bibfnamefont {D.}~\bibnamefont {Midtvedt}},
  \bibinfo {author} {\bibfnamefont {A.}~\bibnamefont {Isacsson}}, \ and\
  \bibinfo {author} {\bibfnamefont {J.~M.}\ \bibnamefont {Kinaret}},\ }\href
  {\doibase 10.1103/PhysRevB.86.235435} {\bibfield  {journal} {\bibinfo
  {journal} {Phys. Rev. B}\ }\textbf {\bibinfo {volume} {86}},\ \bibinfo
  {pages} {235435} (\bibinfo {year} {2012})}\BibitemShut {NoStop}%
\bibitem [{\citenamefont {Lifshitz}\ and\ \citenamefont
  {Roukes}(2000)}]{Lifshitz2000}%
  \BibitemOpen
  \bibfield  {author} {\bibinfo {author} {\bibfnamefont {R.}~\bibnamefont
  {Lifshitz}}\ and\ \bibinfo {author} {\bibfnamefont {M.~L.}\ \bibnamefont
  {Roukes}},\ }\href {\doibase 10.1103/PhysRevB.61.5600} {\bibfield  {journal}
  {\bibinfo  {journal} {Phys. Rev. B}\ }\textbf {\bibinfo {volume} {61}},\
  \bibinfo {pages} {5600} (\bibinfo {year} {2000})}\BibitemShut {NoStop}%
\bibitem [{\citenamefont {De~Martino}\ \emph {et~al.}(2009)\citenamefont
  {De~Martino}, \citenamefont {Egger},\ and\ \citenamefont
  {Gogolin}}]{DeMartino2009}%
  \BibitemOpen
  \bibfield  {author} {\bibinfo {author} {\bibfnamefont {A.}~\bibnamefont
  {De~Martino}}, \bibinfo {author} {\bibfnamefont {R.}~\bibnamefont {Egger}}, \
  and\ \bibinfo {author} {\bibfnamefont {A.~O.}\ \bibnamefont {Gogolin}},\
  }\href@noop {} {\bibfield  {journal} {\bibinfo  {journal} {Phys. Rev. B}\
  }\textbf {\bibinfo {volume} {79}},\ \bibinfo {pages} {205408} (\bibinfo
  {year} {2009})}\BibitemShut {NoStop}%
\bibitem [{\citenamefont {Kunal}\ and\ \citenamefont
  {Aluru}(2011)}]{Kunal2011}%
  \BibitemOpen
  \bibfield  {author} {\bibinfo {author} {\bibfnamefont {K.}~\bibnamefont
  {Kunal}}\ and\ \bibinfo {author} {\bibfnamefont {N.~R.}\ \bibnamefont
  {Aluru}},\ }\href {\doibase 10.1103/PhysRevB.84.245450} {\bibfield  {journal}
  {\bibinfo  {journal} {Phys. Rev. B}\ }\textbf {\bibinfo {volume} {84}},\
  \bibinfo {pages} {245450} (\bibinfo {year} {2011})}\BibitemShut {NoStop}%
\bibitem [{\citenamefont {Ghaffari}\ and\ \citenamefont
  {Kenny}(2015)}]{Ghaffari2015}%
  \BibitemOpen
  \bibfield  {author} {\bibinfo {author} {\bibfnamefont {S.}~\bibnamefont
  {Ghaffari}}\ and\ \bibinfo {author} {\bibfnamefont {T.~W.}\ \bibnamefont
  {Kenny}},\ }\enquote {\bibinfo {title} {Resonant mems: Fundamentals,
  implementation, and application},}\ \ (\bibinfo  {publisher} {Wiley VCH
  (Weinheim, Germany)},\ \bibinfo {year} {2015})\ Chap.\ \bibinfo {chapter}
  {Damping in Resonant MEMS}, pp.\ \bibinfo {pages} {55--71}\BibitemShut
  {NoStop}%
\bibitem [{\citenamefont {Mahboob}\ \emph {et~al.}(2015)\citenamefont
  {Mahboob}, \citenamefont {Perrissin}, \citenamefont {Nishiguchi},
  \citenamefont {Hatanaka}, \citenamefont {Okazaki}, \citenamefont {Fujiwara},\
  and\ \citenamefont {Yamaguchi}}]{Mahboob2015}%
  \BibitemOpen
  \bibfield  {author} {\bibinfo {author} {\bibfnamefont {I.}~\bibnamefont
  {Mahboob}}, \bibinfo {author} {\bibfnamefont {N.}~\bibnamefont {Perrissin}},
  \bibinfo {author} {\bibfnamefont {K.}~\bibnamefont {Nishiguchi}}, \bibinfo
  {author} {\bibfnamefont {D.}~\bibnamefont {Hatanaka}}, \bibinfo {author}
  {\bibfnamefont {Y.}~\bibnamefont {Okazaki}}, \bibinfo {author} {\bibfnamefont
  {A.}~\bibnamefont {Fujiwara}}, \ and\ \bibinfo {author} {\bibfnamefont
  {H.}~\bibnamefont {Yamaguchi}},\ }\href@noop {} {\bibfield  {journal}
  {\bibinfo  {journal} {Nano Lett.}\ }\textbf {\bibinfo {volume} {15}},\
  \bibinfo {pages} {2312} (\bibinfo {year} {2015})}\BibitemShut {NoStop}%
\bibitem [{\citenamefont {Usmani}\ \emph {et~al.}(2007)\citenamefont {Usmani},
  \citenamefont {Blanter},\ and\ \citenamefont {Nazarov}}]{Usmani2007}%
  \BibitemOpen
  \bibfield  {author} {\bibinfo {author} {\bibfnamefont {O.}~\bibnamefont
  {Usmani}}, \bibinfo {author} {\bibfnamefont {Y.~M.}\ \bibnamefont {Blanter}},
  \ and\ \bibinfo {author} {\bibfnamefont {Y.~V.}\ \bibnamefont {Nazarov}},\
  }\href@noop {} {\bibfield  {journal} {\bibinfo  {journal} {Phys. Rev. B}\
  }\textbf {\bibinfo {volume} {75}},\ \bibinfo {pages} {195312} (\bibinfo
  {year} {2007})}\BibitemShut {NoStop}%
\bibitem [{\citenamefont {Steele}\ \emph {et~al.}(2009)\citenamefont {Steele},
  \citenamefont {Huttel}, \citenamefont {Witkamp}, \citenamefont {Poot},
  \citenamefont {Meerwaldt}, \citenamefont {Kouwenhoven},\ and\ \citenamefont
  {van~der Zant}}]{Steele2009}%
  \BibitemOpen
  \bibfield  {author} {\bibinfo {author} {\bibfnamefont {G.~A.}\ \bibnamefont
  {Steele}}, \bibinfo {author} {\bibfnamefont {A.~K.}\ \bibnamefont {Huttel}},
  \bibinfo {author} {\bibfnamefont {B.}~\bibnamefont {Witkamp}}, \bibinfo
  {author} {\bibfnamefont {M.}~\bibnamefont {Poot}}, \bibinfo {author}
  {\bibfnamefont {H.~B.}\ \bibnamefont {Meerwaldt}}, \bibinfo {author}
  {\bibfnamefont {L.~P.}\ \bibnamefont {Kouwenhoven}}, \ and\ \bibinfo {author}
  {\bibfnamefont {H.~S.~J.}\ \bibnamefont {van~der Zant}},\ }\href@noop {}
  {\bibfield  {journal} {\bibinfo  {journal} {Science}\ }\textbf {\bibinfo
  {volume} {325}},\ \bibinfo {pages} {1103} (\bibinfo {year}
  {2009})}\BibitemShut {NoStop}%
\bibitem [{\citenamefont {Lassagne}\ \emph {et~al.}(2009)\citenamefont
  {Lassagne}, \citenamefont {Tarakanov}, \citenamefont {Kinaret}, \citenamefont
  {Garcia-Sanchez},\ and\ \citenamefont {Bachtold}}]{Lassagne2009}%
  \BibitemOpen
  \bibfield  {author} {\bibinfo {author} {\bibfnamefont {B.}~\bibnamefont
  {Lassagne}}, \bibinfo {author} {\bibfnamefont {Y.}~\bibnamefont {Tarakanov}},
  \bibinfo {author} {\bibfnamefont {J.}~\bibnamefont {Kinaret}}, \bibinfo
  {author} {\bibfnamefont {D.}~\bibnamefont {Garcia-Sanchez}}, \ and\ \bibinfo
  {author} {\bibfnamefont {A.}~\bibnamefont {Bachtold}},\ }\href@noop {}
  {\bibfield  {journal} {\bibinfo  {journal} {Science}\ }\textbf {\bibinfo
  {volume} {325}},\ \bibinfo {pages} {1107} (\bibinfo {year}
  {2009})}\BibitemShut {NoStop}%
\bibitem [{\citenamefont {Lulla}\ \emph {et~al.}(2013)\citenamefont {Lulla},
  \citenamefont {Defoort}, \citenamefont {Blanc}, \citenamefont {Bourgeois},\
  and\ \citenamefont {Collin}}]{Lulla2013}%
  \BibitemOpen
  \bibfield  {author} {\bibinfo {author} {\bibfnamefont {K.~J.}\ \bibnamefont
  {Lulla}}, \bibinfo {author} {\bibfnamefont {M.}~\bibnamefont {Defoort}},
  \bibinfo {author} {\bibfnamefont {C.}~\bibnamefont {Blanc}}, \bibinfo
  {author} {\bibfnamefont {O.}~\bibnamefont {Bourgeois}}, \ and\ \bibinfo
  {author} {\bibfnamefont {E.}~\bibnamefont {Collin}},\ }\href@noop {}
  {\bibfield  {journal} {\bibinfo  {journal} {Phys. Rev. Lett.}\ }\textbf
  {\bibinfo {volume} {110}},\ \bibinfo {pages} {177206} (\bibinfo {year}
  {2013})}\BibitemShut {NoStop}%
\bibitem [{\citenamefont {Hoehne}\ \emph {et~al.}(2010)\citenamefont {Hoehne},
  \citenamefont {Pashkin}, \citenamefont {Astafiev}, \citenamefont {Faoro},
  \citenamefont {Ioffe}, \citenamefont {Nakamura},\ and\ \citenamefont
  {Tsai}}]{Hoehne2010}%
  \BibitemOpen
  \bibfield  {author} {\bibinfo {author} {\bibfnamefont {F.}~\bibnamefont
  {Hoehne}}, \bibinfo {author} {\bibfnamefont {Y.~A.}\ \bibnamefont {Pashkin}},
  \bibinfo {author} {\bibfnamefont {O.}~\bibnamefont {Astafiev}}, \bibinfo
  {author} {\bibfnamefont {L.}~\bibnamefont {Faoro}}, \bibinfo {author}
  {\bibfnamefont {L.~B.}\ \bibnamefont {Ioffe}}, \bibinfo {author}
  {\bibfnamefont {Y.}~\bibnamefont {Nakamura}}, \ and\ \bibinfo {author}
  {\bibfnamefont {J.~S.}\ \bibnamefont {Tsai}},\ }\href@noop {} {\bibfield
  {journal} {\bibinfo  {journal} {Phys. Rev. B}\ }\textbf {\bibinfo {volume}
  {81}},\ \bibinfo {pages} {184112} (\bibinfo {year} {2010})}\BibitemShut
  {NoStop}%
\bibitem [{\citenamefont {Venkatesan}\ \emph {et~al.}(2010)\citenamefont
  {Venkatesan}, \citenamefont {Lulla}, \citenamefont {Patton}, \citenamefont
  {Armour}, \citenamefont {Mellor},\ and\ \citenamefont
  {Owers-Bradley}}]{Venkatesan2010}%
  \BibitemOpen
  \bibfield  {author} {\bibinfo {author} {\bibfnamefont {A.}~\bibnamefont
  {Venkatesan}}, \bibinfo {author} {\bibfnamefont {K.~J.}\ \bibnamefont
  {Lulla}}, \bibinfo {author} {\bibfnamefont {M.~J.}\ \bibnamefont {Patton}},
  \bibinfo {author} {\bibfnamefont {A.~D.}\ \bibnamefont {Armour}}, \bibinfo
  {author} {\bibfnamefont {C.~J.}\ \bibnamefont {Mellor}}, \ and\ \bibinfo
  {author} {\bibfnamefont {J.~R.}\ \bibnamefont {Owers-Bradley}},\ }\href@noop
  {} {\bibfield  {journal} {\bibinfo  {journal} {Phys. Rev. B}\ }\textbf
  {\bibinfo {volume} {81}},\ \bibinfo {pages} {073410} (\bibinfo {year}
  {2010})}\BibitemShut {NoStop}%
\bibitem [{\citenamefont {Faust}\ \emph {et~al.}(2014)\citenamefont {Faust},
  \citenamefont {Rieger}, \citenamefont {Seitner}, \citenamefont {Kotthaus},\
  and\ \citenamefont {Weig}}]{Faust2014}%
  \BibitemOpen
  \bibfield  {author} {\bibinfo {author} {\bibfnamefont {T.}~\bibnamefont
  {Faust}}, \bibinfo {author} {\bibfnamefont {J.}~\bibnamefont {Rieger}},
  \bibinfo {author} {\bibfnamefont {M.~J.}\ \bibnamefont {Seitner}}, \bibinfo
  {author} {\bibfnamefont {J.~P.}\ \bibnamefont {Kotthaus}}, \ and\ \bibinfo
  {author} {\bibfnamefont {E.~M.}\ \bibnamefont {Weig}},\ }\href@noop {}
  {\bibfield  {journal} {\bibinfo  {journal} {Phys. Rev. B}\ }\textbf {\bibinfo
  {volume} {89}},\ \bibinfo {pages} {100102} (\bibinfo {year}
  {2014})}\BibitemShut {NoStop}%
\bibitem [{\citenamefont {Eichler}\ \emph {et~al.}(2011)\citenamefont
  {Eichler}, \citenamefont {Moser}, \citenamefont {Chaste}, \citenamefont
  {Zdrojek}, \citenamefont {Wilson-Rae},\ and\ \citenamefont
  {Bachtold}}]{Eichler2011a}%
  \BibitemOpen
  \bibfield  {author} {\bibinfo {author} {\bibfnamefont {A.}~\bibnamefont
  {Eichler}}, \bibinfo {author} {\bibfnamefont {J.}~\bibnamefont {Moser}},
  \bibinfo {author} {\bibfnamefont {J.}~\bibnamefont {Chaste}}, \bibinfo
  {author} {\bibfnamefont {M.}~\bibnamefont {Zdrojek}}, \bibinfo {author}
  {\bibfnamefont {I.}~\bibnamefont {Wilson-Rae}}, \ and\ \bibinfo {author}
  {\bibfnamefont {A.}~\bibnamefont {Bachtold}},\ }\href@noop {} {\bibfield
  {journal} {\bibinfo  {journal} {Nature Nanotech.}\ }\textbf {\bibinfo
  {volume} {6}},\ \bibinfo {pages} {339} (\bibinfo {year} {2011})}\BibitemShut
  {NoStop}%
\bibitem [{\citenamefont {Zaitsev}\ \emph {et~al.}(2012)\citenamefont
  {Zaitsev}, \citenamefont {Shtempluck}, \citenamefont {Buks},\ and\
  \citenamefont {Gottlieb}}]{Zaitsev2012}%
  \BibitemOpen
  \bibfield  {author} {\bibinfo {author} {\bibfnamefont {S.}~\bibnamefont
  {Zaitsev}}, \bibinfo {author} {\bibfnamefont {O.}~\bibnamefont {Shtempluck}},
  \bibinfo {author} {\bibfnamefont {E.}~\bibnamefont {Buks}}, \ and\ \bibinfo
  {author} {\bibfnamefont {O.}~\bibnamefont {Gottlieb}},\ }\bibfield
  {booktitle} {\emph {\bibinfo {booktitle} {Nonlinear Dynamics}},\ }\href@noop
  {} {\ \textbf {\bibinfo {volume} {67}},\ \bibinfo {pages} {859} (\bibinfo
  {year} {2012})}\BibitemShut {NoStop}%
\bibitem [{\citenamefont {Imboden}\ \emph {et~al.}(2013)\citenamefont
  {Imboden}, \citenamefont {Williams},\ and\ \citenamefont
  {Mohanty}}]{Imboden2013}%
  \BibitemOpen
  \bibfield  {author} {\bibinfo {author} {\bibfnamefont {M.}~\bibnamefont
  {Imboden}}, \bibinfo {author} {\bibfnamefont {O.~A.}\ \bibnamefont
  {Williams}}, \ and\ \bibinfo {author} {\bibfnamefont {P.}~\bibnamefont
  {Mohanty}},\ }\href {\doibase 10.1021/nl401978p} {\bibfield  {journal}
  {\bibinfo  {journal} {Nano Lett.}\ }\textbf {\bibinfo {volume} {13}},\
  \bibinfo {pages} {4014} (\bibinfo {year} {2013})}\BibitemShut {NoStop}%
\bibitem [{\citenamefont {Miao}\ \emph {et~al.}(2014)\citenamefont {Miao},
  \citenamefont {Yeom}, \citenamefont {Wang}, \citenamefont {Standley},\ and\
  \citenamefont {Bockrath}}]{Miao2014}%
  \BibitemOpen
  \bibfield  {author} {\bibinfo {author} {\bibfnamefont {T.~F.}\ \bibnamefont
  {Miao}}, \bibinfo {author} {\bibfnamefont {S.}~\bibnamefont {Yeom}}, \bibinfo
  {author} {\bibfnamefont {P.}~\bibnamefont {Wang}}, \bibinfo {author}
  {\bibfnamefont {B.}~\bibnamefont {Standley}}, \ and\ \bibinfo {author}
  {\bibfnamefont {M.}~\bibnamefont {Bockrath}},\ }\href {\doibase
  10.1021/nl403936a} {\bibfield  {journal} {\bibinfo  {journal} {Nano Lett.}\
  }\textbf {\bibinfo {volume} {14}},\ \bibinfo {pages} {2982} (\bibinfo {year}
  {2014})}\BibitemShut {NoStop}%
\bibitem [{\citenamefont {Leghtas}\ \emph {et~al.}(2015)\citenamefont
  {Leghtas}, \citenamefont {Touzard}, \citenamefont {Pop}, \citenamefont {Kou},
  \citenamefont {Vlastakis}, \citenamefont {Petrenko}, \citenamefont {Sliwa},
  \citenamefont {Narla}, \citenamefont {Shankar}, \citenamefont {Hatridge},
  \citenamefont {Reagor}, \citenamefont {Frunzio}, \citenamefont {Schoelkopf},
  \citenamefont {Mirrahimi},\ and\ \citenamefont {Devoret}}]{Leghtas2015}%
  \BibitemOpen
  \bibfield  {author} {\bibinfo {author} {\bibfnamefont {Z.}~\bibnamefont
  {Leghtas}}, \bibinfo {author} {\bibfnamefont {S.}~\bibnamefont {Touzard}},
  \bibinfo {author} {\bibfnamefont {I.~M.}\ \bibnamefont {Pop}}, \bibinfo
  {author} {\bibfnamefont {A.}~\bibnamefont {Kou}}, \bibinfo {author}
  {\bibfnamefont {B.}~\bibnamefont {Vlastakis}}, \bibinfo {author}
  {\bibfnamefont {A.}~\bibnamefont {Petrenko}}, \bibinfo {author}
  {\bibfnamefont {K.~M.}\ \bibnamefont {Sliwa}}, \bibinfo {author}
  {\bibfnamefont {A.}~\bibnamefont {Narla}}, \bibinfo {author} {\bibfnamefont
  {S.}~\bibnamefont {Shankar}}, \bibinfo {author} {\bibfnamefont {M.~J.}\
  \bibnamefont {Hatridge}}, \bibinfo {author} {\bibfnamefont {M.}~\bibnamefont
  {Reagor}}, \bibinfo {author} {\bibfnamefont {L.}~\bibnamefont {Frunzio}},
  \bibinfo {author} {\bibfnamefont {R.~J.}\ \bibnamefont {Schoelkopf}},
  \bibinfo {author} {\bibfnamefont {M.}~\bibnamefont {Mirrahimi}}, \ and\
  \bibinfo {author} {\bibfnamefont {M.~H.}\ \bibnamefont {Devoret}},\ }\href
  {\doibase 10.1126/science.aaa2085} {\bibfield  {journal} {\bibinfo  {journal}
  {Science}\ }\textbf {\bibinfo {volume} {347}},\ \bibinfo {pages} {853}
  (\bibinfo {year} {2015})}\BibitemShut {NoStop}%
\bibitem [{\citenamefont {Roy}\ \emph {et~al.}(2015)\citenamefont {Roy},
  \citenamefont {Leghtas}, \citenamefont {Stone}, \citenamefont {Devoret},\
  and\ \citenamefont {Mirrahimi}}]{A_Roy2015}%
  \BibitemOpen
  \bibfield  {author} {\bibinfo {author} {\bibfnamefont {A.}~\bibnamefont
  {Roy}}, \bibinfo {author} {\bibfnamefont {Z.}~\bibnamefont {Leghtas}},
  \bibinfo {author} {\bibfnamefont {A.~D.}\ \bibnamefont {Stone}}, \bibinfo
  {author} {\bibfnamefont {M.}~\bibnamefont {Devoret}}, \ and\ \bibinfo
  {author} {\bibfnamefont {M.}~\bibnamefont {Mirrahimi}},\ }\href@noop {}
  {\bibfield  {journal} {\bibinfo  {journal} {Phys. Rev. A}\ }\textbf {\bibinfo
  {volume} {91}},\ \bibinfo {pages} {013810} (\bibinfo {year}
  {2015})}\BibitemShut {NoStop}%
\bibitem [{\citenamefont {Mandel}\ and\ \citenamefont
  {Wolf}(1995)}]{Mandel1995}%
  \BibitemOpen
  \bibfield  {author} {\bibinfo {author} {\bibfnamefont {L.}~\bibnamefont
  {Mandel}}\ and\ \bibinfo {author} {\bibfnamefont {E.}~\bibnamefont {Wolf}},\
  }\href@noop {} {\emph {\bibinfo {title} {Optical Coherence and Quantum
  Optics}}}\ (\bibinfo  {publisher} {Cambirdge University Press},\ \bibinfo
  {year} {Cambridge, 1995})\BibitemShut {NoStop}%
\bibitem [{\citenamefont {Dykman}\ and\ \citenamefont
  {Krivoglaz}(1975)}]{Dykman1975a}%
  \BibitemOpen
  \bibfield  {author} {\bibinfo {author} {\bibfnamefont {M.~I.}\ \bibnamefont
  {Dykman}}\ and\ \bibinfo {author} {\bibfnamefont {M.~A.}\ \bibnamefont
  {Krivoglaz}},\ }\href@noop {} {\bibfield  {journal} {\bibinfo  {journal}
  {Phys. Stat. Sol. (b)}\ }\textbf {\bibinfo {volume} {68}},\ \bibinfo {pages}
  {111} (\bibinfo {year} {1975})}\BibitemShut {NoStop}%
\bibitem [{\citenamefont {Meyer}\ \emph {et~al.}(2007)\citenamefont {Meyer},
  \citenamefont {Geim}, \citenamefont {Katsnelson}, \citenamefont {Novoselov},
  \citenamefont {Booth},\ and\ \citenamefont {Roth}}]{Meyer2007}%
  \BibitemOpen
  \bibfield  {author} {\bibinfo {author} {\bibfnamefont {J.~C.}\ \bibnamefont
  {Meyer}}, \bibinfo {author} {\bibfnamefont {A.~K.}\ \bibnamefont {Geim}},
  \bibinfo {author} {\bibfnamefont {M.~I.}\ \bibnamefont {Katsnelson}},
  \bibinfo {author} {\bibfnamefont {K.~S.}\ \bibnamefont {Novoselov}}, \bibinfo
  {author} {\bibfnamefont {T.~J.}\ \bibnamefont {Booth}}, \ and\ \bibinfo
  {author} {\bibfnamefont {S.}~\bibnamefont {Roth}},\ }\href
  {http://dx.doi.org/10.1038/nature05545} {\bibfield  {journal} {\bibinfo
  {journal} {Nature}\ }\textbf {\bibinfo {volume} {446}},\ \bibinfo {pages}
  {60} (\bibinfo {year} {2007})}\BibitemShut {NoStop}%
\bibitem [{\citenamefont {Gurevich}(1988)}]{Gurevich1988}%
  \BibitemOpen
  \bibfield  {author} {\bibinfo {author} {\bibfnamefont {V.~L.}\ \bibnamefont
  {Gurevich}},\ }\href@noop {} {\emph {\bibinfo {title} {Transport in Phonon
  Systems}}}\ (\bibinfo  {publisher} {Elsevier Science Ltd},\ \bibinfo {year}
  {1988})\BibitemShut {NoStop}%
\bibitem [{\citenamefont {Nika}\ \emph {et~al.}(2009)\citenamefont {Nika},
  \citenamefont {Pokatilov}, \citenamefont {Askerov},\ and\ \citenamefont
  {Balandin}}]{Nika2009}%
  \BibitemOpen
  \bibfield  {author} {\bibinfo {author} {\bibfnamefont {D.~L.}\ \bibnamefont
  {Nika}}, \bibinfo {author} {\bibfnamefont {E.~P.}\ \bibnamefont {Pokatilov}},
  \bibinfo {author} {\bibfnamefont {A.~S.}\ \bibnamefont {Askerov}}, \ and\
  \bibinfo {author} {\bibfnamefont {A.~A.}\ \bibnamefont {Balandin}},\ }\href
  {\doibase 10.1103/PhysRevB.79.155413} {\bibfield  {journal} {\bibinfo
  {journal} {Phys. Rev. B}\ }\textbf {\bibinfo {volume} {79}},\ \bibinfo
  {pages} {155413} (\bibinfo {year} {2009})}\BibitemShut {NoStop}%
\bibitem [{\citenamefont {Landau}\ and\ \citenamefont
  {Lifshitz}(1986)}]{LL_Elasticity}%
  \BibitemOpen
  \bibfield  {author} {\bibinfo {author} {\bibfnamefont {L.}~\bibnamefont
  {Landau}}\ and\ \bibinfo {author} {\bibfnamefont {E.}~\bibnamefont
  {Lifshitz}},\ }\href@noop {} {\emph {\bibinfo {title} {Theory of
  elasticity}}},\ \bibinfo {edition} {3rd}\ ed.\ (\bibinfo  {publisher}
  {Butterworth-Heinemann Ltd., Oxford},\ \bibinfo {year} {1986})\BibitemShut
  {NoStop}%
\bibitem [{\citenamefont {Lifshitz}\ and\ \citenamefont
  {Pitaevskii}(1981)}]{Lifshitz1981a}%
  \BibitemOpen
  \bibfield  {author} {\bibinfo {author} {\bibfnamefont {E.}~\bibnamefont
  {Lifshitz}}\ and\ \bibinfo {author} {\bibfnamefont {L.}~\bibnamefont
  {Pitaevskii}},\ }\href@noop {} {\emph {\bibinfo {title} {Physical
  kinetics}}}\ (\bibinfo  {publisher} {Butterworth-Heinemann Ltd., Oxford},\
  \bibinfo {year} {1981})\BibitemShut {NoStop}%
\bibitem [{\citenamefont {Landau}\ and\ \citenamefont
  {Khalatnikov}(1949{\natexlab{a}})}]{Landau1949}%
  \BibitemOpen
  \bibfield  {author} {\bibinfo {author} {\bibfnamefont {L.~D.}\ \bibnamefont
  {Landau}}\ and\ \bibinfo {author} {\bibfnamefont {I.~M.}\ \bibnamefont
  {Khalatnikov}},\ }\href@noop {} {\bibfield  {journal} {\bibinfo  {journal}
  {Zh. Exper. Theor. Phys.}\ }\textbf {\bibinfo {volume} {19}},\ \bibinfo
  {pages} {637} (\bibinfo {year} {1949}{\natexlab{a}})}\BibitemShut {NoStop}%
\bibitem [{\citenamefont {Landau}\ and\ \citenamefont
  {Khalatnikov}(1949{\natexlab{b}})}]{Landau1949a}%
  \BibitemOpen
  \bibfield  {author} {\bibinfo {author} {\bibfnamefont {L.~D.}\ \bibnamefont
  {Landau}}\ and\ \bibinfo {author} {\bibfnamefont {I.~M.}\ \bibnamefont
  {Khalatnikov}},\ }\href@noop {} {\bibfield  {journal} {\bibinfo  {journal}
  {Zh. Exper. Theor. Phys.}\ }\textbf {\bibinfo {volume} {19}},\ \bibinfo
  {pages} {709} (\bibinfo {year} {1949}{\natexlab{b}})}\BibitemShut {NoStop}%
\bibitem [{\citenamefont {Midtvedt}\ \emph {et~al.}(2014)\citenamefont
  {Midtvedt}, \citenamefont {Croy}, \citenamefont {Isacsson}, \citenamefont
  {Qi},\ and\ \citenamefont {Park}}]{Midtvedt2014}%
  \BibitemOpen
  \bibfield  {author} {\bibinfo {author} {\bibfnamefont {D.}~\bibnamefont
  {Midtvedt}}, \bibinfo {author} {\bibfnamefont {A.}~\bibnamefont {Croy}},
  \bibinfo {author} {\bibfnamefont {A.}~\bibnamefont {Isacsson}}, \bibinfo
  {author} {\bibfnamefont {Z.}~\bibnamefont {Qi}}, \ and\ \bibinfo {author}
  {\bibfnamefont {H.~S.}\ \bibnamefont {Park}},\ }\href@noop {} {\bibfield
  {journal} {\bibinfo  {journal} {Phys. Rev. Lett.}\ }\textbf {\bibinfo
  {volume} {112}},\ \bibinfo {pages} {145503} (\bibinfo {year}
  {2014})}\BibitemShut {NoStop}%
\bibitem [{\citenamefont {Ivanov}\ \emph {et~al.}(1965)\citenamefont {Ivanov},
  \citenamefont {Kvashnina},\ and\ \citenamefont {Krivoglaz}}]{Ivanov1965}%
  \BibitemOpen
  \bibfield  {author} {\bibinfo {author} {\bibfnamefont {M.~A.}\ \bibnamefont
  {Ivanov}}, \bibinfo {author} {\bibfnamefont {L.~B.}\ \bibnamefont
  {Kvashnina}}, \ and\ \bibinfo {author} {\bibfnamefont {M.~A.}\ \bibnamefont
  {Krivoglaz}},\ }\href@noop {} {\bibfield  {journal} {\bibinfo  {journal}
  {Sov. Phys. Solid State}\ }\textbf {\bibinfo {volume} {7}},\ \bibinfo {pages}
  {1652} (\bibinfo {year} {1965})}\BibitemShut {NoStop}%
\bibitem [{\citenamefont {Elliott}\ \emph {et~al.}(1965)\citenamefont
  {Elliott}, \citenamefont {Hayes}, \citenamefont {Jones}, \citenamefont
  {MacDonald},\ and\ \citenamefont {Sennett}}]{Elliott1965}%
  \BibitemOpen
  \bibfield  {author} {\bibinfo {author} {\bibfnamefont {R.~J.}\ \bibnamefont
  {Elliott}}, \bibinfo {author} {\bibfnamefont {W.}~\bibnamefont {Hayes}},
  \bibinfo {author} {\bibfnamefont {G.~D.}\ \bibnamefont {Jones}}, \bibinfo
  {author} {\bibfnamefont {H.~F.}\ \bibnamefont {MacDonald}}, \ and\ \bibinfo
  {author} {\bibfnamefont {C.~T.}\ \bibnamefont {Sennett}},\ }\href@noop {}
  {\bibfield  {journal} {\bibinfo  {journal} {Proc. Roy. Soc. London}\ }\textbf
  {\bibinfo {volume} {A289}},\ \bibinfo {pages} {1} (\bibinfo {year}
  {1965})}\BibitemShut {NoStop}%
\bibitem [{\citenamefont {Landau}\ and\ \citenamefont
  {Rumer}(1937)}]{Landau1937}%
  \BibitemOpen
  \bibfield  {author} {\bibinfo {author} {\bibfnamefont {L.~D.}\ \bibnamefont
  {Landau}}\ and\ \bibinfo {author} {\bibfnamefont {G.}~\bibnamefont {Rumer}},\
  }\href@noop {} {\bibfield  {journal} {\bibinfo  {journal} {Phys. Z. Sowjet.}\
  }\textbf {\bibinfo {volume} {11}},\ \bibinfo {pages} {18} (\bibinfo {year}
  {1937})}\BibitemShut {NoStop}%
\bibitem [{\citenamefont {Akhiezer}(1938)}]{Akhiezer1938}%
  \BibitemOpen
  \bibfield  {author} {\bibinfo {author} {\bibfnamefont {A.~I.}\ \bibnamefont
  {Akhiezer}},\ }\href@noop {} {\bibfield  {journal} {\bibinfo  {journal} {Zh.
  Eksp. Teor. Fiz.}\ }\textbf {\bibinfo {volume} {8}},\ \bibinfo {pages} {1318}
  (\bibinfo {year} {1938})}\BibitemShut {NoStop}%
\bibitem [{\citenamefont {Maris}(1966)}]{Maris1966}%
  \BibitemOpen
  \bibfield  {author} {\bibinfo {author} {\bibfnamefont {H.~J.}\ \bibnamefont
  {Maris}},\ }\href {\doibase 10.1080/14786436608212640} {\bibfield  {journal}
  {\bibinfo  {journal} {Phil. Mag.}\ }\textbf {\bibinfo {volume} {13}},\
  \bibinfo {pages} {465} (\bibinfo {year} {1966})}\BibitemShut {NoStop}%
\bibitem [{\citenamefont {Garanin}\ and\ \citenamefont
  {Lutovinov}(1992)}]{Garanin1992}%
  \BibitemOpen
  \bibfield  {author} {\bibinfo {author} {\bibfnamefont {D.}~\bibnamefont
  {Garanin}}\ and\ \bibinfo {author} {\bibfnamefont {V.}~\bibnamefont
  {Lutovinov}},\ }\href@noop {} {\bibfield  {journal} {\bibinfo  {journal}
  {Ann. Phys.}\ }\textbf {\bibinfo {volume} {218}},\ \bibinfo {pages} {293}
  (\bibinfo {year} {1992})}\BibitemShut {NoStop}%
\bibitem [{\citenamefont {Collins}\ \emph {et~al.}(2013)\citenamefont
  {Collins}, \citenamefont {Maznev}, \citenamefont {Tian}, \citenamefont
  {Esfarjani}, \citenamefont {Nelson},\ and\ \citenamefont
  {Chen}}]{Collins2013}%
  \BibitemOpen
  \bibfield  {author} {\bibinfo {author} {\bibfnamefont {K.~C.}\ \bibnamefont
  {Collins}}, \bibinfo {author} {\bibfnamefont {A.}~\bibnamefont {Maznev}},
  \bibinfo {author} {\bibfnamefont {Z.}~\bibnamefont {Tian}}, \bibinfo {author}
  {\bibfnamefont {K.}~\bibnamefont {Esfarjani}}, \bibinfo {author}
  {\bibfnamefont {K.~A.}\ \bibnamefont {Nelson}}, \ and\ \bibinfo {author}
  {\bibfnamefont {G.}~\bibnamefont {Chen}},\ }\href {\doibase
  http://dx.doi.org/10.1063/1.4820572} {\bibfield  {journal} {\bibinfo
  {journal} {J. Appl. Phys.}\ }\textbf {\bibinfo {volume} {114}},\ \bibinfo
  {pages} {104302} (\bibinfo {year} {2013})}\BibitemShut {NoStop}%
\bibitem [{\citenamefont {Lindenfeld}\ and\ \citenamefont
  {Lifshitz}(2013)}]{Lindenfeld2013}%
  \BibitemOpen
  \bibfield  {author} {\bibinfo {author} {\bibfnamefont {Z.}~\bibnamefont
  {Lindenfeld}}\ and\ \bibinfo {author} {\bibfnamefont {R.}~\bibnamefont
  {Lifshitz}},\ }\href@noop {} {\bibfield  {journal} {\bibinfo  {journal}
  {Phys. Rev. B}\ }\textbf {\bibinfo {volume} {87}},\ \bibinfo {pages} {085448}
  (\bibinfo {year} {2013})}\BibitemShut {NoStop}%
\bibitem [{\citenamefont {Feng}\ \emph {et~al.}(2015)\citenamefont {Feng},
  \citenamefont {Qiu},\ and\ \citenamefont {Ruan}}]{Feng2015}%
  \BibitemOpen
  \bibfield  {author} {\bibinfo {author} {\bibfnamefont {T.}~\bibnamefont
  {Feng}}, \bibinfo {author} {\bibfnamefont {B.}~\bibnamefont {Qiu}}, \ and\
  \bibinfo {author} {\bibfnamefont {X.}~\bibnamefont {Ruan}},\ }\href {\doibase
  10.1103/PhysRevB.92.235206} {\bibfield  {journal} {\bibinfo  {journal} {Phys.
  Rev. B}\ }\textbf {\bibinfo {volume} {92}},\ \bibinfo {pages} {235206}
  (\bibinfo {year} {2015})}\BibitemShut {NoStop}%
\bibitem [{\citenamefont {Antonio}\ \emph {et~al.}(2012)\citenamefont
  {Antonio}, \citenamefont {Zanette},\ and\ \citenamefont
  {Lopez}}]{Antonio2012}%
  \BibitemOpen
  \bibfield  {author} {\bibinfo {author} {\bibfnamefont {D.}~\bibnamefont
  {Antonio}}, \bibinfo {author} {\bibfnamefont {D.~H.}\ \bibnamefont
  {Zanette}}, \ and\ \bibinfo {author} {\bibfnamefont {D.}~\bibnamefont
  {Lopez}},\ }\href@noop {} {\bibfield  {journal} {\bibinfo  {journal} {Nature
  Communications}\ }\textbf {\bibinfo {volume} {3}},\ \bibinfo {pages} {806}
  (\bibinfo {year} {2012})}\BibitemShut {NoStop}%
\bibitem [{\citenamefont {Dykman}(2012{\natexlab{b}})}]{Dykman2012}%
  \BibitemOpen
  \bibfield  {author} {\bibinfo {author} {\bibfnamefont {M.~I.}\ \bibnamefont
  {Dykman}},\ }in\ \href@noop {} {\emph {\bibinfo {booktitle} {Fluctuating
  Nonlinear Oscillators: from Nanomechanics to Quantum Superconducting
  Circuits}}},\ \bibinfo {editor} {edited by\ \bibinfo {editor} {\bibfnamefont
  {M.~I.}\ \bibnamefont {Dykman}}}\ (\bibinfo  {publisher} {OUP, Oxford},\
  \bibinfo {year} {2012})\ pp.\ \bibinfo {pages} {165--197}\BibitemShut
  {NoStop}%
\bibitem [{\citenamefont {{Zhang}}\ \emph {et~al.}(2014)\citenamefont
  {{Zhang}}, \citenamefont {{Moser}}, \citenamefont {G\"uttinger},
  \citenamefont {{Bachtold}},\ and\ \citenamefont {{Dykman}}}]{Zhang2014}%
  \BibitemOpen
  \bibfield  {author} {\bibinfo {author} {\bibfnamefont {Y.}~\bibnamefont
  {{Zhang}}}, \bibinfo {author} {\bibfnamefont {J.}~\bibnamefont {{Moser}}},
  \bibinfo {author} {\bibfnamefont {J.}~\bibnamefont {G\"uttinger}}, \bibinfo
  {author} {\bibfnamefont {A.}~\bibnamefont {{Bachtold}}}, \ and\ \bibinfo
  {author} {\bibfnamefont {M.~I.}\ \bibnamefont {{Dykman}}},\ }\href@noop {}
  {\bibfield  {journal} {\bibinfo  {journal} {Phys. Rev. Lett.}\ }\textbf
  {\bibinfo {volume} {113}},\ \bibinfo {pages} {255502} (\bibinfo {year}
  {2014})}\BibitemShut {NoStop}%
\bibitem [{\citenamefont {Eichler}\ \emph {et~al.}(2012)\citenamefont
  {Eichler}, \citenamefont {del {\AA}lamo~Ruiz}, \citenamefont {Plaza},\ and\
  \citenamefont {Bachtold}}]{Eichler2012}%
  \BibitemOpen
  \bibfield  {author} {\bibinfo {author} {\bibfnamefont {A.}~\bibnamefont
  {Eichler}}, \bibinfo {author} {\bibfnamefont {M.}~\bibnamefont {del
  {\AA}lamo~Ruiz}}, \bibinfo {author} {\bibfnamefont {J.~A.}\ \bibnamefont
  {Plaza}}, \ and\ \bibinfo {author} {\bibfnamefont {A.}~\bibnamefont
  {Bachtold}},\ }\href@noop {} {\bibfield  {journal} {\bibinfo  {journal}
  {Phys. Rev. Lett.}\ }\textbf {\bibinfo {volume} {109}},\ \bibinfo {pages}
  {025503} (\bibinfo {year} {2012})}\BibitemShut {NoStop}%
\bibitem [{\citenamefont {Born}\ and\ \citenamefont {Wolf}(1999)}]{Born1999}%
  \BibitemOpen
  \bibfield  {author} {\bibinfo {author} {\bibfnamefont {M.}~\bibnamefont
  {Born}}\ and\ \bibinfo {author} {\bibfnamefont {E.}~\bibnamefont {Wolf}},\
  }\href@noop {} {\emph {\bibinfo {title} {Principles of Optics}}},\ \bibinfo
  {edition} {7th}\ ed.\ (\bibinfo  {publisher} {Cambirdge University Press},\
  \bibinfo {year} {1999})\BibitemShut {NoStop}%
\bibitem [{\citenamefont {Zener}(1937)}]{Zener1937}%
  \BibitemOpen
  \bibfield  {author} {\bibinfo {author} {\bibfnamefont {C.}~\bibnamefont
  {Zener}},\ }\href {\doibase 10.1103/PhysRev.52.230} {\bibfield  {journal}
  {\bibinfo  {journal} {Phys. Rev.}\ }\textbf {\bibinfo {volume} {52}},\
  \bibinfo {pages} {230} (\bibinfo {year} {1937})}\BibitemShut {NoStop}%
\bibitem [{\citenamefont {Graff}(1975)}]{Graff1975}%
  \BibitemOpen
  \bibfield  {author} {\bibinfo {author} {\bibfnamefont {K.~F.}\ \bibnamefont
  {Graff}},\ }\href@noop {} {\emph {\bibinfo {title} {Wave motion in elastic
  solids}}}\ (\bibinfo  {publisher} {Dover Publications, Inc.},\ \bibinfo
  {year} {1975})\BibitemShut {NoStop}%
\bibitem [{\citenamefont {Chandorkar}\ \emph {et~al.}(2009)\citenamefont
  {Chandorkar}, \citenamefont {Candler}, \citenamefont {Duwel}, \citenamefont
  {Melamud}, \citenamefont {Agarwal}, \citenamefont {Goodson},\ and\
  \citenamefont {Kenny}}]{Chandorkar2009}%
  \BibitemOpen
  \bibfield  {author} {\bibinfo {author} {\bibfnamefont {S.~A.}\ \bibnamefont
  {Chandorkar}}, \bibinfo {author} {\bibfnamefont {R.~N.}\ \bibnamefont
  {Candler}}, \bibinfo {author} {\bibfnamefont {A.}~\bibnamefont {Duwel}},
  \bibinfo {author} {\bibfnamefont {R.}~\bibnamefont {Melamud}}, \bibinfo
  {author} {\bibfnamefont {M.}~\bibnamefont {Agarwal}}, \bibinfo {author}
  {\bibfnamefont {K.~E.}\ \bibnamefont {Goodson}}, \ and\ \bibinfo {author}
  {\bibfnamefont {T.~W.}\ \bibnamefont {Kenny}},\ }\href@noop {} {\bibfield
  {journal} {\bibinfo  {journal} {J. Appl. Phys.}\ }\textbf
  {\bibinfo {volume} {105}} (\bibinfo {year} {2009})}\BibitemShut {NoStop}%
\bibitem [{\citenamefont {Landau}\ and\ \citenamefont
  {Lifshitz}(2004)}]{LL_Mechanics2004}%
  \BibitemOpen
  \bibfield  {author} {\bibinfo {author} {\bibfnamefont {L.~D.}\ \bibnamefont
  {Landau}}\ and\ \bibinfo {author} {\bibfnamefont {E.~M.}\ \bibnamefont
  {Lifshitz}},\ }\href@noop {} {\emph {\bibinfo {title} {Mechanics}}},\
  \bibinfo {edition} {3rd}\ ed.\ (\bibinfo  {publisher} {Elsevier, Amsterdam},\
  \bibinfo {year} {2004})\BibitemShut {NoStop}%
\bibitem [{\citenamefont {Iyer}\ and\ \citenamefont
  {Candler}(2016)}]{Iyer2016}%
  \BibitemOpen
  \bibfield  {author} {\bibinfo {author} {\bibfnamefont {S.~S.}\ \bibnamefont
  {Iyer}}\ and\ \bibinfo {author} {\bibfnamefont {R.~N.}\ \bibnamefont
  {Candler}},\ }\href {\doibase 10.1103/PhysRevApplied.5.034002} {\bibfield
  {journal} {\bibinfo  {journal} {Phys. Rev. Applied}\ }\textbf {\bibinfo
  {volume} {5}},\ \bibinfo {pages} {034002} (\bibinfo {year}
  {2016})}\BibitemShut {NoStop}%
\bibitem [{\citenamefont {Dykman}\ and\ \citenamefont
  {Ivanov}(1990)}]{Dykman1990h}%
  \BibitemOpen
  \bibfield  {author} {\bibinfo {author} {\bibfnamefont {M.~I.}\ \bibnamefont
  {Dykman}}\ and\ \bibinfo {author} {\bibfnamefont {M.~A.}\ \bibnamefont
  {Ivanov}},\ }\href@noop {} {\bibfield  {journal} {\bibinfo  {journal} {Sov.
  Phys. Solid State}\ }\textbf {\bibinfo {volume} {32}},\ \bibinfo {pages} {87}
  (\bibinfo {year} {1990})}\BibitemShut {NoStop}%
\bibitem [{\citenamefont {Ivanov}\ \emph {et~al.}(1966)\citenamefont {Ivanov},
  \citenamefont {Krivoglaz}, \citenamefont {Mirlin},\ and\ \citenamefont
  {Reshina}}]{Ivanov1966}%
  \BibitemOpen
  \bibfield  {author} {\bibinfo {author} {\bibfnamefont {M.~A.}\ \bibnamefont
  {Ivanov}}, \bibinfo {author} {\bibfnamefont {M.~A.}\ \bibnamefont
  {Krivoglaz}}, \bibinfo {author} {\bibfnamefont {D.~N.}\ \bibnamefont
  {Mirlin}}, \ and\ \bibinfo {author} {\bibfnamefont {I.~I.}\ \bibnamefont
  {Reshina}},\ }\href@noop {} {\bibfield  {journal} {\bibinfo  {journal} {Sov.
  Phys. Solid State}\ }\textbf {\bibinfo {volume} {8}},\ \bibinfo {pages} {150}
  (\bibinfo {year} {1966})}\BibitemShut {NoStop}%
\end{thebibliography}
\end{document}